\documentclass[pre,aps,reprint]{revtex4-1}

 \usepackage[latin1]{inputenc}
 \usepackage[T1]{fontenc}
 \usepackage{lmodern} 
 
\usepackage{graphicx} 

 \usepackage{subfigure}
 \usepackage[english]{babel}
 \usepackage{amsmath}
 \usepackage{color}
 \usepackage{amssymb}
 \usepackage{verbatim}
 \usepackage{psfrag}
 \usepackage{graphicx}
 \usepackage{dcolumn}%
 \usepackage{pstricks}%
 \usepackage{subfigure}%
 \usepackage{pst-node}
 \usepackage{pst-plot}
 \usepackage{pst-circ}
 \usepackage{moredefs}
 \usepackage{natbib}
 \usepackage{bm}%
 
 \DeclareMathOperator*{\sgn}{sgn}
 
\begin{document}

\title{A variational framework for flow optimization using semi-norm constraints}
\author{D.P.G. Foures$^{1}$, C.P. Caulfield$^{2,1}$ and P.J. Schmid$^3$}
\affiliation{$^1$DAMTP, Centre for Mathematical Sciences, Wilberforce Road, Cambridge CB3 0WA, United Kindgom}
\affiliation{$^2$BP Institute, Madingley Rise, Madingley Road,
  Cambridge, CB3 0EZ, United Kingdom}
\affiliation{$^3$LadHyX, Ecole Polytechnique, 91128, Palaiseau cedex France}

\date{\today}

\begin{abstract}
We present a general variational framework designed to consider
constrained optimization and sensitivity analysis of spatially and temporally
evolving flows defined as solutions of  partial differential
equations, 
where the quantity to be optimized
is defined
in terms of a nontrivial semi-norm of the state vector ${\mathbf q}$,
i.e. a functional $f({\mathbf q})$ which satisfies the triangle
inequality and also $f(c {\mathbf q})=|c| f({\mathbf q})$
for scalar $c$, while having a nontrivial null space (or kernel)
 $f({\mathbf q}_0)=0$. We show that optimizing initial perturbations
 which maximize values of such
 a nontrivial semi-norm over a finite time interval requires implicitly that constraints
be placed on the magnitude of complementary semi-norms of initial
perturbations
such that the sum of these complementary semi-norms
defines a  total ``true'' norm of the state vector (i.e. the unique null space
of the true norm is the zero state vector). 
 Therefore, use of
 this framework requires the introduction of new parameters which describe the
 relative magnitude of the initial perturbation state vector calculated
using 
the various partitioning 
constrained complementary semi-norms to
the magnitude calculated using the true norm, even for linear problems.
We demonstrate that any particular required  optimization has to be carried out
fixing these new parameters as initial conditions on the allowable
perturbations, and the influence and significance of the contributions
of each semi-norm
component partitioning the initial total norm of the perturbation can then be considered
quantitatively.

To demonstrate the utility of this framework, we consider
an idealized problem, the (linear) non-modal stability analysis of a
mean flow given by a ``Reynolds averaging'' of the one-dimensional stochastically forced
Burgers equation. We close  the mean flow equation by introducing a
turbulent viscosity to model the turbulent mixing, which we allow to evolve
subject to a new transport equation. Since we 
are interested in optimizing the relative amplification
of the perturbation kinetic energy (i.e. the perturbation's ``gain'') 
this problem naturally requires the use of our new
framework, as the kinetic energy is a semi-norm
of the full state velocity-viscosity vector,
with a new adjustable parameter,  describing the ratio of an
appropriate viscosity   semi-norm to the sum of this
viscosity semi-norm and the kinetic energy semi-norm.
Using this framework,
we demonstrate that
the dynamics of the full system, allowing the turbulent viscosity to evolve
subject to its transport equation, is qualitatively
different from the behaviour when the 
turbulent viscosity is ``frozen'' at a fixed, mean value,
since a new
mechanism of 
perturbation energy production appears, through the coupling of the
evolving turbulent viscosity perturbation and the mean velocity field.

\end{abstract}

\maketitle


\section{Introduction}\label{INT}

Much fluid dynamical research, dating from the pioneering work of Osborne Reynolds \cite{reynolds},
has been focussed on the identification of a critical value of the ``Reynolds number'' of a flow for
the onset of unsteadiness, significant perturbation growth, or indeed the transition to turbulence
of initially laminar flows.  
The classical approach involves linearizing the governing equations around a steady state (also referred to as a ``base flow''), and then investigating the properties  of  the eigenvalues of the corresponding operator. This modal analysis approach yields good agreement with experiments for a  variety of flows (a prime example being  Rayleigh-Bénard convection), but typically
fails for shear flows. Such a modal stability analysis predicts a critical Reynolds number of 5772 for plane Poiseuille flow \cite{orszag} and predicts no (infinitesimal) instability at all for the case of Couette flow \cite{romanov},
although experiments show that the transition to turbulence actually occurs for Reynolds number around 1000 for plane Poiseuille \cite{pois_exp} flow and around 360 for Couette flow \cite{couette_exp}.

The concept of non-modal stability analysis emerged more recently and allowed for a description of the perturbation for intermediate times, instead of focussing on the infinite time interval implicitly considered in a standard modal analysis. Indeed, because of the non-normality of the Navier-Stokes operator \cite{trefethen}, transient growth of the energy is possible for short times, even though all the (normal) modes are exponentially decaying. This phenomenon has been widely studied in shear flows \cite{gustavsson}, \cite{butler}, \cite{reddy}, \cite{schmid}, and it is now well-known that an optimal perturbation can experience transient energy gain (i.e. the ratio of the kinetic
energy at the end ``target time'' of a finite time interval to the 
kinetic
energy at the start time) of several orders of magnitude. Exploiting 
the properties of the underlying linear operator, the gain can be calculated through a singular value decomposition of the evolution operator \cite{schmid}. This very rich linear process could perhaps explain how a linearly stable flow experiences a sufficiently large energy increase for nonlinear
effects to become significant, and 
thus possibly trigger a transition to turbulence \cite{trefethen}.

In order to find this optimal gain, an alternative Lagrangian variational formulation was proposed \cite{hill}, \cite{gunzburger}, allowing more flexibility in the way we describe and consider ``optimal'' perturbations. Indeed, this formulation can take into account non-autonomous operators (for example a time-dependent base flow), nonlinear operators
\cite{pringle}, and non-quadratic measures of the perturbation energy. Moreover the adjoint variables, which are dual variables of state variables, (Lagrange multipliers in the variational formulation imposing 
the requirement that
the state variables satisfy the underlying evolution partial differential equation)
 yield information concerning the sensitivity of the optimized quantity to any input of the problem,
including for example the sensitivity to the chosen
initial conditions, boundary conditions, physical or modeling coefficients, or even the 
flow geometry and chosen base flow.  Therefore, the particular objective functional can be chosen specifically for sensitivity analysis \cite{marquet}, rather than for optimizing  an initial
perturbation over a finite time interval.  Using a variational formulation, it is possible to derive the sensitivity of an eigenvalue (or a singular value) of the system with respect to any variable of the problem, which is a far more efficient way to gain some insight into the impact of a parameter on the dynamics of a flow 
than performing a more time-consuming  finite-difference analysis.

Formally, the conventional problem of investigating perturbation kinetic energy gain 
in simple cases (for example, incompressible constant density fluid flow) is effectively a problem to determine the initial
conditions which maximize the value (at the end of the time interval of interest) of the 2-norm of the state vector. This
has undoubted technical attractions, as the amplitudes of all the components of the state vector are simultaneously
constrained in magnitude within the objective functional of the 
optimization problem when the 
objective functional is the kinetic energy gain. 
At this stage, it is important to remember the three defining
properties of a norm $\|  {\mathbf q}  \|$ acting on the 
state vectors ${\mathbf q}$, members of  a vector space with appropriate differentiability properties
so that the state vectors ${\mathbf q}$ are solutions of the underlying partial differential equation. The first property is ``scalability'',
i.e. for scalar $c$ (in the cases of interest $c$ is a member of the real number field)
\begin{equation}
\|  c{\mathbf q}  \|= |c| \| {\mathbf q} \| ,\label{eq:norm1}
\end{equation}
while the second property is that norms satisfy the triangle inequality, i.e. for two state
vectors ${\mathbf q}_1$ and ${\mathbf q}_2$, 
\begin{equation}
\|  {\mathbf q}_1 + {\mathbf q}_2 \| \leq    \| {\mathbf q}_1  \| +
\|  {\mathbf q}_2  \| .
\label{eq:norm2}
\end{equation}
The third property is the key property that ensures that the amplitudes of all components
of the state vector are constrained, i.e.
\begin{equation}
\|  {\mathbf q}  \| = 0 \Leftrightarrow {\mathbf q} = {\mathbf 0} ,
\label{eq:norm3}
\end{equation}
and so by definition the null space (or kernel) of a norm on a vector space has
a unique element, the zero vector of that vector space. Although it is perhaps tautological,
for clarity we will refer to such a functional as a ``true'' norm.

We draw this extra distinction since there are many physical circumstances of undoubted fluid-dynamical interest where
the natural objective functional is not a true norm but is 
actually defined in terms of a (nontrivial) ``semi-norm'' on the state vector space.
Our qualification of ``nontriviality'' means that the
semi-norm is a functional of the state vector which has the first
two properties (\ref{eq:norm1})-(\ref{eq:norm2}) of a true norm
but categorically not the third, and so the null space  or kernel of a nontrivial
semi-norm has strictly more
than one element. (Within our nomenclature, 
a ``true'' norm is thus a ``trivial'' semi-norm.)

Two simple examples of optimization problems where the 
objective functional is  a semi-norm are where there is a partitioning in space, and 
we are interested in maximizing the perturbation energy growth strictly in
a subregion of the flow domain, and partitioning of the state vector, where
we are interested in maximizing the gain of some (but strictly not all) components of the state vector.
The former example might arise in an industrial context, where
we might be interested in maximizing perturbation growth in the immediate
vicinity of an injector, while the latter
example might arise in situations where the density of the fluid
is not constant (due to compositional, thermal or compressible
effects) and so the state vector does not
exclusively involve the flow velocity components, but 
also involves the density field. We might be interested in maximizing
over a finite time interval the gain in
the kinetic energy or the potential energy of a perturbation in a stratified yet incompressible flow,
or alternatively maximizing the acoustic energy in a compressible flow, each
of which would mean that the objective functional is most naturally defined 
in terms of a semi-norm of the state vector. 
For such classes of problems, we are then faced with the challenge of identifying 
a way in which to constrain the elements
of the state vector which are in the kernel (i.e. the null space) of the semi-norm defining
the objective functional. A central aim of this paper is to present 
an algorithmic framework to address this challenge. The key idea
is to impose ``complementary semi-norm'' constraints
on the allowable initial conditions for the state vector.

As explained more precisely below in section (\ref{sec:VAR}), 
the  ``complementary semi-norms'' are defined so that they 
have two useful properties. Firstly, some
set of them must appropriately constrain
the amplitudes
of state
vectors in the kernel of the objective functional.
Secondly, the kernels
of all the semi-norms
are distinct (except for the zero 
state vector)
such that their direct sum
must completely span the state vector
space. This latter property effectively
means that the initial constraints
imposed by the complementary semi-norms
can be imposed independently.
Therefore,   
the relative importance
of the dynamics
associated with the initial values of these complementary semi-norms 
and the initial value of the objective 
functional itself can be investigated
in a self-consistent and clear manner by considering
parameters quantifying the relative size of these initial values.

A particular attraction of the proposed framework is its flexibility,
allowing the problems which are considered to extend beyond the obvious (at best weighted) 2-norm
of the state vector \cite{guegan}. Although we present our framework in a quite
general fashion, we also demonstrate its utility by considering a simple idealized fluid dynamical problem
considering parameterized turbulence flow modeling 
where use of a framework such as this is necessary
to yield the correct results 
for the natural perturbation kinetic energy gain optimization problem.

A classical approach to parameterized turbulence flow modeling has been to use the averaging method first proposed by Reynolds,
which  leads to the set of equations now commonly referred to as the Reynolds Averaged Navier-Stokes (RANS) equations. Naturally, due to the quadratic nonlinearity of the advection term in the 
underlying Navier-Stokes equations, a turbulence
``closure'' is required to close the system of equations, and one of the simplest (and most commonly used)
closures is
to assume that the (second-order in velocity) Reynolds stress tensor
can be related to the (first-order) mean stress tensor through an (in general) temporally
and spatially varying coefficient, the ``eddy'' or ``turbulent viscosity''. 
Of course, such a closure naturally leads
to the further question of how the turbulent viscosity should be modelled, and in particular
if it is allowed to vary in space and time, one or indeed several
extra empirical equations may be required to describe the physical processes acting on this new quantity. 

Recently, turbulence modeling techniques have been applied in various
stability problems, and it appears that stability analysis of a Reynolds-averaged set of equations, coupled with an appropriate simple  turbulence model, can be successful in predicting the onset of instabilities affecting mean flows \cite{crouch},
allowing the  appropriate description of large-scale (compared to the turbulence length scale) instability processes in such turbulent flows. In the more general case of transient temporal perturbation growth, due to the non-normality of the underlying linearized Navier-Stokes operator, some research has been conducted on the turbulent boundary layer, assuming a RANS base flow, yet critically perturbations in the velocity and pressure variable only, while fixing the turbulent viscosity at
a constant value throughout the flow evolution (the so-called frozen turbulent viscosity approach,  see \cite{cossu} for more details). Crucially, however,  the influence of the closure on the actual flow evolution is still largely unknown and in particular the robustness of the results
to relaxing the  frozen turbulent viscosity assumption is an open question. 

If the turbulent viscosity is rather allowed to vary spatially and temporally (subject to an appropriately
constructed evolution equation) then the state vector of the 
system formally involves not only
the velocity components, but also the turbulent viscosity. Therefore, even if the problem
of interest is the conventional problem  of maximization of the perturbation kinetic
energy gain over some finite time, the objective functional
for the optimization problem naturally becomes 
a semi-norm of the state vector, and so we obtain
a  relatively simple example of the type of
optimization problem for which we have developed our generalized framework.
In this problem, we have to impose a constraint (using a complementary semi-norm)
on the initial magnitude of the turbulent viscosity, and so this problem has (in a very simple way) the central characteristics
of interest illustrating the utility of our framework.

 Indeed, we  wish to consider an extremely simple one-dimensional problem which nevertheless contains  the salient features of turbulence: time dependence, nonlinearity, enhanced diffusivity and stochastic forcing. An appropriate choice is the stochastically forced Burgers equation. This equation is a good one-dimensional analogue  of the Navier-Stokes equations, where the analogue of ``turbulence'' is artificially introduced by a (stochastic)  forcing term. Furthermore, it has  been shown that there exists an equivalence between the Kuramoto-Sivashinsky equation (which is one of the more famous one-dimensional  turbulence model equations \cite{kuramoto}) and such a stochastically forced Burgers equation \cite{zaleski}, suggesting that this is an appropriate model system to consider. 

Therefore, the rest of this paper 
is organised as follows. In section (\ref{sec:VAR}), we describe our variational framework involving the required
use of complementary
semi-norm constraints in some generality. In section (\ref{sec:RAB}), we then demonstrate the application of 
this framework to the model problem described above. Specifically, we  
derive the Reynolds-Averaged-Burgers (RAB) equations and apply a turbulent viscosity closure with an evolution equation for the turbulent  viscosity, based on the well-known Spalart-Allmaras turbulence model \cite{spalart}. We will then consider the problem of the identification of ``optimal'' perturbations (where optimality is defined in various ways) as an example to show the potential usefulness of our variational framework not only for identification of optimal initial conditions but also for sensitivity analysis \cite{marquet},\cite{brandt}. In section (\ref{sec:RES}) we present our results, focussing
in particular on demonstrating the flexibility (and superiority when compared to other methods) of this framework for 
considering different objective functionals
to optimize when there is no ``natural'' choice of an objective functional corresponding to a ``true'' norm
of the state vector space. 
In section (\ref{sec:EXT}), we 
briefly
discuss other potential fluid-dynamical applications of this framework, 
and 
finally, in section (\ref{sec:CON}), we draw
our conclusions.


\section{Variational framework}\label{sec:VAR}
	
	\subsection{Governing equations}

	We consider  an arbitrary state vector $\mathbf{q}$  from a vector space $\Omega$, defined on the time interval $[0,T]$. We  choose $\mathbf{q}\in H^2(\Omega)$, a Sobolev space of order 2. We choose this space so that $\mathbf{q}$ and its gradient on $\Omega$ are both in $L^2(\Omega)$ (space of square integrable functions on $\Omega$), which 
means that the state
vectors are appropriately
well-behaved for the types
of differential operations we wish to consider.
We now consider a hierarchy of constraints which we wish to impose upon $\mathbf{q}$.
The first constraint is that we wish $\mathbf{q}$  to satisfy a partial differential equation, the most general
form of which is
\begin{equation}
\partial_t \mathbf{q} - \mathcal{N}(\mathbf{q}) = \mathbf{q}_f,\label{eq:nl}
\end{equation}
where $\mathcal{N}$ is a nonlinear operator acting on the variable $\mathbf{q}$, and $\mathbf{q}_f$ a forcing term.
In this section for simplicity and clarity, we will however focus on linear homogeneous equations, although it is
important to stress that this framework can be applied straightforwardly to the case of forced and/or nonlinear equations.
In this simpler case,  (\ref{eq:nl}) reduces to
\begin{equation}\label{eq:GENERAL_EQ}
\partial_t \mathbf{q} - \mathbf{L}\mathbf{q} = 0,
\end{equation}
where $\mathbf{L}$ is a linear operator. 
For a well-posed problem, we must of course impose (as constraints) initial conditions
\begin{equation}\label{eq:IC_CONSTR}
\mathbf{q}(\mathbf{x},0)-\mathbf{q}_0 = 0 \mbox{\ \ ,\ \ } \forall \ \mathbf{x} \in \Omega,
\end{equation}
and  boundary conditions defined on $\partial\Omega$, for all $t$
\begin{equation}\label{eq:IC_BC}
\mathbf{q}(\mathbf{x},t) - \mathbf{q}_{\partial\Omega} = 0  \mbox{\ \ ,\ \ } \forall \ (\mathbf{x},t) \in \partial\Omega \times [0,T].
\end{equation}
Once again, for clarity, we restrict ourselves to Dirichlet boundary conditions, although Neumann boundary conditions can be treated in the same way in the following framework,
subject to conventional
consistency conditions (for example associated
with the divergence theorem) being satisfied.

\subsection{Objective functional and semi-norm considerations}
	
Depending on the particular problem studied, we will define ``the objective functional'' $\mathcal{J}(\mathbf{q})$ which
takes as its input the state vector $\mathbf{q}$. The particular functional form of $\mathcal{J(\mathbf{q})}$ is unique to any problem and can be of many different forms corresponding to some physical quantity of interest. Obvious
examples include (some measure of) the flow's energy, enstrophy, drag, or mixing efficiency. In all cases,
the functional $\mathcal{J}$ outputs  a real number, which we may
want to optimize, or alternatively we
may wish to 
investigate
the sensitivity of that
output real number to
small 
variations of some parameters of the problem. 
Variational frameworks of this form have conventionally been used to optimize the energy growth
over some finite time interval by identifying an optimal initial condition 
for the state vector, which can also be identified (for linear operators) by 
considering a singular value decomposition (see \cite{schmid} for further details).

However, a variational
framework is much more flexible, and there is no formal requirement to restrict
attention to optimization of perturbation energy gain. Indeed, the  objective functional  can describe the receptivity of a system to an external forcing, the sensitivity of the least stable eigenvalue (in the case of a linearized equation only) with respect to parameters or to a base flow modification (\cite{marquet}, \cite{brandt}), and more generally, any (real) quantity derived from the state vector $\mathbf{q}$. Another
specific interesting application of  a variational framework is data-assimilation and consists of minimizing (ideally of
course reducing to zero) the difference between a calculated state vector solution
of the underlying partial differential equation and a (measured) target vector, and thus to identify ``optimal'' choices for coefficients or parameter-functions within the governing equation \cite{data-assimilation}.
	
Although the goal of this particular section is to present a variational framework in as general a fashion as possible,  actual calculations cannot be carried out without specifying the kind of problem we are considering, because the objective functional as well as the various constraints we will consider naturally change depending on the chosen problem.
In order to demonstrate the framework, we will therefore focus on the case of the identification of optimal perturbations, i.e. finding the optimal initial condition $\mathbf{q}_0$ which maximizes (the output of) an objective functional. Such an optimal perturbation is sometimes referred as the most dangerous perturbation (in the sense of what is optimized). It is important to note that we will use true norms or
semi-norms for the objective functional, but in general, the positive definiteness is not required, and any functional can be used. We will in this paper consider the following generic objective functional:
\begin{equation}\label{eq:general_norm}
\mathcal{J}(\mathbf{q}) = \left\| \mathbf{q}(T) \right\|_O^2,
\end{equation}
which defines a quantity of interest given by an objective (in general semi-) norm at the target time $T$ (without loss of generality we will always assume
that the time interval for optimization starts at $t=0$ and so the target time is $T$ and the time interval for optimization is also $T$). We stress again that this objective functional is not uniquely defined, and the norm (or semi-norm) can be  changed depending on the specific problem being considered. For example, the objective functional often describes the kinetic energy of a perturbation evolving around a base flow state. However, it can also describe the total energy, summing the kinetic energy and some form of potential energy. For example the internal energy in a gas or a fluid can be quantified as a function of the temperature of the system \cite{hanifi}, the potential energy density in a stratified fluid can
be straightforwardly calculated from the density distribution, and the electrostatic energy due to the presence of an electrostatic field (\cite{Castellanos}, \cite{fulvio}) or magnetic energy associated with magnetic field \cite{chen} can similarly be evaluated in space and time.

These are only a few examples of the other types of  ``energies'' which  can be defined, and indeed  the objective functional does not have to be  a conventional energy of the physical system under consideration. For example, to find the energy threshold leading to a turbulent state in a Couette flow configuration, an objective functional defined as the time and space average of the viscous dissipation has been used successfully in \cite{monokrousos}.

Of particular interest are problems where the objective functional is actually defined in terms of a ``semi-norm'', as
discussed in detail in the introduction. Such objective functionals
naturally arise when we are interested only in some partitioning of the state vector, either in space
where
we are interested in optimizing the energy in some compact set of the domain, or in terms of
components of the state vector where (for example) we are interested
in only some part of the total energy of the system.
As noted in the introduction, a (nontrivial) semi-norm has a nontrivial null space or kernel, defined for the particular
vector space which we are considering  as the set of state
vectors $\mathbf{q}$ such that the semi-norm
$\left\|\cdot\right\|$ returns a zero value, i.e.
\begin{equation}
\ker\left( \left\| \cdot \right\| \right)= \left\{ \mathbf{q} \in H^2(\Omega) \mbox{\ } ; \mbox{\ } \left\| \mathbf{q} \right\| = 0 \right\}.
\end{equation}
For a ``true'' norm the kernel is trivial, containing only the zero state vector. We then define the complementary space to this kernel (henceforth referred to as the ``cokernel'') as:
\begin{equation}
\begin{array}{ll}
\ker^*\left( \left\| \cdot \right\| \right) & = H^2(\Omega) \backslash \ker\left( \left\| \cdot \right\| \right) \\
																 & = \left\{ \mathbf{q} \in H^2(\Omega) \mbox{\ } ; \mbox{\ } \left\| \mathbf{q} \right\| \neq 0 \right\}.
\end{array}																 
\end{equation}
For reasons of convenience, we also add to the cokernel, the null vector $\mathbf{0}$ such that we have the property:
\begin{equation}
\ker\left( \left\| \cdot \right\| \right) \oplus \ker^*\left( \left\| \cdot \right\| \right) = H^2(\Omega),
\end{equation}
for any (semi-) norm, where $\oplus$ stands for the space direct sum which has for definition for three arbitrary ensembles $A$, $B$ and $C$:
\begin{equation}
A\oplus B = C \Leftrightarrow
\left\{
\begin{array}{ll}
 A  +  B &= C,\\
A \cap B &= \{\mathbf{0}\},
\end{array}
\right.
\label{eq:direct_sum_def}
\end{equation}
with $\{\mathbf{0}\}$ the appropriate zero state vector. The cokernel is thus in fact the restriction of the space $H^2(\Omega)$ for which the semi-norm $\left\| \cdot \right\|$ (on $H^2(\Omega)$) becomes a norm.

As we discuss in the following subsection, optimization of gain defined by such a nontrivial semi-norm requires a special treatment of further constraints.
In order to address the development of a variational framework where the objective functional
may potentially use a semi-norm, we define a particularly simple expression capable of describing both norms and semi-norms. Our objective (in general semi-) norm is then defined as:
\begin{equation}
\left\| \mathbf{q}(T) \right\|_O^2 = \dfrac{1}{2} \displaystyle\int_{\Omega}{\mathbf{q}(T)^H \mathbf{W}_O \mathbf{q}(T)\ d\Omega},
\label{eq:enorm}
\end{equation}
where the superscript $^H$ denotes the transpose conjugate and the matrix $\mathbf{W}_O$ is a weight matrix. If $\mathbf{W}_O$ is singular (non invertible) then $\left\| \mathbf{q}(T) \right\|_O$ is a (nontrivial) semi-norm, while if $\mathbf{W}_O$ is invertible then this expression defines a (true) norm. This weight matrix can in general be a function of position, and so one obvious
way in which it can be non-invertible is if it is non-zero in only a compact sub-region of the flow domain (i.e. space partitioning). Another obvious way 
in which it may be non-invertible is if $\mathbf{W}_O$ is nonzero
only in a block, so that certain components of the state vector (for example the fluid's density, or as we shall see below, a spatially and temporally varying eddy or turbulent viscosity) do not have any effect on the value of the ``energy'' norm (i.e. state partitioning).
This class of parameterized (through the weight  matrix $\mathbf{W}_O$) quadratic norms will be the only one considered in this paper.
However,  a more general semi-norm could take into account the total time-evolving flow \cite{monokrousos}, i.e., the full space-time evolution of the state vector and be for example the evaluation of (at least some component of) the energy integrated over space and time. Moreover, we are not in general constrained to choose a quadratic norm.

	\subsection{Lagrangian framework using constraints}\label{sec:VAR_lagrangian}
	
	A sensitivity analysis identifies  the impact of a small variation of an input of the optimization problem on the value of the objective functional, and so in a particularly natural way, a Lagrangian variational framework enables  the performance of a sensitivity analysis 
subject to  constraints. 
Indeed, the Lagrangian framework allows us to add as many dimensions to the problem as we have constraints. By adding these extra degrees of freedom, we are then able to investigate the impact of variations of the constraints on the returned value of the objective functional. As a consequence, if the variables whose magnitude we wish to optimize (or to consider within a sensitivity analysis)  are part of the formulation of the constraints acting on the system, we then have to embed them in an augmented functional which takes into account the objective functional and the constraints at the same time. In other words, when we allow the constraints to vary, we have to include them in the augmented functional (i.e. the Lagrangian) of the problem, in order to retrieve the sensitivity information.

In many situations, we are interested in optimizing a given quantity (for example the initial condition, the external forcing or the boundary conditions) which will have an impact on the space-time evolution of the state vector $\mathbf{q}$, and as a consequence the objective functional $\mathcal{J}$ not only depends implicitly on the optimized quantity, but also inevitably
on the full space-time evolution of the state vector $\mathbf{q}$. Therefore, as already noted, 
the  first constraint which we must impose  is that the state vector must satisfy the evolution equation (\ref{eq:GENERAL_EQ}). 
Then, depending on the problem we are solving, different constraints must be imposed. 
In general, a correct, well-posed problem statement involves appropriate boundary conditions; although it
is of course possible to optimize with respect to such boundary conditions, in an entirely equivalent way 
to optimizing with respect to initial conditions (see \cite{bewley} for a fuller discussion), for clarity in this paper we opt to restrict our attention to problems
where the boundary conditions are chosen conveniently and appropriately
to not enter explicitly into the variational problem of interest.
Rather we wish to focus on identifying optimal perturbations, and so initial conditions play a central role, so that we add an appropriate (and essentially self-evident) initial condition constraint (\ref{eq:IC_CONSTR}). Furthermore, in order to avoid the final state vector amplitude becoming arbitrarily large during the optimization process, we have to impose
a normalization (and hence scalar) constraint on the initial condition,
i.e.
\begin{equation}
\begin{array}{c}
\left\| \mathbf{q}_0 \right\|_N^2 - N_0 = 0,\\
\left\| \mathbf{q}_0 \right\|_N^2 = \dfrac{1}{2} \displaystyle\int_{\Omega}{\mathbf{q}_0^H \mathbf{W}_N \mathbf{q}_0\ d\Omega} ,
\end{array}\label{eq:nnorm}
\end{equation}
where the subscript $N$ stands for normalization and emphasizes the fact that this (true) norm is used for a normalization purpose. This (true) norm is  defined in an analogous way to  $\|\cdot\|_{O}$  defined in (\ref{eq:enorm}) but is in general defined by a weight matrix $\mathbf{W}_N$ different from the energy weight matrix $\mathbf{W}_O$.
In particular, since we wish all possible state vectors to be constrained, we require $\mathbf{W}_N$ to be non-singular, and so  
the (true) norm used for the normalization of the initial condition can be different from the objective norm (or semi-norm) used to define the optimized quantity $\mathcal{J}$. Indeed, in general, we can optimize the value of a certain semi-norm, given that the initial perturbation is normalized with respect to a different (true) norm.

In the specific case where we wish to optimize
a ``gain''
 (i.e. the ratio of final to initial objective value), we need to optimize (value at the final time) and constrain (value at the initial time) the same quantity. It is therefore natural to choose the same norm for optimization and normalization and so $\mathbf{W}_O = \mathbf{W}_N$, such that the normalization constraint is simply:
\begin{equation}\label{eq:eq0}
\left\| \mathbf{q}_0 \right\|_O^2 - O_0 = 0,
\end{equation}
with $O_0$ describing an initial value of the objective functional.
The gain in the objective functional is then straightforwardly defined as:
\begin{equation}
G_O=\dfrac{\left\|\mathbf{q}(T) \right\|_O}{O_0}. 
\label{eq:gain_o}
\end{equation}
Since $\left\|\mathbf{q}(T) \right\|_O$ is the optimized quantity and $O_0$ is fixed, at the end of the optimization process, the gain found will be optimal.

In general, the normalization constraint has to act on the totality of the state vector $\mathbf{q}$ in order to have a well-posed optimization problem.
In particular, imposing constraint  (\ref{eq:eq0}) with a singular matrix $\mathbf{W}_O$ is not an appropriate constraint, as this constraint will not affect any vector which is part of the kernel of the semi-norm involved in the definition of the objective functional, and will as a consequence remain unbounded. Although an optimization, investigating the optimal final state objective value $\left\|\mathbf{q}(T)\right\|_O$ under a semi-norm constraint defining the initial semi-norm of the state vector  $O_0$ can still be conducted, and an optimal $\mathbf{q}_0$ can in principle be identified,  it is very likely that the objective functional will diverge with the magnitude of the non-constrained part of the state vector, a typically  undesired and unrealistic behaviour.

As a consequence, in the more general situation (for which we wish to construct a framework) where we want to constrain the state vector with the help of a semi-norm, to define a well-posed problem we have to add (at least) a further constraint on the part of the state vector which is in the kernel of the semi-norm. 

From now on, the semi-norm which we wish to impose as a constraint 
will be denoted with a subscript 
$K_0$.
A natural way to do this is to appreciate that there is some flexibility in the construction of the constraint, and especially that we can have several normalization constraints.
Therefore, to be able to constrain the semi-norm $\left\| \cdot \right\|_{K_0}$ of interest and the magnitude of all possible state vectors at the same time, we are then led to the necessity of (at least) a second initial condition constraint beyond the normalization through $\left\| \cdot \right\|_{K_0}$ in order to impose an appropriate constraint equivalent to (\ref{eq:nnorm}). A very convenient way to do this is through 
defining a set of  ``complementary semi-norms'' 
$\left\| \cdot \right\|_{K_{i}}$
\begin{equation}\label{eq:semi_norm_constraints}
\left\{
\begin{array}{l}
\left\|\mathbf{q}_0\right\|_{K_0}^2 - K_{00} = 0,\\
\left\|\mathbf{q}_0\right\|_{K_1}^2 - K_{10} = 0,
\end{array}
\right.
\end{equation}
where $\left\|\cdot\right\|_{K_1}$ is constructed such that the norms $\left\|\cdot\right\|_{K_0}$ and $\left\|\cdot\right\|_{K_1}$ are ``complementary''. 
In this context, we wish to refer to a set of semi-norms as being complementary
if the direct sum of the cokernels of these two semi-norms is the entire state vector space (\ref{eq:direct_sum_def}), i.e.
\begin{equation}\label{eq:direct_sum}
\ker^*\left(\left\| . \right\|_{K_0}\right)  \oplus   \ker^*\left(\left\| . \right\|_{K_1}\right) = H^2(\Omega).
\end{equation}

By construction, this complementary semi-norm $\left\| \cdot \right\|_{K_1}$ constrains the initial magnitude of the state vectors in the kernel of the semi-norm $\left\| \cdot \right\|_{K_0}$ and vice versa, such that the full space $H^2(\Omega)$ is constrained without any interference between the two normalizations.
Therefore, for a general state vector $\mathbf{q}$, we define the total normalization norm through $\left\| \cdot \right\|_N$
\begin{equation}
\left\| \mathbf{q}\right\|_N^2 =
\left\| \mathbf{q}\right\|_{K_0}^2 + 
\left\| \mathbf{q}\right\|_{K_1}^2.
\label{eq:normsum}
\end{equation}
This is clearly a straightforward construction when the first semi-norm constraint considers only a compact subregion of the flow domain (i.e. when the problem is partitioned in space) or partitions the state vector by its components (e.g. when $\left\| \cdot \right\|_{K_0}$ only considers the kinetic energy of a stratified flow).
Therefore, we can define the initial (true) norm value as
\begin{equation}
N_0=K_{00}+K_{10}.
\end{equation}

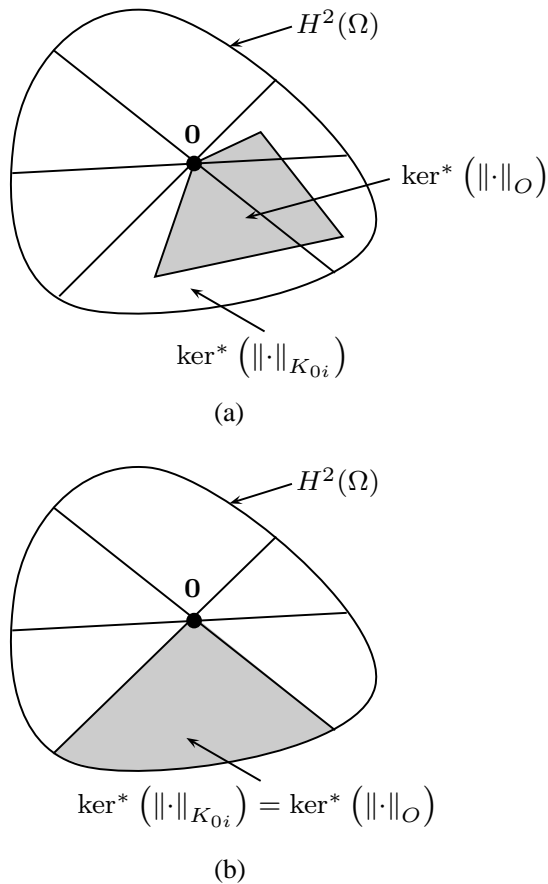
\begin{figure}\label{fig:semi_norm_cases}
\hspace*{-2.5cm}
\scalebox{1.3} 
{
\begin{pspicture}(0,-4.5642185)(10.707188,4.534219)
\definecolor{color67b}{rgb}{0.8,0.8,0.8}
\pspolygon[linewidth=0.02,fillstyle=solid,fillcolor=color67b](5.343125,3.2042189)(6.183125,2.1342187)(4.263125,1.7242187)(4.663125,2.8842187)
\psbezier[linewidth=0.02](4.198817,4.4517927)(4.9221296,4.3793674)(6.6831207,2.910635)(6.513418,2.2065928)(6.3437157,1.5025504)(4.142108,1.2047006)(3.4755037,1.4099088)(2.8088994,1.615117)(2.7236068,2.4309537)(2.824522,3.1481283)(2.9254375,3.8653028)(3.4755037,4.5242186)(4.198817,4.4517927)
\psline[linewidth=0.02cm](2.803125,2.7842188)(6.223125,2.9642189)
\psdots[dotsize=0.16](4.663125,2.8842187)
\usefont{T1}{ptm}{m}{n}
\rput(4.646719,3.1792188){\footnotesize $\mathbf{0}$}
\psline[linewidth=0.02cm](3.223125,4.0442185)(6.103125,1.7642188)
\psline[linewidth=0.02cm](3.283125,1.5242188)(5.503125,3.7442188)
\psline[linewidth=0.02cm,arrowsize=0.05291667cm 2.0,arrowlength=1.4,arrowinset=0.4]{<-}(4.623125,1.6042187)(5.383125,1.1642188)
\usefont{T1}{ptm}{m}{n}
\rput(5.366719,0.89921874){\footnotesize $\ker^*\left(\left\|\cdot\right\|_{K_{0i}}\right)$}
\psline[linewidth=0.02cm,arrowsize=0.05291667cm 2.0,arrowlength=1.4,arrowinset=0.4]{<-}(5.183125,2.3242188)(6.663125,2.7242188)
\usefont{T1}{ptm}{m}{n}
\rput(7.5467186,2.7392187){\footnotesize $\ker^*\left(\left\|\cdot\right\|_{O}\right)$}
\psline[linewidth=0.02cm](2.803125,-1.8957813)(6.223125,-1.7157812)
\usefont{T1}{ptm}{m}{n}
\rput(4.646719,-1.4607812){\footnotesize $\mathbf{0}$}
\pspolygon[linewidth=0.0020,linecolor=color67b,fillstyle=solid,fillcolor=color67b](4.663125,-1.7557813)(6.103125,-2.9157813)(4.503125,-3.3157814)(3.223125,-3.1557813)
\psdots[dotsize=0.16](4.663125,-1.7957813)
\psline[linewidth=0.02cm](3.223125,-0.6357812)(6.103125,-2.9157813)
\psline[linewidth=0.02cm](3.223125,-3.1557813)(5.503125,-0.93578124)
\psbezier[linewidth=0.02,linecolor=color67b,fillstyle=solid,fillcolor=color67b](3.223125,-3.1557813)(3.823125,-3.3957813)(4.023125,-3.3157814)(4.343125,-3.3157814)(4.663125,-3.3157814)(5.423125,-3.1957812)(6.063125,-2.9157813)
\psbezier[linewidth=0.02](4.198817,-0.22820707)(4.9221296,-0.30063286)(6.6831207,-1.769365)(6.513418,-2.4734073)(6.3437157,-3.1774497)(4.142108,-3.4752994)(3.4755037,-3.2700913)(2.8088994,-3.064883)(2.7236068,-2.249046)(2.824522,-1.5318717)(2.9254375,-0.81469715)(3.4755037,-0.15578125)(4.198817,-0.22820707)
\psline[linewidth=0.02cm,arrowsize=0.05291667cm 2.0,arrowlength=1.4,arrowinset=0.4]{<-}(4.583125,-2.9957812)(5.343125,-3.4357812)
\usefont{T1}{ptm}{m}{n}
\rput(5.3167186,-3.7007813){\footnotesize $\ker^*\left(\left\|\cdot\right\|_{K_{0i}}\right) = \ker^*\left(\left\|\cdot\right\|_{O}\right)$}
\psline[linewidth=0.02cm,arrowsize=0.05291667cm 2.0,arrowlength=1.4,arrowinset=0.4]{<-}(5.023125,4.084219)(5.663125,4.284219)
\usefont{T1}{ptm}{m}{n}
\rput(6.1367188,4.2992187){\footnotesize $H^2(\Omega)$}
\psline[linewidth=0.02cm,arrowsize=0.05291667cm 2.0,arrowlength=1.4,arrowinset=0.4]{<-}(5.023125,-0.59578127)(5.663125,-0.39578125)
\usefont{T1}{ptm}{m}{n}
\rput(6.1367188,-0.38078126){\footnotesize $H^2(\Omega)$}
\usefont{T1}{ptm}{m}{n}
\rput(5.023281,0.31921875){\footnotesize (a)}
\usefont{T1}{ptm}{m}{n}
\rput(5.0332813,-4.360781){\footnotesize (b)}
\end{pspicture} 
}
\caption{Schematic representation of the partition of the space $H^2(\Omega)$ through the choice of $n_c+1$ (here $n_c=5$) semi-norm constraints. (a) Schematic representation of the generic case of final value optimization, 
where the objective semi-norm is different from all the (initial) constraint semi-norms. 
(b) 
Schematic representation of the special case of gain optimization, where the objective semi-norm coincides with one of the constraint semi-norms (and as a consequence, so do their cokernels).}
\label{fig:blob}
\end{figure}

Consequently, a new (adjustable) parameter arises which quantifies the relative size of the initial magnitude given by the energy semi-norm to the total normalization norm i.e. 
\begin{equation}
R_0=\frac{K_{00}}{N_0}=\frac{K_{00}}{K_{00}+K_{10}}=1-\frac{K_{10}}{N_0}. \label{eq:r0def}
\end{equation}
In order actually to find the optimal perturbation, we also have to optimize with respect to this parameter (and not with respect to the total norm $N_0$ since the problem is linear). Indeed, this framework offers the possibility to perform a multi-scale stability analysis where the initial amplitude of the perturbation is different in each component of the state vector. Optimizing on the parameter $R_0$ would then maximize the corresponding objective functional. However, in some cases, the ratio $R_0$ will be fixed physically or be an input if one wants to investigate a certain case. For example in the case where we want to constrain the initial condition to lie only within a compact subregion of the domain, we would enforce the initial condition on this subregion and on the complementary subregion independently with semi-norms weighted by spatial (mask) functions, and would set the ratio $R_0$ to be zero which forces the initial condition to be completely
free of any component of the kernel of the semi-norm $\left\|\cdot\right\|_{K_0}$, and ensure the initial localization of the perturbation in the desired subregion.

The situation is somewhat more straightforward if the problem of interest is one of
optimization of a gain defined by a semi-norm.
In this particular case, the optimization semi-norm has to coincide with one of the constraint semi-norms, and so $\left\| \cdot \right\|_{K_0} \equiv \left\| \cdot \right\|_O$ and so we may write the complementary semi-norm as $\left\| \cdot \right\|_K$. The associated semi-norm initial values are denoted $O_0$ and $K_0$, and the corresponding full-norm initial value $N_0$ is still the sum of these two values. The new single parameter arising is defined in the same way as in (\ref{eq:r0def}):
\begin{equation}
R_0=\dfrac{K_0}{N_0}=\dfrac{K_0}{O_0+K_0}=1-\dfrac{O_0}{N_0}. \label{eq:r0def_gain}
\end{equation}
The gain is then defined in the exact same way as in (\ref{eq:gain_o}) as the ratio of the final value of the objective functional norm $\left\| \mathbf{q}(T) \right\|_O$ to its initial value $O_0$.

More generally, if there are other (multiple) physically motivated constraints we wish to impose upon the problem (for example by requiring the initial conditions to have specific magnitudes in different subregions of the flow domain) we can impose a larger complete yet complementary set of initial constraints:
\begin{equation}
\left\{
\begin{array}{l}
\left\|\mathbf{q}_0\right\|_{K_i}^2 - K_{0_i} = 0,\mbox{\ \ \ for $i$ from $0$ to $n_c$},\\
\displaystyle\bigoplus_{i=0}^{n_c}{\ker^*\left(\left\| . \right\|_{K_i}\right)}= H^2(\Omega),
\end{array}
\right.
\end{equation}
where the symbol $\oplus$ denotes the direct sum (explicitly written in (\ref{eq:direct_sum}) and where the number of  complementary constraints is $n_c+1$,
with implicitly $n_c+1$ different complementary semi-norms 
which satisfy
\begin{equation} 
\sum^{n_c}_{i=0}
 \left \| \mathbf{q}_0 \right \| ^2_{K_i} 
=\sum^{n_c}_{i=0} K_{0_i}
=\left \| \mathbf{q}_0 \right \| ^2_{N} 
= N_0. 
\end{equation} 
The number of new parameters to optimize over is $n_c$ (because the system is linear) and can be defined as (generalizing (\ref{eq:r0def}))   
\begin{equation}
R_{0_i}=\frac{K_{0_i}}{N_0}, \
i=1 \dots n_c,
\label{eq:ratios}
\end{equation}
and we will retain this general form for the constraints to construct our general framework. 
This general situation is shown in figure (\ref{fig:blob}a), making it explicit that the semi-norm
used to define the objective functional does not need to correspond to any of these constraint 
semi-norms.

As before, the situation is simpler if the problem of interest corresponds to a problem where we wish to optimize a gain, because 
then one of the constraint semi-norms has to coincide with the objective semi-norm, and so without loss
of generality, we define $\left\|\cdot\right\|_{K_0}=\left\|\cdot\right\|_O$ (see figure (\ref{fig:blob}b)). We have decided to express the objective functional with a norm denoted with a subscript $O$ for ``objective'' to highlight that final energy, or energy gain optimization is only a single possibility allowed by this framework. 
In an energy gain optimization case, we choose to write $\left\|\cdot\right\|_O \equiv \left\|\cdot\right\|_E$ and $O_0 \equiv E_0$. If optimized, the gain will then be an energy gain and denoted
\begin{equation}
G_E(T)=\dfrac{\left\|\mathbf{q}(T)\right\|_E}{ E_0}.
\end{equation}

We are now able to express the appropriate
Lagrangian functional for our optimization problem embedding the constraints, provided we define the different scalar products we will need to use. We will use three different scalar products in this study: one related to  space; one to time and one to both space and time.
Respectively, these scalar products are
\begin{equation}
\begin{array}{c}
\displaystyle<f,g>(t)=\int_\Omega{f^H g\ d\Omega},\\
\displaystyle[f,g](x)=\int_0^T{f^H g\ dt},\\
\displaystyle(f,g)=\int_\Omega\int_0^T{f^H g\ d\Omega dt}.
\end{array}
\end{equation}
Using these definitions, the augmented Lagrangian functional for our optimal perturbation problem can now be written in a rather general way:
\begin{equation}
\begin{split}
\mathcal{L}(\mathbf{q},\mathbf{q}_0,\mathbf{q}^\dagger  ,\mathbf{q}_0^\dagger,\lambda_i) & = \left\| \mathbf{q}(T) \right\|_O^2
\\
																										 & - \left(\mathbf{q}^\dagger,\partial_t \mathbf{q} - \mathbf{L}\mathbf{q}\right) \\
																										 & -  \left< \mathbf{q}_0^\dagger, \mathbf{q}(0) - \mathbf{q}_0 \right>\\
																										 & - \displaystyle\sum_{i=0}^{n_c}{\lambda_i \left(\left\| \mathbf{q}_0 \right\|_{K_i}^2 - K_{0_i} \right)},
 \end{split}
\end{equation}
where the objective functional $\mathcal{J}(\mathbf{q})=\left\| \mathbf{q}(T) \right\|_O^2$ consistently with (\ref{eq:general_norm}).

\subsection{Optimality conditions}

We wish to find an extremum of the augmented Lagrangian functional $\mathcal{L}$ by ensuring that the variations with respect to all the considered variables vanish. The total variation of the (augmented) Lagrangian is:
\begin{equation}
\begin{split}
\delta \mathcal{L}& =\dfrac{\delta \mathcal{L}}{\delta \mathbf{q}}\delta \mathbf{q} + \dfrac{\delta \mathcal{L}}{\delta \mathbf{q}_0}\delta \mathbf{q}_0  +  \dfrac{\delta \mathcal{L}}{\delta \mathbf{q}^\dagger}\delta \mathbf{q}^\dagger + \dfrac{\delta \mathcal{L}}{\delta \mathbf{q}_0^\dagger}\delta \mathbf{q}_0^\dagger\\
									& + \displaystyle\sum_{i=0}^{n_c}{\dfrac{\partial \mathcal{L}}{\partial \lambda_i}\delta \lambda_i}=0.
\end{split}
\end{equation}
Since all the variables of the problem are independent, all the terms in the previous equation have to vanish. Variations with respect to $\mathbf{q}^\dagger$ and $\mathbf{q}_0^\dagger$ yield  the ``direct'' or ``forward'' partial differential  equation 
(\ref{eq:GENERAL_EQ}) and the initial conditions for $\mathbf{q}$, while the first variation with respect to the $\lambda_i$ will simply yield the constraints on the normalization of the initial perturbation.

Requiring  variations with respect to the direct variable $\mathbf{q}$ to be zero leads (typically after some integration by parts, and application of appropriate boundary conditions) to the adjoint evolution equation, defined as
\begin{equation}
-\partial_t \mathbf{q}^\dagger + \mathbf{L}^\dagger \mathbf{q}^\dagger  = 0.
\end{equation}
The integration by parts of the time derivative yields the final condition
\begin{equation}\label{eq:final_adj}
\mathbf{q}^\dagger(T)= \dfrac{\delta \mathcal{J}}{\delta \mathbf{q}} = \mathbf{W}_O \mathbf{q}(T).
\end{equation}
Because of the Laplacian structure of the diffusive term in equations of interest, the adjoint equation turns out to be an anti-diffusive equation which, for well-posedness reasons, has to be integrated backward in time from $t=T$ to $t=0$ to calculate $\mathbf{q}^\dagger(0)$ which can then be used  to find the sensitivity of the Lagrangian  to the chosen initial condition of the state vector. By requiring that the boundary terms play no role, (and hence are homogeneous) the natural boundary conditions for the adjoint are found straightforwardly to be $\mathbf{q}^\dagger(\partial\Omega)=0$.

\begin{figure}
\hspace*{-1cm}
\scalebox{1} 
{
\begin{pspicture}(0,-2.3292189)(10.742813,2.3692188)
\psframe[linewidth=0.04,dimen=outer](7.099531,1.2707813)(3.8995314,0.07078125)
\psframe[linewidth=0.04,dimen=outer](7.099531,-1.1292187)(3.8995314,-2.3292189)
\psline[linewidth=0.04cm,arrowsize=0.05291667cm 2.0,arrowlength=1.4,arrowinset=0.4]{->}(2.6995313,0.67078125)(2.6995313,1.8707813)
\psline[linewidth=0.04cm,arrowsize=0.05291667cm 2.0,arrowlength=1.4,arrowinset=0.4]{->}(1.4995313,0.67078125)(2.6995313,0.67078125)
\psline[linewidth=0.04cm,arrowsize=0.05291667cm 2.0,arrowlength=1.4,arrowinset=0.4]{->}(8.299532,-1.7292187)(7.099531,-1.7292187)
\psline[linewidth=0.04cm,arrowsize=0.05291667cm 2.0,arrowlength=1.4,arrowinset=0.4]{->}(2.6995313,0.67078125)(3.8995314,0.67078125)
\psline[linewidth=0.04cm](8.299532,-1.7292187)(8.299532,0.67078125)
\psline[linewidth=0.04cm](7.099531,0.67078125)(8.299532,0.67078125)
\psline[linewidth=0.04cm](2.6995313,0.67078125)(2.6995313,-1.7292187)
\psline[linewidth=0.04cm](2.6995313,-1.7292187)(3.8995314,-1.7292187)
\usefont{T1}{ptm}{m}{n}
\rput(1.1523438,0.68078125){$\mathbf{q}_0$}
\usefont{T1}{ptm}{m}{n}
\rput(2.682344,2.1807814){$\mathbf{q}_0^*$}
\usefont{T1}{ptm}{m}{n}
\rput(3.442344,0.98078126){$\mathbf{q}(0)$}
\usefont{T1}{ptm}{m}{n}
\rput(7.572344,0.98078126){$\mathbf{q}(T)$}
\usefont{T1}{ptm}{m}{n}
\rput(3.3823438,-1.4192188){$\mathbf{q}^\dagger(0)$}
\usefont{T1}{ptm}{m}{n}
\rput(7.652344,-1.4192188){$\mathbf{q}^\dagger(T)$}
\pscircle[linewidth=0.04,dimen=outer,fillstyle=solid](8.299532,-0.52921873){0.6}
\pscircle[linewidth=0.04,dimen=outer,fillstyle=solid](2.6995313,-0.52921873){0.6}
\usefont{T1}{ptm}{m}{n}
\rput(2.717969,-0.55921876){opt}
\psline[linewidth=0.04cm](7.9995313,-0.82921875)(8.599531,-0.22921875)
\psline[linewidth=0.04cm](7.9995313,-0.22921875)(8.599531,-0.82921875)
\usefont{T1}{ptm}{m}{n}
\rput(5.5351562,0.68078125){Direct: $\mathbf{L}$}
\usefont{T1}{ptm}{m}{n}
\rput(5.528594,-1.7192187){Adjoint: $\mathbf{L}^\dagger$}
\usefont{T1}{ptm}{m}{n}
\rput(7.222344,-0.5392187){$\dfrac{\delta\mathcal{J}}{\delta \mathbf{q}}$}
\usefont{T1}{ptm}{m}{n}
\rput(3.912344,-0.51921874){$\nabla_{\mathbf{q}_0}\mathcal{J}$}
\end{pspicture} 
}
\caption{\label{fig:loop}
A schematic representation of the ``Direct/Adjoint'' loop process in order to find the optimal perturbation. We start with a guess $\mathbf{q}_0$, apply the initial condition constraint (\ref{eq:IC_CONSTR}), then integrate the direct equation forward in time. This gives the direct state vector at time $T$ which allows us to define the ``final'' condition for the adjoint state vector using (\ref{eq:final_adj}). We then integrate the adjoint equation backward in time from this final condition to obtain the ``initial'' adjoint state vector which allows us to compute the sensitivity with respect to the chosen initial condition on the (forward) state vector $\mathbf{q}_0$ using (\ref{eq:gradient}). We then use an appropriate optimization method in order to find the ``best'' initial condition achieving the maximum value of the objective functional defined in (\ref{eq:general_norm}).}
\end{figure}
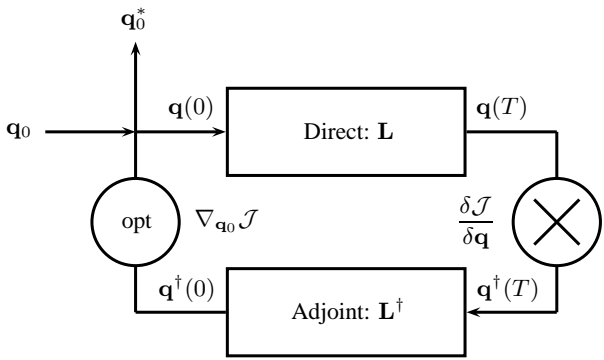

Taking variations with respect to the initial condition $\mathbf{q}_0$ leads to the following expression for the gradient of the objective functional with respect to the initial condition:
\begin{equation}\label{eq:gradient}
\nabla_{\mathbf{q}_0} \mathcal{J} =  \mathbf{q}_0^\dagger - \displaystyle\sum_{i=0}^{n_c}{\lambda_i \mathbf{W}_{K_i} \mathbf{q}_0},
\end{equation}
where $\mathbf{q}_0^\dagger= \mathbf{q}^\dagger(0)$.
Ideally, at the stationary point of the Lagrangian, (when the solution to the underlying variational problem has then been identified) this gradient vanishes. However, this is not true for generic initial conditions, and we have to employ an optimization technique in order to reach the (solution) condition. The $\lambda_i$, the Lagrange multipliers imposing the various
amplitude complementary semi-norm constraints on the initial state
vector,  will be determined at each iteration of the optimization algorithm by satisfying the initial normalization conditions. This determination will however depend on the particular iterative optimization algorithm used. The whole loop process is represented schematically in figure (\ref{fig:loop}).

Eventually, at the end of the optimization, we have the optimal value of the objective functional  $\mathcal{J}^{*}=\mathcal{J}(\mathbf{q}^{*}(T))$, associated with the optimal set of direct and adjoint state vectors $(\mathbf{q}^{*},\mathbf{q}^{*\dagger})$ at all times, and in particular at the initial time $t=0$,
thus identifying the optimal initial condition $\mathbf{q}_0^{*}$, for which the gradient given by (\ref{eq:gradient}) vanishes by definition. We will now see that the adjoint state vector can also yield information on the sensitivity of the objective functional with respect to every varying field or coefficient taken into account in our Lagrangian framework.

\subsection{Sensitivity analysis}

In this section, we will describe the sensitivity analysis possibilities that our variational framework allows. In some sense, the optimal perturbation framework presented above is already a sensitivity analysis, with the appropriate sensitivity information (the gradient in
(\ref{eq:gradient})) with respect to the choice of initial conditions
being used to find the optimal perturbation. 
A general sensitivity analysis will allow us to find what is the impact of a small variation of a parameter $p$ on the value of a functional at the optimal state vector point $\mathbf{q}^{*}$.
As the sensitivity analysis can be performed on a functional which is totally different from the (optimized) objective functional, we will define a general functional $\mathcal{I}$ which is \textit{a priori} different from the original optimized objective functional $\mathcal{J}$. 

The sensitivity may then be defined as 
\begin{equation}\label{eq:sensitivity}
\nabla_p \mathcal{I}(\mathbf{q}^{*}(p),p) = \dfrac{\partial \mathcal{I}}{\partial p}(\mathbf{q}^{*}) + \left(\dfrac{\delta \mathcal{I}}{\delta \mathbf{q}},\dfrac{\partial \mathbf{q}^{*}}{\partial p}\right),
\end{equation}
where $\nabla_p \mathcal{I}$ is just a condensed way to write the total derivative of $\mathcal{I}$ with respect to $p$, and where the chain rule appears under the form of a scalar product on the state vector space. The first term on the right-hand side of equation represents the explicit contribution of $p$ to the  functional while the second term is the implicit contribution of $p$ to $\mathcal{I}$ through the (optimal) state vector $\mathbf{q}^{*}$.

We consider   two qualitatively different cases, depending on the particular properties of the parameter $p$. We can define two broad classes of parameters: constraint parameters $p_c$ which will modify the constraints while keeping the functional $\mathcal{I}$ unchanged;
and external parameters $p_e$ which will change the functional $\mathcal{I}$ without changing the constraints. An example
of a  constraint parameter is a coefficient of the underlying partial differential equation satisfied by the state vector, such as a viscosity coefficient or a modeling parameter, while an example
of an external parameter is a parameter directly involved in the definition of the energy semi-norm.

Focusing first on sensitivity with respect to constraint parameters ($p_c$), the first term on the right-hand side of equation (\ref{eq:sensitivity}) is zero by definition of a constraint parameter, as it does not appear directly in the  functional $\mathcal{I}$. Therefore, 
\begin{equation}
\nabla_{p_c} \mathcal{I}(\mathbf{q}^{*}(p_c),p_c) = \left(\dfrac{\delta \mathcal{I}}{\delta \mathbf{q}},\dfrac{\partial \mathbf{q}^{*}}{\partial p_c}\right).
\end{equation}
This implicit contribution can be expressed, analogously to before using  a Lagrangian framework. We can add the constraint  into a yet further new augmented functional $\mathcal{K}$ combining the functional $\mathcal{I}$ as well as the dynamical PDE constraint on the (optimal) state vector $\mathbf{q}^{*}$:
\begin{equation}
\mathcal{K}(\mathbf{q}^{*},\mathbf{q}^{\dagger*}) = \mathcal{I}(\mathbf{q}^{*})  - \left(\mathbf{q}^{* \dagger_\mathcal{I}},\partial_t \mathbf{q}^{*} - \mathbf{L}\mathbf{q}^{*}\right),
\end{equation}
where we have added a  subscript $\mathcal{I}$ since the adjoint will depend on the functional $\mathcal{I}$ and is 
in general different from the adjoint state vector $\mathbf{q}^{*\dagger}$ associated with the optimization of the underlying objective functional $\mathcal{J}$.

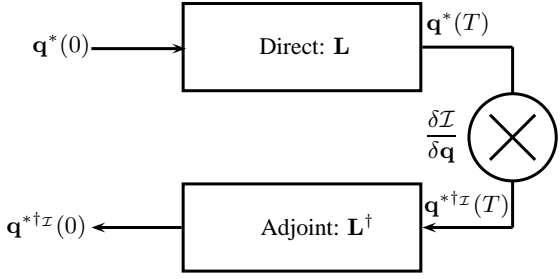
\begin{figure}
\hspace*{-2cm}
\scalebox{1} 
{
\begin{pspicture}(0,-1.8)(11.752812,1.8)
\psframe[linewidth=0.04,dimen=outer](7.9895315,1.8)(4.789531,0.6)
\psframe[linewidth=0.04,dimen=outer](7.9895315,-0.6)(4.789531,-1.8)
\psline[linewidth=0.04cm,arrowsize=0.05291667cm 2.0,arrowlength=1.4,arrowinset=0.4]{->}(3.5895312,1.18)(4.849531,1.18)
\psline[linewidth=0.04cm,arrowsize=0.05291667cm 2.0,arrowlength=1.4,arrowinset=0.4]{->}(9.189531,-1.2)(7.9895315,-1.2)
\psline[linewidth=0.04cm](9.189531,-1.2)(9.189531,1.2)
\psline[linewidth=0.04cm](7.9895315,1.2)(9.189531,1.2)
\psline[linewidth=0.04cm,arrowsize=0.05291667cm 2.0,arrowlength=1.4,arrowinset=0.4]{<-}(3.5895312,-1.2)(4.789531,-1.2)
\usefont{T1}{ptm}{m}{n}
\rput(3.1923437,1.21){$\mathbf{q}^*(0)$}
\usefont{T1}{ptm}{m}{n}
\rput(8.462343,1.51){$\mathbf{q}^*(T)$}
\usefont{T1}{ptm}{m}{n}
\rput(2.9923437,-1.17){$\mathbf{q}^{*\dagger_\mathcal{I}}(0)$}
\usefont{T1}{ptm}{m}{n}
\rput(8.582344,-0.89){$\mathbf{q}^{*\dagger_\mathcal{I}}(T)$}
\pscircle[linewidth=0.04,dimen=outer,fillstyle=solid](9.189531,0.0){0.6}
\psline[linewidth=0.04cm](8.889531,-0.3)(9.4895315,0.3)
\psline[linewidth=0.04cm](8.889531,0.3)(9.4895315,-0.3)
\usefont{T1}{ptm}{m}{n}
\rput(6.425156,1.21){Direct: $\mathbf{L}$}
\usefont{T1}{ptm}{m}{n}
\rput(6.598594,-1.19){Adjoint: $\mathbf{L}^{\dagger}$}
\usefont{T1}{ptm}{m}{n}
\rput(8.242344,0.01){$\dfrac{\delta\mathcal{I}}{\delta \mathbf{q}}$}
\end{pspicture} 
}
\caption{ A schematic representation of the algorithm used to calculate the sensitivity of a general functional $\mathcal{I}$ of the optimal initial condition state vector $\mathbf{q}_0^{*}$ to constraint parameters $p_c$. We start from the optimal initial condition state vector $\mathbf{q}_0^{*}$ we obtained using the optimization framework and integrate the direct equation to obtain the ``final'' state vector $\mathbf{q}^{*}(T)$. (This step may not actually be required if the final 
state of the optimal direct state vector 
$\mathbf{q}^{*}$ has been saved in the last iteration of the optimization framework.) We then construct a new final adjoint state
vector $\mathbf{q}^{* \dagger_\mathcal{I}}(T)$, which construction
depends on the particular choice of the functional $\mathcal{I}$. Finally a backward integration of the adjoint state vector leads to a new ``initial'' adjoint state vector $\mathbf{q}^{* \dagger_\mathcal{I}}(0)$
 which is needed in order to determine the required sensitivity.}
 \label{eq:semi_loop}
\end{figure}

The required implicit derivative can be obtained  by calculating the partial derivative of the augmented Lagrangian functional $\mathcal{K}$ with respect to $p_c$ since the constraints have been embedded in this augmented functional. The direct state vector $\mathbf{q}^{*}$ is defined by its initial condition $\mathbf{q}^{*}_0$ (the optimal for maximizing the original, underlying objective functional $\mathcal{J}$) and the adjoint state vector $\mathbf{q}^{*\dagger_\mathcal{I}}$ (which carries the sensitivity information) will be retrieved through the backward integration of the adjoint equations, the structure of which is not changed by this algorithm. However, the chosen starting form of the ``final'' adjoint state vector $\mathbf{q}^{*\dagger_\mathcal{I}}(T)$ is now determined by the gradient of the new functional $\mathcal{I}$, and so in general is different from the final adjoint state vector $\mathbf{q}^{* \dagger}$ associated with the optimization of the original underlying objective functional $\mathcal{J}$. Using this new final adjoint state vector $\mathbf{q}^{* \dagger_\mathcal{I}}(T)$, a single backward-in-time evolution using the adjoint equations yields the sensitivity information. This means that the sensitivity to a constraint parameter of a functional (potentially different from the original optimized objective functional)  satisfies
\begin{equation}
\dfrac{ \partial \mathcal{K} }{ \partial p_c } = \left(\dfrac{\delta \mathcal{I}}{\delta \mathbf{q}},\dfrac{\partial \mathbf{q}^{*}}{\partial p_c}\right).
\end{equation}
As a direct consequence,
\begin{equation}
\nabla_{p_c} \mathcal{I}(\mathbf{q}^{*}(p_c),p_c) = \dfrac{\partial \mathcal{K}}{\partial p_c},
\end{equation}
where  $\mathcal{K}$ is the secondary augmented  Lagrangian functional. This expression is a scalar product between a function of the (optimal) direct state vector $\mathbf{q}^{*}$ and the adjoint state vector $\mathbf{q}^{* \dagger_\mathcal{I}}$ corresponding to the  functional $\mathcal{I}$.
A schematic representation of this particular algorithm is shown in figure (\ref{eq:semi_loop}).
For the particular special case where $\mathcal{I}$ actually is the original optimized functional $\mathcal{J}$, then the gradient is given by the same equation, where the adjoint vector $\mathbf{q}^{* \dagger_\mathcal{J}}=\mathbf{q}^{* \dagger}$ was already evaluated during the optimization problem.

In the other case of an external parameter, the objective functional $\mathcal{I}$ depends explicitly on the parameter $p_e$, so the first term  on the right-hand side of  equation (\ref{eq:sensitivity}) will be different from zero. The gradient of the functional with respect to an explicit parameter $p_e$ can be found in many cases analytically, (for example for functionals defined in terms of integrals) and so the principal issue remains to evaluate the second term on the right-hand side of equation (\ref{eq:sensitivity}). We believe that the calculation of  the second term of the product, (i.e. the gradient of the optimal state vector $\mathbf{q}^{*}$ with respect to $p_e$) requires the use of a simple, yet computationally costly, finite-difference method. Indeed, we have to utilize our variational
framework to identify the optimal state vector $\mathbf{q}^{*}$  to the 
problem  for a particular value of $p_e$, then for $p_e+\delta p_e$ and then evaluate:
\begin{equation}
\dfrac{\partial \mathbf{q}^{*}}{\partial p}\simeq\dfrac{\mathbf{q}^{*}(p_e+\delta p_e)-\mathbf{q}^{*}(p_e)}{\delta p_e}.
\end{equation}
Once there is a need to use finite differences however, in general there is no need to evaluate the terms in (\ref{eq:sensitivity}) 
independently, because sensitivity can of course also be directly
estimated using finite-difference:
\begin{equation}
\begin{array}{r l}
\nabla_{p_e} \mathcal{I} & (\mathbf{q}^{*}(p_e),p_e) \simeq  \\
\\
&\dfrac{\mathcal{I}(\mathbf{q}^{*}(p_e+\delta p_e),p_e+\delta p_e)-\mathcal{I}(\mathbf{q}^{*}(p_e),p_e)}{\delta p_e}.
\end{array}
\end{equation}

The situation is substantially more straightforward in the special
case when the functional $\mathcal{I}$ whose sensitivity is being investigated is  actually the same as the underlying optimized functional $\mathcal{J}$. In this specific case, we observe that the second term on the right-hand side of (\ref{eq:sensitivity}) actually vanishes. 
Since the objective functional $\mathcal{J}$ is (by definition) optimized, variations of the state vector while
still satisfying all the imposed constraints cannot improve
the value of the objective functional $\mathcal{J}$. 

Formally,  the gradient of the objective functional with respect to the state vector is perpendicular to the subspace defined by all the imposed constraints. Equivalently, the level lines of $\mathcal{J}$ are parallel to the constraint subspace at the optimal point $\mathbf{q}^{*}$. On the other hand, $\partial_{p_e} \mathbf{q}^{*}$ is tangent to the subspace
defined by the constraints (since the optimized state vector must always satisfy
all the constraints by definition) which subspace does not change as $p_e$ varies, by the definition of an external parameter.
Therefore, combining these two observations,  the gradient of $\mathcal{J}$ with respect to the state vector  is normal to the variation of $\mathbf{q}^{*}$ (confined to the subspace defined by the constraints) with respect to the external parameter, and so the second term on the right-hand side of (\ref{eq:sensitivity}) (which is simply the scalar product of these two quantities) is exactly zero. We can then simply express
the sensitivity of the optimized objective functional $\mathcal{J}$ to variations in an external parameter as
\begin{equation}
\nabla_{p_e} \mathcal{J}(\mathbf{q}^*(p_e),p_e) = \dfrac{\partial \mathcal{J}}{\partial p_e}.
\end{equation}

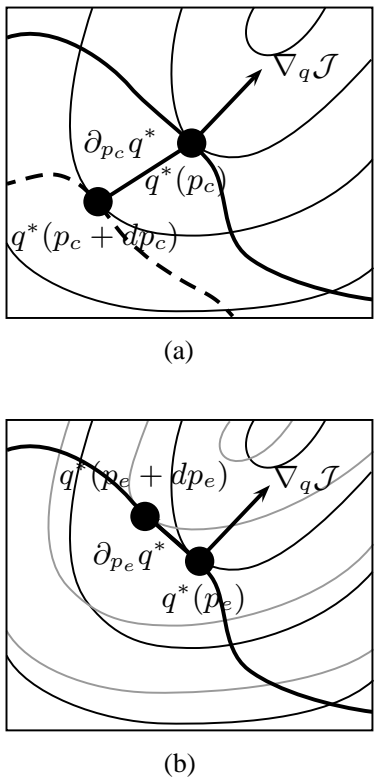
\begin{figure}\label{fig:sens_schemes}
\hspace*{1cm}
\scalebox{1.3} 
{
\begin{pspicture}(0,-4.003031)(5.3028126,3.9663706)
\definecolor{color1466}{rgb}{0.6,0.6,0.6}
\psbezier[linewidth=0.02](4.0695305,-1.8207331)(3.558208,-2.4086056)(2.7028966,-2.6019554)(2.2527237,-2.6019554)(1.8025509,-2.6019554)(1.4809983,-2.551318)(1.1594462,-2.1462176)(0.8378941,-1.7411171)(0.9879517,-0.9055977)(1.2023197,-0.2726284)
\psbezier[linewidth=0.02](4.0549216,-0.80060136)(3.4997714,-1.4230547)(3.10638,-1.6873853)(2.8423567,-1.768862)(2.5783336,-1.8503386)(2.316425,-1.768862)(2.1703331,-1.5959584)(2.024241,-1.4230547)(1.9252979,-0.85496014)(2.1182292,-0.2726284)
\psbezier[linewidth=0.02](0.32539335,-2.6779118)(0.5183246,-3.0576932)(1.2768883,-3.3615186)(1.9331541,-3.374178)(2.5894198,-3.3868372)(3.615944,-3.3488588)(4.076835,-2.9310994)
\psbezier[linewidth=0.02](2.856966,-0.26459998)(2.5538168,-0.7617495)(3.1431825,-1.0617752)(3.5874267,-0.26459998)
\psbezier[linewidth=0.02,linecolor=color1466](4.076835,-1.5385671)(3.5979424,-2.151628)(2.531531,-2.3725038)(2.0691574,-2.3725038)(1.6067834,-2.3725038)(1.2765163,-2.3274813)(0.94624937,-1.967301)(0.61598235,-1.6071209)(0.7701069,-0.86424947)(0.96849763,-0.2726284)
\psbezier[linewidth=0.02,linecolor=color1466](4.076835,-0.589113)(3.531889,-1.0753244)(2.911644,-1.2962003)(2.6298504,-1.3486762)(2.348057,-1.401152)(1.9495119,-1.4355047)(1.718786,-1.2220824)(1.4880601,-1.0086601)(1.5656528,-0.819227)(1.772378,-0.2726284)
\psbezier[linewidth=0.02,linecolor=color1466](0.32539335,-2.4247239)(0.5235536,-2.7777064)(1.1419002,-3.0478415)(1.8159527,-3.059097)(2.490005,-3.0703526)(3.5979424,-2.9732888)(4.076835,-2.6146147)
\psbezier[linewidth=0.02,linecolor=color1466](2.6086092,-0.28189036)(2.274957,-0.7059482)(2.8131485,-0.9375526)(3.2514148,-0.26459998)
\psbezier[linewidth=0.04](0.32957226,-0.5758266)(0.8262854,-0.40292296)(1.3943727,-0.85051066)(1.5859644,-1.0772473)(1.7775563,-1.3039838)(2.2422214,-1.6261041)(2.4186897,-1.8726043)(2.595158,-2.1191044)(2.5063448,-2.5987997)(2.7985291,-2.8235745)(3.0907133,-3.0483491)(3.6604726,-3.2039626)(4.0695305,-3.2558336)
\usefont{T1}{ptm}{m}{n}
\rput(1.5923438,-1.6645932){$\partial_{p_e} q^*$}
\psdots[dotsize=0.3](2.299531,-1.7145933)
\usefont{T1}{ptm}{m}{n}
\rput(1.7323438,-0.84459335){$q^*(p_e+dp_e)$}
\usefont{T1}{ptm}{m}{n}
\rput(3.3923438,-0.8845933){$\nabla_{q}\mathcal{J}$}
\usefont{T1}{ptm}{m}{n}
\rput(2.3423438,-2.1045933){$q^*(p_e)$}
\psdots[dotsize=0.3](1.7395312,-1.2545934)
\psline[linewidth=0.04cm,arrowsize=0.05291667cm 2.0,arrowlength=1.4,arrowinset=0.4]{<-}(1.7195312,-1.2145933)(2.2795312,-1.7145933)
\psframe[linewidth=0.02,dimen=outer](4.076835,-0.25793996)(0.31459722,-3.4547653)
\psline[linewidth=0.04cm,arrowsize=0.05291667cm 2.0,arrowlength=1.4,arrowinset=0.4]{<-}(3.0195312,-0.9345934)(2.2595313,-1.7345934)
\psbezier[linewidth=0.02](4.069557,2.4001572)(3.558234,1.8122544)(2.702921,1.6188947)(2.2527473,1.6188947)(1.8025736,1.6188947)(1.481021,1.6695348)(1.1594683,2.074656)(0.8379156,2.4797773)(0.9879735,3.3153398)(1.2023419,3.9483416)
\psbezier[linewidth=0.02](4.054948,3.4203415)(3.4997966,2.797856)(3.1064048,2.5335116)(2.8423815,2.4520311)(2.5783582,2.3705504)(2.3164487,2.4520311)(2.1703568,2.6249435)(2.0242646,2.797856)(1.925321,3.3659801)(2.1182528,3.9483416)
\psbezier[linewidth=0.02](0.32541424,1.5429344)(0.51834583,1.1631334)(1.2769105,0.85929245)(1.9331774,0.8466324)(2.5894442,0.8339724)(3.6159692,0.8719525)(4.076862,1.2897336)
\psbezier[linewidth=0.02](2.8569906,3.9563706)(2.5538406,3.4591954)(3.1432076,3.1591544)(3.5874524,3.9563706)
\psbezier[linewidth=0.04](0.32959318,3.6451278)(0.8263071,3.8180404)(1.3943951,3.3704298)(1.5859872,3.1436813)(1.7775793,2.9169328)(2.2422452,2.5947962)(2.4187136,2.3482835)(2.595182,2.1017709)(2.506369,1.6220505)(2.7985537,1.3972642)(3.0907383,1.172478)(3.6604986,1.0168567)(4.069557,0.9649828)
\psbezier[linewidth=0.04,linestyle=dashed,dash=0.17638889cm 0.10583334cm](0.32951665,2.1487799)(0.7020631,2.2517147)(0.8116356,2.2946043)(1.0307806,2.1659358)(1.2499256,2.0372674)(1.4386934,1.6642395)(1.6151671,1.5225928)(1.791641,1.3809459)(1.7028251,1.4282358)(2.016933,1.222366)(2.3310406,1.0164962)(2.4186988,1.0508077)(2.6378436,0.79347056)
\usefont{T1}{ptm}{m}{n}
\rput(1.4923438,2.5554066){$\partial_{p_c} q^*$}
\psdots[dotsize=0.3](1.2595313,1.9654068)
\usefont{T1}{ptm}{m}{n}
\rput(1.2323438,1.5954068){$q^*(p_c+dp_c)$}
\usefont{T1}{ptm}{m}{n}
\rput(3.3823438,3.3354065){$\nabla_{q} \mathcal{J}$}
\usefont{T1}{ptm}{m}{n}
\rput(2.1623437,2.1754067){$q^*(p_c)$}
\psdots[dotsize=0.3](2.2195313,2.5654066)
\psline[linewidth=0.04cm,arrowsize=0.05291667cm 2.0,arrowlength=1.4,arrowinset=0.4]{<-}(1.2195312,1.9254067)(2.2195313,2.5654066)
\psframe[linewidth=0.02,dimen=outer](4.076862,3.9630308)(0.3146181,0.7660409)
\psline[linewidth=0.04cm,arrowsize=0.05291667cm 2.0,arrowlength=1.4,arrowinset=0.4]{<-}(2.939531,3.3254066)(2.179531,2.5254066)
\usefont{T1}{ptm}{m}{n}
\rput(2.0898438,0.42040664){\footnotesize (a)}
\usefont{T1}{ptm}{m}{n}
\rput(2.0998437,-3.7995934){\footnotesize (b)}
\end{pspicture} 
}
\caption{(a) A schematic representation illustrating  the sensitivity of the optimized functional $\mathcal{J}$ to a constraint parameter. 
(b) A
schematic representation illustrating the sensitivity of the optimized functional $\mathcal{J}$  to an external parameter. 
Black lines are the level lines of the objective functional $\mathcal{J}$ (grey lines of part (b) of the figure  correspond to the level lines of the functional for $p_e=p_e+\delta p_e$). Thick black lines are the constraints (thick dashed line is the constraint for $p_c=p_c+\delta p_c$). Black dots represent the  optimal locations in solution space for the state vector. In the case of the sensitivity with respect to an external parameter we can see that the terms $\frac{\delta \mathcal{J}}{\delta \mathbf{q}}$ and $\frac{\partial \mathbf{q}^*}{\partial p}$ are orthogonal whereas they are not in the case of a constraint parameter.}
\end{figure}

Here, we have only discussed variations with respect to a parameter. However, it could also be of interest to consider the sensitivity of a functional to a function, either associated with the definition of the objective functional or the constraints. For example if the operator $\mathbf{L}$ describes the linear evolution of a small perturbation evolving on a base flow defined by a base state vector $\mathbf{q}_B$ (which is a function governing the dynamics of the perturbation), it is possible to derive the sensitivity of a functional  to this base flow  in 
an analogous fashion to the algorithm described above to  investigate
sensitivity to parameters.

In the particular example of considering the sensitivity to the base flow state vector, the base flow must satisfy base flow equations which can be expressed in the same form as equation (\ref{eq:GENERAL_EQ}), where the implicit coefficients in the operators (such as  the flow's
Reynolds number) are ``constraint parameters''  $p_c$ and are  shared by the base flow and perturbation equations (since the perturbation equation is derived from the base flow equation). In general, small variations in these coefficients will affect both the perturbation state
vector $\mathbf{q}$ and the base state vector $\mathbf{q}_B$. As a consequence,
the requirement (effectively another constraint) that
the base state vector satisfies the base flow equation 
must be 
embedded within the Lagrangian functional, with the constraint imposed by  a new Lagrange multiplier $\mathbf{q}_B^\dagger$.

\section{Reynolds-Averaged Burgers equation (RAB) optimal perturbation problem formulation}\label{sec:RAB}

\subsection{Derivation of the Reynolds-Averaged Burgers equations}

As a relatively simple demonstration example of our variational framework, we will in this section construct a model problem of interest, where optimization
of the perturbation (kinetic) energy gain inevitably leads to an objective functional
which is defined in terms of a (nontrivial) semi-norm of the state vector.
We study the evolution of a velocity-like variable defined on $[0,1]\times[0,T]$ and governed by the stochastically forced Burgers equation, entirely defined by the viscosity coefficient $\nu$, with Dirichlet boundary conditions and a well-posed (in particular appropriately
smooth) initial condition. This can be formulated as
\begin{equation}
\partial_t u + u\partial_x u - \nu \partial_{xx} u = s,
\end{equation}
with $u(0,t)=u_l$, $u(1,t)=u_r$ and $u(x,0)=u_0$ and $s$ a stochastic forcing of zero ensemble average (which is needed in order to later 
be consistent with (\ref{eq:GENERAL_EQ}) where no forcing term is present), and vanishing at the boundary.
To obtain nontrivial energy production dynamics, we consider a symmetric focussing base flow, and so  we choose the boundary conditions to be $u_l=-u_r=1$.
The solution $u(x,t)$ is stochastic because of the nature of the forcing, but can be expressed as the superposition of a coherent field $\left<u\right>$ and a stochastic field $u_s$, i.e.
\begin{equation}
\begin{array}{c}
u(x,t)=\left<u\right>(x,t)+u_s(x,t),\\
\mbox{such that\ \ \ } \left<u_s\right>=0,
\end{array}
\end{equation}
where $\left<\cdot\right>$ denotes ensemble averaging. We interpret $u_s$ as
the ``turbulent'' component of the flow, and so 
this decomposition  of the flow into two variables with different spatial and temporal scales of variation constitutes a so-called ``Reynolds decomposition''.

We introduce this decomposition into the governing equation, and then ensemble-average to obtain the mean 
flow equation for
$\left<u\right>$, which is
\begin{equation}\label{eq:RAB_open}
\partial_t \left<u\right> + \left<u\right>\partial_x \left<u\right> + \left<u_s\partial_x u_s\right>- \nu \partial_{xx} \left<u\right> = 0.
\end{equation}
 In this equation all the terms except the third one are expressed in terms of the ensemble-average velocity of the flow. Indeed, when Reynolds-averaging a nonlinear state equation, higher order terms inevitably appear which cannot directly be expressed as a function of the first-order ``mean'' quantities, leading to a classic ``closure'' problem.  Here, this term is the equivalent of the gradient of the Reynolds stress tensor in the Reynolds-Averaged Navier-Stokes equations, a second-order quantity in a first-order equation.  In this particularly simple one-dimensional context, we can rewrite this  term as the spatial derivative of the turbulent kinetic energy, defined as
\begin{equation}\label{eq:closure}
\left< u_s \partial_x u_s \right> =\partial_x e_t, \mbox{\ \ \ \ with\ \ \ } e_t=\dfrac{1}{2}\left<u_s^2\right>.
\end{equation}
Therefore, to close the evolution equation for the mean velocity (\ref{eq:RAB_open}), 
we need to add a model in order to express the turbulent kinetic energy density (defined in (\ref{eq:closure})) as a function of mean quantities.

\subsection{Turbulent viscosity closure}

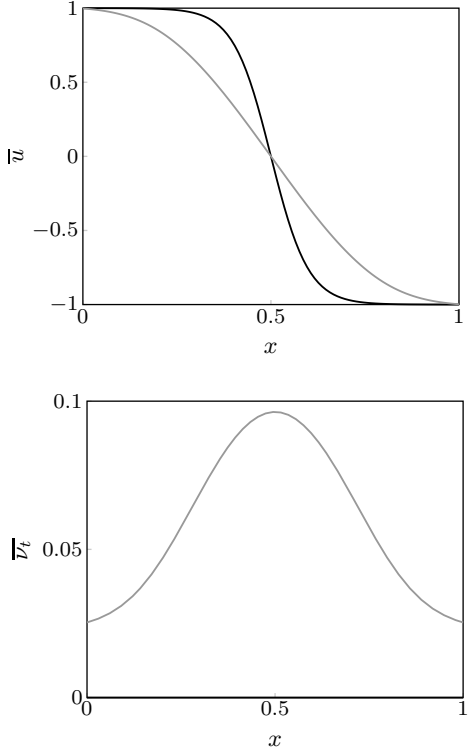
\begin{figure}\label{fig:steady}
\begin{center}
%
%
%
\providelength{\AxesLineWidth}       \setlength{\AxesLineWidth}{0.5pt}%
\providelength{\plotwidth}           \setlength{\plotwidth}{5cm}
\providelength{\LineWidth}           \setlength{\LineWidth}{0.7pt}%
\providelength{\MarkerSize}          \setlength{\MarkerSize}{4pt}%
\newrgbcolor{GridColor}{0.8 0.8 0.8}%
%
\psset{xunit=1.000000\plotwidth,yunit=0.394355\plotwidth}%
\begin{pspicture}(-0.210000,-1.380368)(1.000000,1.000000)%


\psline[linewidth=\AxesLineWidth,linecolor=GridColor](0.000000,-1.000000)(0.000000,-0.969571)
\psline[linewidth=\AxesLineWidth,linecolor=GridColor](0.500000,-1.000000)(0.500000,-0.969571)
\psline[linewidth=\AxesLineWidth,linecolor=GridColor](1.000000,-1.000000)(1.000000,-0.969571)
\psline[linewidth=\AxesLineWidth,linecolor=GridColor](0.000000,-1.000000)(0.012000,-1.000000)
\psline[linewidth=\AxesLineWidth,linecolor=GridColor](0.000000,-0.500000)(0.012000,-0.500000)
\psline[linewidth=\AxesLineWidth,linecolor=GridColor](0.000000,0.000000)(0.012000,0.000000)
\psline[linewidth=\AxesLineWidth,linecolor=GridColor](0.000000,0.500000)(0.012000,0.500000)
\psline[linewidth=\AxesLineWidth,linecolor=GridColor](0.000000,1.000000)(0.012000,1.000000)

{ \footnotesize 
\rput[t](0.000000,-1.030429){$0$}
\rput[t](0.500000,-1.030429){$0.5$}
\rput[t](1.000000,-1.030429){$1$}
\rput[r](-0.012000,-1.000000){$-1$}
\rput[r](-0.012000,-0.500000){$-0.5$}
\rput[r](-0.012000,0.000000){$0$}
\rput[r](-0.012000,0.500000){$0.5$}
\rput[r](-0.012000,1.000000){$1$}
} 

\psframe[linewidth=\AxesLineWidth,dimen=middle](0.000000,-1.000000)(1.000000,1.000000)

{ \small 
\rput[b](0.500000,-1.380368){
\begin{tabular}{c}
$x$\\
\end{tabular}
}

\rput[t]{90}(-0.210000,0.000000){
\begin{tabular}{c}
$\overline{u}$\\
\end{tabular}
}
} 

\newrgbcolor{color198.0417}{0  0  0}
\psline[plotstyle=line,linejoin=1,linestyle=solid,linewidth=\LineWidth,linecolor=color198.0417]
(0.000000,0.999984)(0.090909,0.999532)(0.131313,0.998858)(0.161616,0.997841)(0.181818,0.996725)
(0.202020,0.995053)(0.212121,0.993927)(0.222222,0.992549)(0.232323,0.990863)(0.242424,0.988802)
(0.252525,0.986284)(0.262626,0.983207)(0.272727,0.979451)(0.282828,0.974869)(0.292929,0.969286)
(0.303030,0.962489)(0.313131,0.954228)(0.323232,0.944203)(0.333333,0.932063)(0.343434,0.917397)
(0.353535,0.899735)(0.363636,0.878540)(0.373737,0.853214)(0.383838,0.823109)(0.393939,0.787543)
(0.404040,0.745826)(0.414141,0.697309)(0.424242,0.641438)(0.434343,0.577822)(0.444444,0.506319)
(0.454545,0.427104)(0.464646,0.340745)(0.474747,0.248228)(0.484848,0.150951)(0.515152,-0.150951)
(0.525253,-0.248228)(0.535354,-0.340745)(0.545455,-0.427104)(0.555556,-0.506319)(0.565657,-0.577822)
(0.575758,-0.641438)(0.585859,-0.697309)(0.595960,-0.745826)(0.606061,-0.787543)(0.616162,-0.823109)
(0.626263,-0.853214)(0.636364,-0.878540)(0.646465,-0.899735)(0.656566,-0.917397)(0.666667,-0.932063)
(0.676768,-0.944203)(0.686869,-0.954228)(0.696970,-0.962489)(0.707071,-0.969286)(0.717172,-0.974869)
(0.727273,-0.979451)(0.737374,-0.983207)(0.747475,-0.986284)(0.757576,-0.988802)(0.767677,-0.990863)
(0.777778,-0.992549)(0.787879,-0.993927)(0.797980,-0.995053)(0.818182,-0.996725)(0.838384,-0.997841)
(0.868687,-0.998858)(0.919192,-0.999631)(1.000000,-0.999984)

\newrgbcolor{color199.0413}{0.6         0.6         0.6}
\psline[plotstyle=line,linejoin=1,linestyle=solid,linewidth=\LineWidth,linecolor=color199.0413]
(0.000000,0.998078)(0.010101,0.995890)(0.020202,0.993404)(0.030303,0.990585)(0.040404,0.987394)
(0.050505,0.983789)(0.060606,0.979725)(0.070707,0.975155)(0.080808,0.970030)(0.090909,0.964300)
(0.101010,0.957911)(0.111111,0.950810)(0.121212,0.942946)(0.131313,0.934267)(0.141414,0.924724)
(0.151515,0.914271)(0.161616,0.902864)(0.171717,0.890465)(0.181818,0.877043)(0.191919,0.862571)
(0.202020,0.847027)(0.212121,0.830399)(0.222222,0.812678)(0.232323,0.793865)(0.242424,0.773964)
(0.252525,0.752987)(0.262626,0.730952)(0.272727,0.707881)(0.282828,0.683800)(0.292929,0.658742)
(0.303030,0.632740)(0.313131,0.605832)(0.323232,0.578059)(0.333333,0.549461)(0.343434,0.520084)
(0.353535,0.489972)(0.373737,0.427729)(0.393939,0.363109)(0.414141,0.296493)(0.444444,0.193667)
(0.484848,0.053148)(0.545455,-0.158809)(0.575758,-0.262557)(0.595960,-0.330026)(0.616162,-0.395692)
(0.636364,-0.459171)(0.646465,-0.489972)(0.656566,-0.520084)(0.666667,-0.549461)(0.676768,-0.578059)
(0.686869,-0.605832)(0.696970,-0.632740)(0.707071,-0.658742)(0.717172,-0.683800)(0.727273,-0.707881)
(0.737374,-0.730952)(0.747475,-0.752987)(0.757576,-0.773964)(0.767677,-0.793865)(0.777778,-0.812678)
(0.787879,-0.830399)(0.797980,-0.847027)(0.808081,-0.862571)(0.818182,-0.877043)(0.828283,-0.890465)
(0.838384,-0.902864)(0.848485,-0.914271)(0.858586,-0.924724)(0.868687,-0.934267)(0.878788,-0.942946)
(0.888889,-0.950810)(0.898990,-0.957911)(0.909091,-0.964300)(0.919192,-0.970030)(0.929293,-0.975155)
(0.939394,-0.979725)(0.949495,-0.983789)(0.959596,-0.987394)(0.969697,-0.990585)(0.979798,-0.993404)
(0.989899,-0.995890)(1.000000,-0.998078)

\end{pspicture}%
		\vspace{0.5cm}
%
%
%
\providelength{\AxesLineWidth}       \setlength{\AxesLineWidth}{0.5pt}%
\providelength{\plotwidth}           \setlength{\plotwidth}{5cm}
\providelength{\LineWidth}           \setlength{\LineWidth}{0.7pt}%
\providelength{\MarkerSize}          \setlength{\MarkerSize}{4pt}%
\newrgbcolor{GridColor}{0.8 0.8 0.8}%
%
\psset{xunit=1.000000\plotwidth,yunit=7.887097\plotwidth}%
\begin{pspicture}(-0.210000,-0.019018)(1.000000,0.100000)%


\psline[linewidth=\AxesLineWidth,linecolor=GridColor](0.000000,0.000000)(0.000000,0.001521)
\psline[linewidth=\AxesLineWidth,linecolor=GridColor](0.500000,0.000000)(0.500000,0.001521)
\psline[linewidth=\AxesLineWidth,linecolor=GridColor](1.000000,0.000000)(1.000000,0.001521)
\psline[linewidth=\AxesLineWidth,linecolor=GridColor](0.000000,0.000000)(0.012000,0.000000)
\psline[linewidth=\AxesLineWidth,linecolor=GridColor](0.000000,0.050000)(0.012000,0.050000)
\psline[linewidth=\AxesLineWidth,linecolor=GridColor](0.000000,0.100000)(0.012000,0.100000)

{ \footnotesize 
\rput[t](0.000000,-0.001521){$0$}
\rput[t](0.500000,-0.001521){$0.5$}
\rput[t](1.000000,-0.001521){$1$}
\rput[r](-0.012000,0.000000){$0$}
\rput[r](-0.012000,0.050000){$0.05$}
\rput[r](-0.012000,0.100000){$0.1$}
} 

\psframe[linewidth=\AxesLineWidth,dimen=middle](0.000000,0.000000)(1.000000,0.100000)

{ \small 
\rput[b](0.500000,-0.019018){
\begin{tabular}{c}
$x$\\
\end{tabular}
}

\rput[t]{90}(-0.210000,0.050000){
\begin{tabular}{c}
$\overline{\nu_t}$\\
\end{tabular}
}
} 

\newrgbcolor{color218.0349}{0  0  0}
\psline[plotstyle=line,linejoin=1,linestyle=solid,linewidth=\LineWidth,linecolor=color218.0349]
(0.000000,0.000000)(1.000000,0.000000)

\newrgbcolor{color219.0344}{0.6         0.6         0.6}
\psline[plotstyle=line,linejoin=1,linestyle=solid,linewidth=\LineWidth,linecolor=color219.0344]
(0.000000,0.025340)(0.030303,0.026637)(0.060606,0.028452)(0.090909,0.030924)(0.121212,0.034186)
(0.141414,0.036848)(0.171717,0.041599)(0.202020,0.047214)(0.232323,0.053545)(0.323232,0.074168)
(0.353535,0.080515)(0.373737,0.084329)(0.393939,0.087720)(0.414141,0.090618)(0.434343,0.092970)
(0.454545,0.094729)(0.474747,0.095865)(0.494949,0.096355)(0.515152,0.096191)(0.535354,0.095377)
(0.555556,0.093926)(0.575758,0.091865)(0.595960,0.089234)(0.616162,0.086082)(0.636364,0.082470)
(0.666667,0.076353)(0.707071,0.067340)(0.767677,0.053545)(0.797980,0.047214)(0.828283,0.041599)
(0.858586,0.036848)(0.888889,0.033004)(0.919192,0.030018)(0.949495,0.027781)(0.979798,0.026154)
(1.000000,0.025340)

\end{pspicture}%
		\caption{\label{fig:base}(a) Velocity $\overline{u}$ and (b) turbulent viscosity $\overline{\nu_t}$ solutions of the steady RAB equations (\ref{eq:RAB_sys}) with boundary conditions $\overline{u}(0)=1$, $\overline{u}(1)=-1$ and $\overline{\nu_t}(0)=r\nu$, $\overline{\nu_t}(1)=r\nu$. The value of the laminar viscosity is set to $\nu=0.05$. Production and destruction coefficients are chosen to have the illustrative values $c_1=0.75$ and $c_2=2$. For $r=0$, no turbulent viscosity is created, and  the flow is laminar. When $r\neq0$, turbulent viscosity is generated in the zones of high gradient, which are, as a consequence, smoothed for $r=0.5$. }
\end{center}
\end{figure}

We here follow the classical Boussinesq \cite{boussinesq} turbulence hypothesis, by assuming that  $e_t$
is proportional to the gradient of the mean velocity field with  a viscosity-like coefficient of proportionality $\nu_t$,
which in general is itself a function of space and time:
\begin{equation}\label{eq:RAB_closure}
\dfrac{1}{2}\left<u_s^2\right>= - \nu_t \partial_{x} \left<u\right>.
\end{equation}
Simple assumptions of this kind are  widely used as closures for RANS equations. In the highly idealized model situation
we are considering,  it is thoroughly plausible that the stochastic field will have  the effect of increasing the total viscosity of the flow by a certain amount $\nu_t$. Furthermore, in this special case of Burgers equation, the left-hand side of (\ref{eq:RAB_closure}) is always positive. As a consequence, the product $\nu_t \partial_{x} \left<u\right>$ has to be negative, which means that the turbulent (eddy)  viscosity must have the same sign as $-\partial_{x} \left<u\right>$. Therefore, the slope of the ensemble average of $u$ cannot be positive (flow going toward the edges of the domain), as that would require the turbulent viscosity to  be negative,
which is physically inconsistent. We will
always
respect the positivity constraint of the modelled turbulent kinetic energy $e_t$ since
we restrict ourselves to consideration  of a focussing flow (with negative slope, as shown in figure (\ref{fig:base})a).

Combining (\ref{eq:RAB_open}) and (\ref{eq:RAB_closure}) we obtain the following equation for the mean field $\left<u\right>$:
\begin{equation}
\label{RAB_closed}
\partial_t \left<u\right> + \left<u\right>\partial_x \left<u\right> - \partial_x\left[\left(\nu+\nu_t\right) \partial_{x} \left<u\right>\right] = 0.
\end{equation}

At this point, we could simply close the equation by choosing a fixed value of this new viscosity, based, for example, on the evaluation of a mixing length scale.
However, this naive technique is typically not appropriate since the  properties of a turbulent flow naturally vary in space and time. As a consequence,  it is appropriate  to develop a model equation allowing us to describe the spatial and temporal evolution of the new turbulent
(or eddy) viscosity coefficient. 

A widely used approach is to produce a transport equation for the turbulent viscosity.   We follow this approach and assume that $\nu_t$ is a solution of an advection-diffusion equation with production and destruction terms. We assume that the production of turbulent viscosity is driven by the magnitude of the first derivative of the mean field $\left<u\right>$ (equivalent to the shear in a real flow) while we  assume the destruction term is quadratic in the turbulent viscosity (see \cite{spalart}). This leads to the following equation for $\nu_t$:
\begin{equation}
\label{eq:RAB_model}
\begin{split}
\partial_t \nu_t + \left<u\right>\partial_x \nu_t - \partial_x\left[\left(\nu+\nu_t\right) \partial_{x} \nu_t\right]\\
\hfill - c_1 \left|\partial_x \left<u\right>\right| \nu_t + c_2 \nu_t^2 = 0,
\end{split}
\end{equation}
where $c_1$ and $c_2$ are two real coefficients defining the strength of production and destruction mechanisms. We also add a Dirichlet boundary condition for the viscosity: 
\begin{equation}\label{eq:ratio}
\nu_t(0,t)=\nu_t(1,t)=r\nu,
\end{equation}
where $r$ is the ratio between the turbulent viscosity at the boundaries of the flow domain and the laminar viscosity. This parameter just controls the amount of ``turbulence'' we want to introduce at the boundaries. Equation (\ref{eq:RAB_model}) has also the property of preserving the positivity of the 
turbulent viscosity. Equations (\ref{RAB_closed}) and (\ref{eq:RAB_model}) then constitute a closed set for the Reynolds-Averaged-Burgers (RAB) equations, acting on the state vector $\left< \mathbf{q} \right>(x,t) = \left(\left<u\right>,\nu_t\right)^\top$.

\begin{equation}
\label{eq:URAB_sys}
\left\{
\begin{array}{ll}
\partial_t \left<u\right> + \left<u\right>\partial_x \left<u\right> - \partial_x\left(\left(\nu+\nu_t\right) \partial_{x} \left<u\right>\right) & = 0,\\
\partial_t \nu_t + \left<u\right>\partial_x \nu_t - \partial_x\left(\left(\nu+\nu_t\right) \partial_{x} \nu_t\right)&\\
\hfill - c_1 \left|\partial_x \left<u\right>\right| \nu_t + c_2 \nu_t^2 & = 0.
\end{array}
\right.
\end{equation}

By considering steady flows (i.e. $\partial_t=0$), we can search for steady states as a solution of the 
coupled equations. We denote the steady flow solution of such an equation as $\overline{\mathbf{q}}(x)=\left(\overline{u},\overline{\nu_t}\right)^\top$. This steady state 
must satisfy
\begin{equation}
\label{eq:RAB_sys}
\left\{
\begin{array}{ll}
\overline{u}d_x \overline{u} - d_x\left(\left(\nu+\overline{\nu_t}\right) d_{x} \overline{u}\right) & = 0,\\
\overline{u}d_x \overline{\nu_t} - d_x\left(\left(\nu+\overline{\nu_t}\right) d_{x} \overline{\nu_t}\right)&\\
\hfill - c_1 \left|d_x \overline{u}\right| \overline{\nu_t} + c_2 \overline{\nu_t}^2 & = 0.
\end{array}
\right.
\Leftrightarrow \overline{\mathbf{L}} (\overline{\mathbf{q}}) = 0,\hfill
\end{equation}
defining a set of ordinary differential equations.

We plot a solution of this steady set of equations  in figure (\ref{fig:base}). We first notice that in the two cases, $r=0$ (laminar) and $r\neq0$ (turbulent), the flow indeed corresponds to a focussing, with a positive velocity (from left to right) in the left part of the domain and a negative velocity (from right to left) in the right part of the domain. Then, in the turbulent case ($r\neq0$), we can see that the turbulent viscosity equation has the effect of enhancing the turbulent viscosity in the middle of the domain (where the gradients of $\overline{u}$ are strong), which has the secondary effect of smoothing the gradient of velocity in the middle part of the domain.

\subsection{Perturbation equations}

Now that we have constructed a (steady) base flow, we can investigate
its stability properties by introducing small perturbations, which are in general functions
of space and time. The ensemble average of the flow can now be decomposed as follows:
\begin{equation}
\label{eq:tilde}
\left<\mathbf{q}\right>(x,t) = \overline{\mathbf{q}}(x) + \tilde{\mathbf{q}}(x,t),
\end{equation}
with  $\tilde{\mathbf{q}} = \left(\tilde{u},\tilde{\nu} \right)^\top$, so the perturbation 
state vector in general involves  a perturbation to both the velocity and the turbulent viscosity.
The perturbation velocity may be thought of as a ``coherent'' velocity perturbation, as it has a non-zero
ensemble average.
We assume that the magnitude and gradients associated with this perturbation state vector are sufficiently  ``small'' relative to the base flow 
for a linearization to be well-posed. 
Substituting this decomposition into the full equations (\ref{RAB_closed}) and (\ref{eq:RAB_model}), imposing the mean flow equations (\ref{eq:RAB_sys}) and neglecting nonlinear terms, we obtain the (full) linearized perturbation equation:
\begin{equation}
\label{eq:pert_linearized}
\begin{array}{ll}
&\left\{\begin{array}{ll}
\partial_t \tilde{u} + \overline{u}\partial_x \tilde{u} + \tilde{u}\partial_x \overline{u} - \partial_x\left(\left(\nu+\overline{\nu_t}\right)\partial_{x} \tilde{u}\right) = \partial_x\left(\tilde{\nu} \partial_{x} \overline{u}\right)\\
\partial_t \tilde{\nu} + \overline{u}\partial_x \tilde{\nu} - \partial_x\left(\tilde{\nu} \partial_{x} \overline{\nu_t}\right) - \partial_x\left(\left(\nu+\overline{\nu_t}\right) \partial_{x} \tilde{\nu}\right)
\end{array}\right.\\
& \hfill - c_1  \left|\partial_x \overline{u}\right| \tilde{\nu}  + 2 c_2 \tilde{\nu}\overline{\nu_t} = c_1 \sgn(\partial_x \overline{u})\overline{\nu_t}\partial_x \tilde{u} - \tilde{u}\partial_x \overline{\nu_t}.\\ \\
&\Leftrightarrow  \partial_t \tilde{q} - \tilde{\mathbf{L}} \tilde{q} = 0,
\end{array}
\end{equation}
where $\tilde{\mathbf{L}}$ is the (full) linearized operator of the closed RAB equations. We write the differential operator in block matrix form in appendix (\ref{app:op}).

In each evolution equation, 
transport terms of the relevant dependent variable are
on the left-hand sides, while the 
right-hand sides may be interpreted as forcing terms since
they are independent of the relevant dependent variable. 
A particular point to note is that this full linearized system of equations has a forcing term for the mean flow perturbation velocity $\tilde{u}$ equation involving the perturbation turbulent viscosity and the gradient of the base mean flow
\begin{equation}
F_{\nu}= \partial_x\left(\tilde{\nu} \partial_{x} \overline{u}\right),
\end{equation}
a term which plays a crucial role in the stability analysis of the total
flow when the turbulent viscosity is allowed to vary in space and time, and so $\tilde{\nu}$ is non-zero.

It is also mathematically possible to consider a perturbation to the mean flow velocity only. Indeed, the simplest way to deal with stability analysis of mean flows (or more precisely stability analysis of the coherent flow) is to consider a steady mean flow solution of the previous system which constitutes the base flow $\overline{\mathbf{q}} = \left( \overline{u},\overline{\nu_t} \right)^\top$ and then apply a perturbation $\tilde{q}_F = \left(\tilde{u},0 \right)^\top$ which has a perturbation component in the (mean flow as opposed to the stochastic field) velocity, but does not allow any variation in the turbulent viscosity, which is as a consequence ``frozen'' in its base state. This is equivalent to considering only the first equation of system (\ref{eq:pert_linearized}), and imposing $\tilde{\nu}=0$. It leads to the ``frozen turbulent viscosity perturbation equation'', defined as
\begin{equation}\label{eq:pert_frozen}
\begin{split}
& \partial_t \tilde{u} + \overline{u}\partial_x \tilde{u} + \tilde{u}\partial_x \overline{u} - \partial_x\left(\left(\nu+\overline{\nu_t}\right)\partial_{x} \tilde{u}\right) = 0 \\
& \Leftrightarrow \partial_t \tilde{u} - \tilde{\mathbf{L}}_{11} \tilde{u} = 0,
\end{split}
\end{equation}
where the operator $\tilde{\mathbf{L}}_{11}$ is the operator describing the evolution of the perturbation in a frozen turbulent viscosity context. 
By comparison of (\ref{eq:pert_linearized}) 
and (\ref{eq:pert_frozen}), it is apparent that, if represented in matrix
form $\tilde{\mathbf{L}}_{11}$ 
corresponds
to the top left block matrix of the full linearized  perturbation operator $\mathbf{\tilde{L}}$, as written in appendix (\ref{app:op}).

\subsection{Energy evolution equation}

In order to understand the various growth mechanisms, it is useful to consider an evolution equation for an appropriately defined perturbation ``energy''.

A natural choice of course is to define the perturbation
energy 
in terms
of the
coherent velocity perturbation, i.e.
\begin{equation}\label{eq:energy}
E=\dfrac{1}{2}\int_0^1 \tilde{u}^2 dx.
\end{equation}
The perturbation kinetic energy evolution equation can thus be derived by multiplying the first perturbation equation (for $\tilde{u}$) of system (\ref{eq:pert_linearized}) by the perturbation velocity $\tilde{u}$, to obtain (after various integrations by parts)
\begin{equation}\label{eq:pert_ener}
\begin{split}
\displaystyle \partial_t E = &\underbrace{-\int_0^1 \dfrac{1}{2}\tilde{u}^2 \partial_x \overline{u}\ dx}_{P_{E_1}}
													    \underbrace{-\int_0^1 \tilde{\nu} \partial_x \tilde{u} \partial_x \overline{u}\ dx}_{P_{E_2}}\\
													   &\underbrace{-\int_0^1 \left(\nu+\overline{\nu_t}\right) \left(\partial_x \tilde{u}\right)^2\ dx}_{D_E} .	
\end{split}													 
\end{equation}

The first term on the right-hand side (labeled $P_{E_1}$), is associated with the production or destruction of energy due to the interaction between the perturbation in the coherent velocity field and the base mean flow velocity. We notice that the sign of this quantity only depends on the sign of the base mean flow gradient: since this gradient is negative by definition, (required by (\ref{eq:RAB_closure})) this term is a (perturbation) energy production term. The second term (due to the presence of a forcing term in the coherent perturbation equation) also involves the perturbation in the turbulent viscosity, and is as a consequence denoted $P_{E_2}$. Assuming a negative base mean flow gradient, this term will be a source of perturbation kinetic energy when $\tilde{\nu} \partial_x \tilde{u}$ is positive, 
and will be a sink when
$\tilde{\nu} \partial_x \tilde{u}$ is negative. It represents a (in general not sign-definite) catalytic term describing the amount of energy we are able to extract from the base mean flow due to variations in the turbulent viscosity.

Of course, in the case of the frozen turbulent viscosity analysis, this term is not present, which removes a possible mechanism for perturbation kinetic energy production. The last term quantifies
the (appropriately linearized) dissipation of 
perturbation kinetic energy by both laminar
and turbulent viscosity, and 
is always negative.

To complete the ``energy'' analysis, we can also consider the evolution of the squared norm of the second component $\tilde{\nu}$. We define this quantity as:
\begin{equation}
K=\dfrac{1}{2}\int_0^1 \tilde{\nu}^2 dx.
\end{equation}

An evolution equation for this quantity can be derived following the same method described above for the perturbation energy. The evolution of the quantity $K$ is governed by the following equation:

\begin{equation}
\begin{array}{ll}
\partial_t K = &\displaystyle  \underbrace{\dfrac{1}{2} \int_0^1  \tilde{\nu}^2\left(\partial_x\overline{u}+\partial_{xx} \overline{\nu_t}\right)\ dx}_{S_{K_1}}\\
  						 &\displaystyle	 \underbrace{- \int_0^1  \left(\nu+\overline{\nu_t}\right) \left(\partial_x \tilde{\nu}\right)^2\ dx}_{D_{K_1}}
							   							 \underbrace{+ \int_0^1 c_1\left|\partial_x \overline{u}\right|\tilde{\nu}^2\ dx}_{P_{K_1}}\\
							 &\displaystyle  \underbrace{-\int_0^1  2c_2\overline{\nu_t}\nu^2\ dx}_{D_{K_2}}
							 								 \underbrace{+ \int_0^1 c_1\tilde{\nu}\partial_x\tilde{u}\sgn\left(\partial_x \overline{u}\right)\overline{\nu_t}\ dx}_{S_{K_2}}\\
							 &\displaystyle  \underbrace{-\int_0^1 \tilde{u}\tilde{\nu} \partial_x \overline{\nu_t}\ dx}_{P_{K_2}}.
\end{array}
\label{eq:Keq}
\end{equation}

Since $\partial_x\overline{u}+\partial_{xx} \overline{\nu_t}$ is always negative, the first term $S_{K_1}$ trivially acts like a sink for $K$. $D_{K_1}$ is also always a diffusive term and so negative for all time while $P_{K_1}$ and $D_{K_2}$ are trivially associated with production and destruction of turbulent viscosity. At first sight, the term $S_{K_2}$ has no obvious sign. However, the optimization of $E$ suggests from term $P_{E_2}$ that the product $\tilde{\nu}\partial_x \tilde{u}$ is positive. Furthermore, the term $\sgn\left(\partial_x \overline{u}\right)\overline{\nu_t}$ being negative, this term (in the optimization of $E$ context) will act as a new sink of $K$. Finally, the last term $P_{K_2}$ has no obvious sign and will depend on the perturbation symmetry.

Most importantly, there is no equivalent term to $P_{E_2}$ in this equation, meaning that there is no transfer from one component of the state vector to the other. Therefore, the perturbation kinetic energy $E$ can grow substantially by extracting energy from the mean flow with $\tilde{\nu}$ acting as a catalyst, rather than a direct source of energy, while $K$ may well vary more slowly.

\subsection{Optimal perturbation Lagrangian formulation}

We are now in position to
define an optimization problem using a  Lagrangian approach based upon the general framework developed in section (\ref{sec:VAR}). 
We are particularly interested
in the effect on our results
of the application (or not) of a range of increasingly more restrictive assumptions.
We will first formulate the ``FULL'' problem using the  full linearized set of equations, allowing for coherent perturbations
from the base mean flow velocity and the turbulent 
viscosity (i.e. using (\ref{eq:pert_linearized})). We then formulate the ``FROZ'' problem by deriving the equations for the frozen turbulent viscosity analysis from the complete set of equations by not allowing any perturbations for the turbulent viscosity i.e. $\tilde{\nu}=0$. Finally, we consider a specific particularly simple example, the ``LAM'' problem of a completely laminar Burgers equation, removing all turbulent viscosity from the evolution equations ($\overline{\nu_t}=0$). We present a summary of the key features of each of these three cases in table \ref{tab:cases}.

\begin{table}
  \begin{center}
    \begin{tabular}{l c c c c c c}
      \hfill & \phantom{12345} & LAM & \phantom{12} & FROZ &\phantom{12} & FULL  \\
      \\
      \hline
      \hline
      \\
      Mean steady flow $\overline{\mathbf{q}}$ & & $(\overline{u},0)^\top$ & & $(\overline{u},\overline{\nu_t})^\top$ & & $(\overline{u},\overline{\nu_t})^\top$ \\
      Perturbation $\tilde{\mathbf{q}}$ & & $(\tilde{u},0)^\top$ & & $(\tilde{u},0)^\top$ & & $(\tilde{u},\tilde{\nu})^\top$ \\
      \\
      \hline  
      \hline
      \\ \end{tabular}
    \caption{\label{tab:cases}Summary of the different cases considered. LAM: Laminar analysis, FROZ: Frozen turbulent viscosity analysis, FULL: Full linearized analysis.}
  \end{center}
\end{table}

\subsubsection{Semi-norm gain}

In all cases, we are interested in the gain of the perturbation 
kinetic energy over a finite time interval $[0,T]$, and so we define the objective functional which we optimize as
\begin{equation}\label{eq:objfdef}
\mathcal{J}(\tilde{\mathbf{q}})=E(T)=\left\| \tilde{\mathbf{q}}(T) \right\|_E^2,
\end{equation}
where the ``energy'' semi-norm is defined as
\begin{equation}\label{eq:wedef}
\begin{array}{c}
\left\| \tilde{\mathbf{q}}(T) \right\|_E^2 = \dfrac{1}{2}\displaystyle\int_0^1{\tilde{\mathbf{q}}(T)^\top\mathbf{W}_E\tilde{\mathbf{q}}(T)\ dx},
\\
\mathbf{W}_E=\left(
\begin{array}{cc}
1 & 0\\
0 & 0
\end{array}
\right).
\end{array}
\end{equation}
This is clearly a specific (and very simple) example of the energy semi-norms described in section (\ref{sec:VAR}), and
we can also define the appropriate complementary semi-norm, acting on the kernel of $\left\| \cdot \right\|_E$ defined as:
\begin{equation}\label{eq:wkdef}
\begin{array}{c}
\left\| \tilde{\mathbf{q}}(T) \right\|_K^2 = \dfrac{1}{2}\displaystyle\int_0^1{\tilde{\mathbf{q}}(T)^\top \mathbf{W}_K\tilde{\mathbf{q}}(T)\ dx},
\\
\mathbf{W}_K=\mathbf{I}-\mathbf{W}_E, \hfill
\end{array}
\end{equation}
with associated normalization norm defined as
\begin{equation}\label{eq:wndef}
\begin{array}{c}
\left\| \tilde{\mathbf{q}}(T) \right\|_N^2 = \dfrac{1}{2}\displaystyle\int_0^1{\tilde{\mathbf{q}}(T)^\top \mathbf{W}_N\tilde{\mathbf{q}}(T)\ dx},
\\
\mathbf{W}_N=\mathbf{I}=\mathbf{W}_E+\mathbf{W}_K.  \hfill
\end{array}
\end{equation}
Here it is clear that the appropriate ``energy'' is thus defined as a semi-norm of the state vector $\tilde{\mathbf{q}}$. We use this expression, because we have seen in (\ref{eq:energy}) that from a stability point of view, the most relevant quantity to look at is the kinetic energy of the perturbation  velocity $\tilde{u}$ from the base mean flow. The kernel of this energy nor thus exclusively contains perturbations of the 
turbulent viscosity $\tilde{\nu}$. As we have shown in the method developed in (\ref{sec:VAR}), the normalization constraints are thus more subtle, because the initial magnitude of these turbulent viscosity perturbations must also be constrained.

Constraints for the fully linearized system (FULL) are the dynamical constraint (\ref{eq:pert_linearized}) 
requiring that the state vector satisfies the appropriate evolution equation and the initial condition
\begin{equation}
\begin{array}{c}
\tilde{\mathbf{q}}(x,0)-\tilde{\mathbf{q}}_0=0,\\
\tilde{\mathbf{q}}_0 = ( \tilde{u}_0,\tilde{\nu}_0)^\top.
\end{array}
\end{equation}
The required constraints for normalization are  
\begin{equation}
\left\{
\begin{array}{ll}
E_0 - \left\| \tilde{q}_0 \right\|_E^2 = 0,\\
K_0 - \left\| \tilde{q}_0 \right\|_K^2 = 0,
\end{array}
\right.
\end{equation}
with $E_0$ the initial amount of perturbation energy from the (coherent) velocity  and $K_0$ is the initial amount of turbulent viscosity perturbation in the system, i.e.
\begin{equation}
\left\{
\begin{array}{ll}
E_0 - \dfrac{1}{2}\displaystyle\int_0^1{\tilde{u}_0^2\ dx} = 0,\\
K_0 - \dfrac{1}{2}\displaystyle\int_0^1{\tilde{\nu}_0^2\ dx} = 0.
\end{array}
\right.
\end{equation}
As we explained in section (\ref{sec:VAR}), we introduce a new parameter $R_0$
describing the relative contribution of $E_0$ and $K_0$ to the initial normalization of the perturbation state vector
$N_0$, 
where
$R_0$ is defined as
\begin{equation}
R_{0}=\dfrac{K_0}{N_0}= 1 -\dfrac{E_0}{N_0}.
\label{eq:opt_ratios}
\end{equation}
Since our problem is linear, the total norm $N_0$ has no influence on the dynamics of the flow and as a consequence 
the ratio $C_0$ of $K_0$ and $E_0$ is sometimes a more relevant parameter. This ratio can 
be straightforwardly related to the parameter
$R_0$:
\begin{equation}\label{eq:ratio_0}
C_0 = \dfrac{K_0}{E_0} = \dfrac{R_{0}}{1-R_{0}}.
\end{equation}

These quantities represent the initial structure of the state vector. These quantities are however of interest for any time, and we will as a consequence extend their definition for all $t$ such that 
\begin{equation}
\begin{array}{c l c}
R(t) &= \dfrac{K(t)}{N(t)},\ & R(0)=R_0,\\
C(t) &= \dfrac{K(t)}{E(t)},\ & C(0)=C_0.
\end{array}
\label{eq:rtdef}
\end{equation}

We notice that we can also rewrite the coefficient $C(t)$ the following ways:
\begin{equation}
C(t) = \dfrac{R(t)}{1-R(t)} = \dfrac{\displaystyle\int_0^1 \tilde{\nu}(t)^2 dx}{\displaystyle\int_0^1 \tilde{u}(t)^2 dx}.
\label{eq:rtdef2}
\end{equation}

Therefore, we can express the Lagrangian functional of our problem as:
\begin{equation}\label{eq:lagrangian}
\begin{split}
\mathcal{L}(\tilde{\mathbf{q}},\tilde{\mathbf{q}}_0,\tilde{\mathbf{q}}^\dagger,\tilde{\mathbf{q}}_0^\dagger,\lambda_E,\lambda_K) & = \left\| \tilde{\mathbf{q}}(T) \right\|_E^2 \\
																																								 & - \left( \partial_t \tilde{\mathbf{q}} - \tilde{\mathbf{L}} \tilde{\mathbf{q}}  , \tilde{\mathbf{q}}^\dagger \right) \\ 
																	      																				 & - \left< \tilde{\mathbf{q}}_0 - \tilde{\mathbf{q}}(x,0) ,\tilde{\mathbf{q}}_0^\dagger \right>\\
																	      																				 & - \lambda_E\left(E_0 -\left\| \tilde{\mathbf{q}}_0 \right\|_E^2 \right)\\
																	      																				 & - \lambda_K\left(K_0 - \left\| \tilde{\mathbf{q}}_0 \right\|_K^2 \right).
\end{split}
\end{equation}
All the variations of the Lagrangian functional with respect to the parameters have to vanish, i.e. $\delta \mathcal{L} = 0$. Once again, we note that taking variations with respect to the adjoint variables will yield the constraints on the initial
condition, and the underlying evolution equation. Conversely, taking variations with respect to the direct variables yields the adjoint set of equations
\begin{equation}
\label{eq:adjoint}
-\partial_t \tilde{\mathbf{q}}^\dagger + \tilde{\mathbf{L}}^\dagger \tilde{\mathbf{q}}^\dagger = 0.
\end{equation}
The operator $\tilde{\mathbf{L}}^\dagger$ is the adjoint of the direct (full) perturbation RAB operator $\tilde{\mathbf{L}}$
defined  in (\ref{eq:pert_linearized}) and in appendix (\ref{app:op}), where $\tilde{\mathbf{L}}^\dagger$ is also written
out in full.

This adjoint equation also has the ``final'' condition:
\begin{equation}\label{eq:adjoint_IC}
\tilde{\mathbf{q}}^\dagger(T) = \mathbf{W}_E \tilde{\mathbf{q}}(T).
\end{equation}
Since $\mathbf{W}_E \tilde{\mathbf{q}}(T) = \tilde{u}$ by definition, the ``final'' condition for the adjoint turbulent viscosity perturbation is zero. Nevertheless, because of the coupling terms in the adjoint equation (\ref{eq:adjoint}), the adjoint turbulent viscosity does not remain zero during its evolution.
We also obtain homogeneous Dirichlet boundary conditions $\tilde{\mathbf{q}}^\dagger(\partial\Omega)=0$, and a natural compatibility condition linking $\tilde{\mathbf{q}}^\dagger_0$ to the initial condition of the adjoint problem, i.e. $\tilde{\mathbf{q}}^\dagger_0=\tilde{\mathbf{q}}^\dagger(0)$.

Finally, we take variations of $\mathcal{L}$ with respect to the initial condition $\tilde{\mathbf{q}}_0$, which gives us the expression for $\frac{\delta\mathcal{L}}{\delta \tilde{\mathbf{q}}_0}\delta \tilde{\mathbf{q}}_0 $ which immediately yields gradient information to optimize the objective functional $\mathcal{J}$:
\begin{equation}
\nabla_{\tilde{\mathbf{q}}_0} \mathcal{J} = \tilde{\mathbf{q}}^\dagger_0 - \left( \lambda_E \mathbf{W}_E  + \lambda_K \left( \mathbf{I} - \mathbf{W}_E\right) \right) \tilde{\mathbf{q}}_0.
\end{equation}
In order to find the maximum of our functional, we will use this gradient information to find the optimal initial condition realizing  maximum energy at time $t=T$.
Therefore in summary, this framework  allows us to find the optimal initial perturbation associated with the maximum energy gain $G_E(T)$ over an optimization time interval, where $G_E(T)$ is defined as:
\begin{equation}\label{eq:gain_u}
G_E(T)=\dfrac{E(T)}{E_0}.
\end{equation}

\subsubsection{Full norm gain and frozen turbulent viscosity analysis}

In order to compare our framework to already existing tools, we will also perform a classical SVD analysis of the system. In this case, the norm optimized is the total norm defined (as in \ref{eq:wndef}) by:
\begin{equation}
\begin{split}
\left\| \tilde{\mathbf{q}} \right\|_N^2 = & \left\| \tilde{\mathbf{q}} \right\|_E^2 + \left\| \tilde{\mathbf{q}} \right\|_K^2\\
 													     = & \dfrac{1}{2}\displaystyle\int_0^1{\left(\tilde{u}^2+\tilde{\nu}^2\right) dx}
 \end{split}
\end{equation}
As a consequence, the gain we identify is not an energy gain, but a ``total'' gain (which has no particular physical meaning). It is simply defined as:
\begin{equation}\label{eq:total_gain}
G_N(T)=\dfrac{N(T)}{N_0}=G_E(T)g(C_0),
\end{equation}
where
\begin{equation}
g(C_0)=\frac{1+C(T)}{1+C_0}=\frac{1+h(C_0)}{1+C_0},
\end{equation}
where $h$ is a function embedding all the dynamics relating the initial ratio $C_0$ (as defined in (\ref{eq:rtdef})) to the final ratio at the end of the optimization time interval $C(T)$.
This expression shows that the SVD optimization by construction cannot correctly describe the physics of $G_E$ since it implicitly optimizes the product of $G_E$ and a nontrivial function of the initial ratio $C_0$, which product has no physical meaning.

In the frozen turbulent viscosity case (FROZ), the perturbation vector is $\tilde{\mathbf{q}}=\left(\tilde{u},0\right)^\top$, i.e. $\tilde{\nu}=0$ in the system of direct and adjoint equations. The state vector has only one component and the energy norm is once again a true norm, 
$\| \tilde{\mathbf{q}} \|_E=\| \tilde{\mathbf{q}} \|_N$, 
allowing  the use of the well-known singular value decomposition (SVD) analysis technique, which is explained briefly in appendix (\ref{app:svd}).
The laminar case (LAM) is a further special case where the frozen turbulent viscosity 
$\overline{\nu_t}$ is set precisely to zero. In these two cases, $C_0=0$ (also true for all times) so that the two gains $G_E$ and $G_N$ defined in (\ref{eq:gain_u}) and (\ref{eq:total_gain}) respectively are naturally equal.

\subsection{Sensitivity analysis}

We emphasized in section (\ref{sec:VAR}) that our Lagrangian framework was not only a way to perform an optimization subject to constraints, but also a way to analyze the sensitivity of the objective functional to those constraints. In the particular class of problems under consideration,  the state vector is constrained by a partial differential equation which is entirely defined by the base mean flow $\overline{\mathbf{q}}$ and the parameters $\nu$, $c_1$ and $c_2$. The sensitivity of the optimized objective functional to a small change in any of these parameters can be retrieved thanks to the additional sensitivity information in the adjoint state vector.
In terms of the nomenclature of section (\ref{sec:VAR}), all these parameters are constraint parameters, as they do not feature
explicitly in the objective functional defined by (\ref{eq:objfdef}).

\subsubsection{Sensitivity with respect to the mean flow}

We first consider  the sensitivity of the objective functional with respect to the mean flow. This consists of computing the change of the Lagrangian functional $\mathcal{L}$ (defined in (\ref{eq:lagrangian})) when we allow a small variation in the mean flow components. Since the mean flow is time-independent, so are the infinitesimal variations $\delta \overline{\mathbf{q}}=(\delta \overline{u},\delta \overline{\nu_t})^\top$. 

The sensitivities information is computed by taking the functional derivative of $\mathcal{L}$ with respect to the base flow. The full sensitivity vector ($\nabla_{\overline{\mathbf{q}}}\mathcal{J}$) has two components and they can be expressed, after some integration by parts as
\begin{equation}\begin{split}
\nabla_{\overline{u}}\mathcal{J} & = \displaystyle \int_0^T{ \tilde{\mathbf{S}}_{\overline{u}}(\tilde{\mathbf{q}}^\dagger,\tilde{\mathbf{q}})\ dt}, \\
\nabla_{\overline{\nu_t}}\mathcal{J}				 & = \displaystyle \int_0^T{ \tilde{\mathbf{S}}_{\overline{\nu_t}}(\tilde{\mathbf{q}}^\dagger,\tilde{\mathbf{q}})\ dt},
\end{split}
\label{eq:sens_bf}
\end{equation}
where 
the explicit expressions of the sensitivity vectors $\tilde{\mathbf{S}}_{\overline{u}}(\overline{\mathbf{q}})$ and $\tilde{\mathbf{S}}_{\overline{\nu_t}}$ are given in appendix (\ref{app:sens}).
We notice that the sensitivity with respect to the base mean flow is a time scalar product (time integral). This means that  this sensitivity is the cumulative contribution of the base mean flow variation at each time step.
Therefore, in a practical situation, the 
longer the time interval for the optimization, the larger will be the error in the evaluation of the objective functional
if there is any uncertainty in the mean flow state
vector.

\subsubsection{Sensitivity with respect to parameters}

This model has three parameters: $\nu$, $c_1$ and $c_2$. In order to derive the sensitivity of the optimal energy with respect to these parameters, we first need to notice that a change in their value will not only change the dynamics governing the perturbation equation but also the mean flow equation and therefore, the mean flow itself. Thus, we need to define a new functional to account for the change in the base mean flow because of the small variation we allow in the parameters. We then incorporate the base mean flow equations (\ref{eq:RAB_sys}) in the Lagrangian functional. This Lagrangian can be expressed as an extension of the one defined previously in (\ref{eq:lagrangian}):
\begin{equation}
\mathcal{L}'(\tilde{\mathbf{q}},\tilde{\mathbf{q}}^\dagger,\overline{\mathbf{q}},\overline{\mathbf{q}}^\dagger) =  \mathcal{J}(\tilde{\mathbf{q}}) - \left( \tilde{\mathbf{L}}\tilde{\mathbf{q}}, \tilde{\mathbf{q}}^\dagger\right) - \left< \overline{\mathbf{L}}\left(\overline{\mathbf{q}}\right), \overline{\mathbf{q}}^\dagger\right>.
\end{equation}
Consequently, we have a new Lagrange multiplier which is the base flow adjoint state vector $\overline{\mathbf{q}}^\dagger=(\overline{u}^\dagger,\overline{\nu_t}^\dagger)^\top$. In order to fulfill the optimality condition of the problem, we have to satisfy the condition:
\begin{equation}
\dfrac{\partial \mathcal{L}'}{\partial \overline{\mathbf{q}}}\delta \overline{\mathbf{q}}=0.
\end{equation}

This condition leads to the definition of the base flow adjoint variables. We notice that the adjoint base flow system is no longer homogeneous, but is additionally forced by the sensitivity with respect to the base flow:
\begin{equation}
\overline{\mathbf{L}}^\dagger\overline{\mathbf{q}}^\dagger = \nabla_{\overline{\mathbf{q}}} \mathcal{J}.
\end{equation}
Therefore,
\begin{equation}
\begin{split}
\nabla_{\nu}\mathcal{J} & = \left( \tilde{\mathbf{q}}^\dagger,\tilde{\mathbf{S}}_{\nu}\tilde{\mathbf{q}} \right) +  \left< \overline{\mathbf{q}}^\dagger,\overline{\mathbf{S}}_{\nu}(\overline{\mathbf{q}}) \right>,\\
\nabla_{c1}\mathcal{J} & = \left( \tilde{\mathbf{q}}^\dagger,\tilde{\mathbf{S}}_{c1}\tilde{\mathbf{q}} \right) +  \left< \overline{\mathbf{q}}^\dagger,\overline{\mathbf{S}}_{c1}(\overline{\mathbf{q}}) \right>,\\
\nabla_{c2}\mathcal{J} & = \left( \tilde{\mathbf{q}}^\dagger,\tilde{\mathbf{S}}_{c2}\tilde{\mathbf{q}} \right) +  \left< \overline{\mathbf{q}}^\dagger,\overline{\mathbf{S}}_{c2}(\overline{\mathbf{q}}) \right>.\\
\end{split}
\label{eq:sens_param}
\end{equation}
The expression of the sensitivity matrices $\tilde{\mathbf{S}}$ and vectors $\overline{\mathbf{S}}(\overline{\mathbf{q}})$ are also given in appendix (\ref{app:sens}). The sensitivities have two contributions: a space-time scalar product accounting for the sensitivity due to the perturbation equation,  and a space-only scalar product accounting for  the sensitivity due to the base mean flow change induced by variation of the
relevant parameter.
	
\section{Results}\label{sec:RES}

	\subsection{Optimal perturbations}

The results are presented in three parts, in order of increasing complexity. First of all, we will consider the laminar case ``LAM'' as summarized in table (\ref{tab:cases}), i.e. the stability analysis of the Burgers equation, without any coupling with another partial differential equation, and in particular constant viscosity with no turbulent viscosity contribution. We then consider the ``FROZ''  case for a particular
constant nonzero choice $\overline{\nu_t}$ of the   turbulent viscosity, 
and considering the stability of the RAB equations with only a (coherent) perturbation
velocity $\tilde{u}$, which
allows us to understand the impact
of a constant 
turbulent viscosity on the system. Finally, we  consider the behaviour of the full linearized model ``FULL'',
using both our semi-norm based framework and an SVD analysis based on optimizing the total
true norm of the system. By considering the results of our framework and the unphysical SVD in tandem, we are able to 
identify the significance or otherwise of the output of the SVD analysis in a consistent manner.
Fixing the turbulent viscosity to its mean value 
is a common simplifying assumption, and we
are very interested in the robustness of our results
to the application of this assumption.

\subsubsection{Laminar analysis: the LAM case}
Let us consider  Burgers equation, with a constant and uniform eddy viscosity $\nu=0.05$. The equation governing the evolution of a perturbation of the form $\tilde{\mathbf{q}}=(\tilde{u},0)^\top$ is thus given by (\ref{eq:pert_frozen}) with $\overline{\nu_t}=0$. 
In this case, the perturbation kinetic energy  is simply the 2-norm of the state vector, and so we can use an SVD  analysis (as described in appendix (\ref{app:svd}). The optimal gain is then given by the largest singular value of the evolution operator. 
Here, the production of energy can only come from the coupling between the coherent perturbation $\tilde{u}$ and the base mean flow $\overline{u}$, (i.e. via term $P_{E_1}$ of equation (\ref{eq:pert_ener})) and since we are considering only focussing base mean flows (with negative slopes), we will have some energy production in the middle of the domain, due to the focussing of the perturbation.

\begin{figure}
		\centering 
%
%
%
\providelength{\AxesLineWidth}       \setlength{\AxesLineWidth}{0.5pt}%
\providelength{\plotwidth}           \setlength{\plotwidth}{5cm}
\providelength{\LineWidth}           \setlength{\LineWidth}{0.7pt}%
\providelength{\MarkerSize}          \setlength{\MarkerSize}{4pt}%
\newrgbcolor{GridColor}{0.8 0.8 0.8}%
%
\psset{xunit=0.500000\plotwidth,yunit=0.262903\plotwidth}%
\begin{pspicture}(-0.420000,-0.570552)(2.000000,3.000000)%


\psline[linewidth=\AxesLineWidth,linecolor=GridColor](0.000000,0.000000)(0.000000,0.045644)
\psline[linewidth=\AxesLineWidth,linecolor=GridColor](1.000000,0.000000)(1.000000,0.045644)
\psline[linewidth=\AxesLineWidth,linecolor=GridColor](2.000000,0.000000)(2.000000,0.045644)
\psline[linewidth=\AxesLineWidth,linecolor=GridColor](0.000000,0.000000)(0.024000,0.000000)
\psline[linewidth=\AxesLineWidth,linecolor=GridColor](0.000000,1.000000)(0.024000,1.000000)
\psline[linewidth=\AxesLineWidth,linecolor=GridColor](0.000000,2.000000)(0.024000,2.000000)
\psline[linewidth=\AxesLineWidth,linecolor=GridColor](0.000000,3.000000)(0.024000,3.000000)

{ \footnotesize 
\rput[t](0.000000,-0.045644){$0$}
\rput[t](1.000000,-0.045644){$1$}
\rput[t](2.000000,-0.045644){$2$}
\rput[r](-0.024000,0.000000){$0$}
\rput[r](-0.024000,1.000000){$1$}
\rput[r](-0.024000,2.000000){$2$}
\rput[r](-0.024000,3.000000){$3$}
} 

\psframe[linewidth=\AxesLineWidth,dimen=middle](0.000000,0.000000)(2.000000,3.000000)

{ \small 
\rput[b](1.000000,-0.570552){
\begin{tabular}{c}
$T$\\
\end{tabular}
}

\rput[t]{90}(-0.420000,1.500000){
\begin{tabular}{c}
$G_E$\\
\end{tabular}
}
} 

\newrgbcolor{color218.0284}{0  0  0}
\psline[plotstyle=line,linejoin=1,linestyle=solid,linewidth=\LineWidth,linecolor=color218.0284]
(0.005000,1.019110)(0.020000,1.078421)(0.035000,1.140347)(0.055000,1.226266)(0.075000,1.315244)
(0.100000,1.429682)(0.140000,1.617246)(0.185000,1.828678)(0.210000,1.943097)(0.230000,2.031612)
(0.245000,2.095695)(0.260000,2.157460)(0.275000,2.216640)(0.290000,2.273018)(0.305000,2.326431)
(0.320000,2.376766)(0.330000,2.408582)(0.340000,2.438997)(0.350000,2.468014)(0.360000,2.495644)
(0.370000,2.521905)(0.380000,2.546822)(0.390000,2.570424)(0.400000,2.592747)(0.410000,2.613828)
(0.420000,2.633709)(0.430000,2.652434)(0.440000,2.670049)(0.450000,2.686600)(0.460000,2.702134)
(0.470000,2.716699)(0.480000,2.730343)(0.490000,2.743112)(0.500000,2.755053)(0.510000,2.766210)
(0.520000,2.776626)(0.530000,2.786345)(0.540000,2.795407)(0.550000,2.803852)(0.560000,2.811716)
(0.570000,2.819035)(0.580000,2.825844)(0.595000,2.835171)(0.610000,2.843525)(0.625000,2.851000)
(0.640000,2.857682)(0.655000,2.863651)(0.670000,2.868977)(0.685000,2.873727)(0.700000,2.877958)
(0.720000,2.882886)(0.740000,2.887101)(0.760000,2.890700)(0.785000,2.894461)(0.810000,2.897533)
(0.840000,2.900476)(0.875000,2.903090)(0.915000,2.905249)(0.960000,2.906897)(1.015000,2.908124)
(1.085000,2.908871)(1.185000,2.909033)(1.350000,2.908271)(1.885000,2.904037)(2.000000,2.903080)

\newrgbcolor{color219.028}{0.50196     0.50196     0.50196}
\psline[plotstyle=line,linejoin=1,linestyle=solid,linewidth=\LineWidth,linecolor=color219.028]
(0.005000,1.005312)(0.040000,1.042358)(0.060000,1.062812)(0.075000,1.077515)(0.090000,1.091509)
(0.105000,1.104672)(0.120000,1.116899)(0.135000,1.128110)(0.145000,1.134993)(0.155000,1.141389)
(0.165000,1.147295)(0.175000,1.152711)(0.185000,1.157641)(0.195000,1.162092)(0.205000,1.166077)
(0.215000,1.169610)(0.225000,1.172708)(0.240000,1.176579)(0.255000,1.179580)(0.270000,1.181782)
(0.285000,1.183257)(0.300000,1.184076)(0.315000,1.184306)(0.330000,1.184012)(0.350000,1.182907)
(0.370000,1.181102)(0.395000,1.178034)(0.420000,1.174236)(0.450000,1.168929)(0.485000,1.161969)
(0.530000,1.152185)(0.590000,1.138259)(0.710000,1.109188)(0.880000,1.067913)(1.010000,1.037112)
(1.135000,1.008281)(1.260000,0.980237)(1.385000,0.952968)(1.515000,0.925411)(1.645000,0.898651)
(1.775000,0.872665)(1.910000,0.846474)(2.000000,0.829452)

\end{pspicture}%
		\caption{Optimal gain  against time for the ``LAM''  case (plotted with a black line) and the ``FROZ'' case  (plotted with a grey line). The amplification in the LAM case is due to the positive energy production term $P_{E_1}$ of equation (\ref{eq:pert_ener}), which can be seen as a focussing of the perturbation in the middle part of the domain. The gain then decreases very slowly due to the low value of the viscosity. In the FROZ case with $r=0.5$, (as defined in (\ref{eq:ratio})) the transient growth remains but is reduced dramatically because of the larger total viscosity. For sufficiently large values of $r$, any transient growth can be completely suppressed due to  strong viscous damping.}
		\label{fig:gain_lam_fr}
\end{figure}
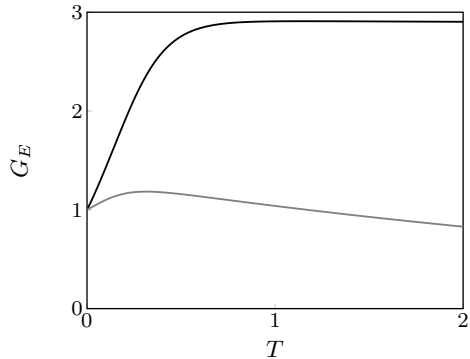

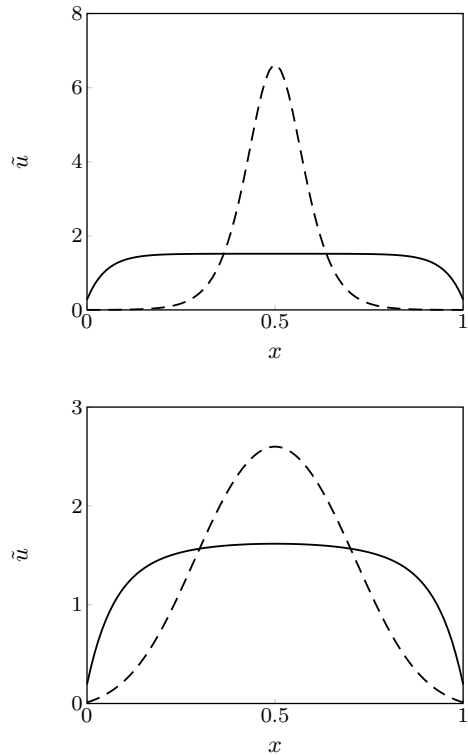
\begin{figure}
\begin{center}
%
%
%
\providelength{\AxesLineWidth}       \setlength{\AxesLineWidth}{0.5pt}%
\providelength{\plotwidth}           \setlength{\plotwidth}{5cm}
\providelength{\LineWidth}           \setlength{\LineWidth}{0.7pt}%
\providelength{\MarkerSize}          \setlength{\MarkerSize}{4pt}%
\newrgbcolor{GridColor}{0.8 0.8 0.8}%
%
\psset{xunit=1.000000\plotwidth,yunit=0.098589\plotwidth}%
\begin{pspicture}(-0.210000,-1.521472)(1.000000,8.000000)%


\psline[linewidth=\AxesLineWidth,linecolor=GridColor](0.000000,0.000000)(0.000000,0.121718)
\psline[linewidth=\AxesLineWidth,linecolor=GridColor](0.500000,0.000000)(0.500000,0.121718)
\psline[linewidth=\AxesLineWidth,linecolor=GridColor](1.000000,0.000000)(1.000000,0.121718)
\psline[linewidth=\AxesLineWidth,linecolor=GridColor](0.000000,0.000000)(0.012000,0.000000)
\psline[linewidth=\AxesLineWidth,linecolor=GridColor](0.000000,2.000000)(0.012000,2.000000)
\psline[linewidth=\AxesLineWidth,linecolor=GridColor](0.000000,4.000000)(0.012000,4.000000)
\psline[linewidth=\AxesLineWidth,linecolor=GridColor](0.000000,6.000000)(0.012000,6.000000)
\psline[linewidth=\AxesLineWidth,linecolor=GridColor](0.000000,8.000000)(0.012000,8.000000)

{ \footnotesize 
\rput[t](0.000000,-0.121718){$0$}
\rput[t](0.500000,-0.121718){$0.5$}
\rput[t](1.000000,-0.121718){$1$}
\rput[r](-0.012000,0.000000){$0$}
\rput[r](-0.012000,2.000000){$2$}
\rput[r](-0.012000,4.000000){$4$}
\rput[r](-0.012000,6.000000){$6$}
\rput[r](-0.012000,8.000000){$8$}
} 

\psframe[linewidth=\AxesLineWidth,dimen=middle](0.000000,0.000000)(1.000000,8.000000)

{ \small 
\rput[b](0.500000,-1.521472){
\begin{tabular}{c}
$x$\\
\end{tabular}
}

\rput[t]{90}(-0.210000,4.000000){
\begin{tabular}{c}
$\tilde{u}$\\
\end{tabular}
}
} 

\newrgbcolor{color206.0393}{0  0  0}
\psline[plotstyle=line,linejoin=1,linestyle=solid,linewidth=\LineWidth,linecolor=color206.0393]
(0.000000,0.277992)(0.010101,0.504979)(0.020202,0.690322)(0.030303,0.841664)(0.040404,0.965246)
(0.050505,1.066161)(0.060606,1.148569)(0.070707,1.215868)(0.080808,1.270828)(0.090909,1.315715)
(0.101010,1.352375)(0.111111,1.382320)(0.121212,1.406780)(0.131313,1.426761)(0.141414,1.443084)
(0.151515,1.456421)(0.161616,1.467319)(0.171717,1.476224)(0.181818,1.483502)(0.191919,1.489450)
(0.202020,1.494313)(0.212121,1.498289)(0.222222,1.501539)(0.232323,1.504198)(0.242424,1.506372)
(0.252525,1.508151)(0.262626,1.509606)(0.272727,1.510796)(0.292929,1.512568)(0.313131,1.513755)
(0.343434,1.514844)(0.383838,1.515571)(0.464646,1.516025)(0.575758,1.515893)(0.636364,1.515280)
(0.666667,1.514550)(0.696970,1.513221)(0.717172,1.511770)(0.727273,1.510796)(0.737374,1.509606)
(0.747475,1.508151)(0.757576,1.506372)(0.767677,1.504198)(0.777778,1.501539)(0.787879,1.498289)
(0.797980,1.494313)(0.808081,1.489450)(0.818182,1.483502)(0.828283,1.476224)(0.838384,1.467319)
(0.848485,1.456421)(0.858586,1.443084)(0.868687,1.426761)(0.878788,1.406780)(0.888889,1.382320)
(0.898990,1.352375)(0.909091,1.315715)(0.919192,1.270828)(0.929293,1.215868)(0.939394,1.148569)
(0.949495,1.066161)(0.959596,0.965246)(0.969697,0.841664)(0.979798,0.690322)(0.989899,0.504979)
(1.000000,0.277992)

\newrgbcolor{color207.0388}{0  0  0}
\psline[plotstyle=line,linejoin=1,linestyle=dashed,linewidth=\LineWidth,linecolor=color207.0388]
(0.000000,0.000216)(0.030303,0.001199)(0.050505,0.002276)(0.070707,0.003890)(0.080808,0.004977)
(0.090909,0.006306)(0.101010,0.007933)(0.111111,0.009924)(0.121212,0.012360)(0.131313,0.015340)
(0.141414,0.018986)(0.151515,0.023446)(0.161616,0.028902)(0.171717,0.035574)(0.181818,0.043733)
(0.191919,0.053708)(0.202020,0.065900)(0.212121,0.080796)(0.222222,0.098988)(0.232323,0.121194)
(0.242424,0.148282)(0.252525,0.181297)(0.262626,0.221496)(0.272727,0.270382)(0.282828,0.329740)
(0.292929,0.401680)(0.303030,0.488668)(0.313131,0.593558)(0.323232,0.719599)(0.333333,0.870424)
(0.343434,1.049987)(0.353535,1.262437)(0.363636,1.511900)(0.373737,1.802143)(0.383838,2.136079)
(0.393939,2.515114)(0.404040,2.938325)(0.414141,3.401527)(0.424242,3.896336)(0.444444,4.922189)
(0.454545,5.411265)(0.464646,5.849864)(0.474747,6.210270)(0.484848,6.467101)(0.494949,6.600834)
(0.505051,6.600834)(0.515152,6.467101)(0.525253,6.210270)(0.535354,5.849864)(0.545455,5.411265)
(0.555556,4.922189)(0.575758,3.896336)(0.585859,3.401527)(0.595960,2.938325)(0.606061,2.515114)
(0.616162,2.136079)(0.626263,1.802143)(0.636364,1.511900)(0.646465,1.262437)(0.656566,1.049987)
(0.666667,0.870424)(0.676768,0.719599)(0.686869,0.593558)(0.696970,0.488668)(0.707071,0.401680)
(0.717172,0.329740)(0.727273,0.270382)(0.737374,0.221496)(0.747475,0.181297)(0.757576,0.148282)
(0.767677,0.121194)(0.777778,0.098988)(0.787879,0.080796)(0.797980,0.065900)(0.808081,0.053708)
(0.818182,0.043733)(0.828283,0.035574)(0.838384,0.028902)(0.848485,0.023446)(0.858586,0.018986)
(0.868687,0.015340)(0.878788,0.012360)(0.888889,0.009924)(0.898990,0.007933)(0.909091,0.006306)
(0.919192,0.004977)(0.939394,0.003002)(0.959596,0.001683)(0.989899,0.000480)(1.000000,0.000216)

\end{pspicture}%
  \vspace{0.5cm}
%
%
%
\providelength{\AxesLineWidth}       \setlength{\AxesLineWidth}{0.5pt}%
\providelength{\plotwidth}           \setlength{\plotwidth}{5cm}
\providelength{\LineWidth}           \setlength{\LineWidth}{0.7pt}%
\providelength{\MarkerSize}          \setlength{\MarkerSize}{4pt}%
\newrgbcolor{GridColor}{0.8 0.8 0.8}%
%
\psset{xunit=1.000000\plotwidth,yunit=0.262903\plotwidth}%
\begin{pspicture}(-0.210000,-0.570552)(1.000000,3.000000)%


\psline[linewidth=\AxesLineWidth,linecolor=GridColor](0.000000,0.000000)(0.000000,0.045644)
\psline[linewidth=\AxesLineWidth,linecolor=GridColor](0.500000,0.000000)(0.500000,0.045644)
\psline[linewidth=\AxesLineWidth,linecolor=GridColor](1.000000,0.000000)(1.000000,0.045644)
\psline[linewidth=\AxesLineWidth,linecolor=GridColor](0.000000,0.000000)(0.012000,0.000000)
\psline[linewidth=\AxesLineWidth,linecolor=GridColor](0.000000,1.000000)(0.012000,1.000000)
\psline[linewidth=\AxesLineWidth,linecolor=GridColor](0.000000,2.000000)(0.012000,2.000000)
\psline[linewidth=\AxesLineWidth,linecolor=GridColor](0.000000,3.000000)(0.012000,3.000000)

{ \footnotesize 
\rput[t](0.000000,-0.045644){$0$}
\rput[t](0.500000,-0.045644){$0.5$}
\rput[t](1.000000,-0.045644){$1$}
\rput[r](-0.012000,0.000000){$0$}
\rput[r](-0.012000,1.000000){$1$}
\rput[r](-0.012000,2.000000){$2$}
\rput[r](-0.012000,3.000000){$3$}
} 

\psframe[linewidth=\AxesLineWidth,dimen=middle](0.000000,0.000000)(1.000000,3.000000)

{ \small 
\rput[b](0.500000,-0.570552){
\begin{tabular}{c}
$x$\\
\end{tabular}
}

\rput[t]{90}(-0.210000,1.500000){
\begin{tabular}{c}
$\tilde{u}$\\
\end{tabular}
}
} 

\newrgbcolor{color194.0386}{0  0  0}
\psline[plotstyle=line,linejoin=1,linestyle=solid,linewidth=\LineWidth,linecolor=color194.0386]
(0.000000,0.194074)(0.010101,0.363075)(0.020202,0.510383)(0.030303,0.638918)(0.040404,0.751204)
(0.050505,0.849422)(0.060606,0.935455)(0.070707,1.010929)(0.080808,1.077246)(0.090909,1.135617)
(0.101010,1.187087)(0.111111,1.232556)(0.121212,1.272803)(0.131313,1.308498)(0.141414,1.340219)
(0.151515,1.368469)(0.161616,1.393678)(0.171717,1.416222)(0.181818,1.436422)(0.191919,1.454560)
(0.202020,1.470878)(0.212121,1.485588)(0.222222,1.498871)(0.232323,1.510888)(0.242424,1.521777)
(0.252525,1.531658)(0.262626,1.540637)(0.272727,1.548805)(0.282828,1.556243)(0.292929,1.563021)
(0.303030,1.569200)(0.313131,1.574834)(0.323232,1.579970)(0.333333,1.584650)(0.343434,1.588909)
(0.353535,1.592781)(0.363636,1.596291)(0.373737,1.599466)(0.383838,1.602326)(0.393939,1.604891)
(0.404040,1.607176)(0.414141,1.609196)(0.424242,1.610964)(0.434343,1.612488)(0.444444,1.613780)
(0.454545,1.614845)(0.464646,1.615690)(0.474747,1.616320)(0.484848,1.616738)(0.494949,1.616947)
(0.505051,1.616947)(0.515152,1.616738)(0.525253,1.616320)(0.535354,1.615690)(0.545455,1.614845)
(0.555556,1.613780)(0.565657,1.612488)(0.575758,1.610964)(0.585859,1.609196)(0.595960,1.607176)
(0.606061,1.604891)(0.616162,1.602326)(0.626263,1.599466)(0.636364,1.596291)(0.646465,1.592781)
(0.656566,1.588909)(0.666667,1.584650)(0.676768,1.579970)(0.686869,1.574834)(0.696970,1.569200)
(0.707071,1.563021)(0.717172,1.556243)(0.727273,1.548805)(0.737374,1.540637)(0.747475,1.531658)
(0.757576,1.521777)(0.767677,1.510888)(0.777778,1.498871)(0.787879,1.485588)(0.797980,1.470878)
(0.808081,1.454560)(0.818182,1.436422)(0.828283,1.416222)(0.838384,1.393678)(0.848485,1.368469)
(0.858586,1.340219)(0.868687,1.308498)(0.878788,1.272803)(0.888889,1.232556)(0.898990,1.187087)
(0.909091,1.135617)(0.919192,1.077246)(0.929293,1.010929)(0.939394,0.935455)(0.949495,0.849422)
(0.959596,0.751204)(0.969697,0.638918)(0.979798,0.510383)(0.989899,0.363075)(1.000000,0.194074)

\newrgbcolor{color195.0381}{0  0  0}
\psline[plotstyle=line,linejoin=1,linestyle=dashed,linewidth=\LineWidth,linecolor=color195.0381]
(0.000000,0.011166)(0.010101,0.023850)(0.020202,0.038215)(0.030303,0.054432)(0.040404,0.072683)
(0.050505,0.093156)(0.060606,0.116045)(0.070707,0.141546)(0.080808,0.169856)(0.090909,0.201165)
(0.101010,0.235654)(0.111111,0.273490)(0.121212,0.314820)(0.131313,0.359765)(0.141414,0.408414)
(0.151515,0.460822)(0.161616,0.517002)(0.171717,0.576921)(0.181818,0.640500)(0.191919,0.707611)
(0.202020,0.778077)(0.212121,0.851672)(0.222222,0.928126)(0.232323,1.007123)(0.242424,1.088312)
(0.262626,1.255692)(0.303030,1.598203)(0.323232,1.765917)(0.333333,1.847259)(0.343434,1.926323)
(0.353535,2.002678)(0.363636,2.075910)(0.373737,2.145625)(0.383838,2.211453)(0.393939,2.273046)
(0.404040,2.330083)(0.414141,2.382269)(0.424242,2.429337)(0.434343,2.471049)(0.444444,2.507194)
(0.454545,2.537591)(0.464646,2.562091)(0.474747,2.580572)(0.484848,2.592943)(0.494949,2.599144)
(0.505051,2.599144)(0.515152,2.592943)(0.525253,2.580572)(0.535354,2.562091)(0.545455,2.537591)
(0.555556,2.507194)(0.565657,2.471049)(0.575758,2.429337)(0.585859,2.382269)(0.595960,2.330083)
(0.606061,2.273046)(0.616162,2.211453)(0.626263,2.145625)(0.636364,2.075910)(0.646465,2.002678)
(0.656566,1.926323)(0.666667,1.847259)(0.676768,1.765917)(0.696970,1.598203)(0.737374,1.255692)
(0.757576,1.088312)(0.767677,1.007123)(0.777778,0.928126)(0.787879,0.851672)(0.797980,0.778077)
(0.808081,0.707611)(0.818182,0.640500)(0.828283,0.576921)(0.838384,0.517002)(0.848485,0.460822)
(0.858586,0.408414)(0.868687,0.359765)(0.878788,0.314820)(0.888889,0.273490)(0.898990,0.235654)
(0.909091,0.201165)(0.919192,0.169856)(0.929293,0.141546)(0.939394,0.116045)(0.949495,0.093156)
(0.959596,0.072683)(0.969697,0.054432)(0.979798,0.038215)(0.989899,0.023850)(1.000000,0.011166)

\end{pspicture}%
		\caption{\label{fig:pert_lam_fr}(a) LAM case; (b)  FROZ case. Optimal initial conditions (in solid lines) and final state (in dashed lines) at global maximum gain
time  $T_{E_{opt}}=1.15$, for $r=0.5$ as defined in (\ref{eq:ratio}). The perturbation is concentrated in the middle of the domain under the action of the base mean flow. For the FROZ case, the focussing is weaker (smaller base flow gradient), and the damping is larger.}
\end{center}
\end{figure}

In figure (\ref{fig:gain_lam_fr}), we show the optimal coherent perturbation energy gain (\ref{eq:gain_u}) against time
for such a LAM case.
We identify  optimal transient growth which reaches its maximum gain for $T_{E_{opt}}=1.15$, subsequent to which the gain decays  slowly due to the relatively low value of the viscosity we have chosen.
In figure (\ref{fig:pert_lam_fr}), we show the structure of  the optimal perturbation state vector both initially and at $T_{E_{opt}}=1.15$, where the energy reaches its maximum value. We can see that the initial perturbation is not localized, but has a constant value over much of the domain, only decreasing at the edge of the domain to satisfy the boundary conditions, while the final perturbation has been strongly localized in the centre of the domain and has a much larger amplitude than the initial state vector.

\subsubsection{Frozen turbulent viscosity analysis: FROZ case}

The laminar analysis we just performed is equivalent to a frozen turbulent viscosity stability analysis (perturbation of the form $\tilde{\mathbf{q}}=(\tilde{u},0)^\top$) for $r=0$ (and so no turbulent viscosity, or indeed any turbulent property in the system).
In the FROZ case, the base mean flow $\overline{\mathbf{q}}$ is a solution of the complete set of base flow equations (\ref{eq:RAB_sys}) with a non-zero viscosity ratio $r$, defined by (\ref{eq:ratio}). 
The parameters in this set of equations
are essentially arbitrary, and we 
choose to use  $\nu=0.05$, $r=0.5$, $c_1=0.75$ and $c_2=2$, which are a good set of parameters in order to produce enough turbulent viscosity to have an effect on the base mean flow, but not to remove all the dynamics of the system.This appeared
to be a  balanced choice of parameters allowing us to examine all the interesting features of the system.

The observed gain for the FROZ case must  be smaller than in the LAM case for two reasons. First of all, the total base mean flow viscosity will be larger than the laminar value because it now includes the space-dependent turbulent viscosity. The damping term $D_E$ in equation (\ref{eq:pert_ener}) will as a consequence be stronger. Moreover, a direct consequence of having more viscosity is a smaller slope for the base mean flow velocity which is directly involved in the production of energy $P_{E_1}$ which will therefore be smaller than in the LAM case. The optimal gain curve for the FROZ case must then be beneath that of the LAM case.
In figure (\ref{fig:gain_lam_fr}), we plot the optimal curves corresponding to $r=0$ (LAM case) and  $r=0.5$ (FROZ case). We  note that even if the ratio of the turbulent viscosity to the laminar viscosity is small, the optimal gain curve is substantially affected. However, this depends in a nontrivial way on the modelling coefficients $c_1$ and $c_2$. We also notice that the optimal horizon time decreases as the amount of turbulence (modelled by the parameter $r$) increases.

\subsubsection{Full linearized analysis: FULL case}

\paragraph{Total norm optimization: SVD analysis}\label{sec:fullsvd}\ \\

The analysis of the full linearized system of equations
(\ref{eq:pert_linearized})
requires the use of a norm for the two-component  perturbation state vector $\tilde{\mathbf{q}}=(\tilde{u},\tilde{\nu})^T$. We will present the results for the time dependence of the objective functional $\mathcal{J}$ of interest defined in (\ref{eq:objfdef}),
first for the total gain $G_N(T)$ 
(defined in (\ref{eq:total_gain})) optimized in the total normalization norm $\|\cdot \|_N$
defined by (\ref{eq:wndef}) using SVD analysis, and 
then optimizing gain with respect to the energy semi-norm $\|\cdot \|_E$ defined
by (\ref{eq:wedef}).
Let us first start with the case of the total norm ($\|\cdot \|_N$)  gain optimization. 
For this total norm optimization, we will consider the gain in the energy semi-norm for comparison with the other cases
(in particular with the results obtained using our variational framework based on semi-norm constraints)
although the coherent perturbation kinetic energy is not actually the quantity being optimized.
Other quantities which are also of interest to characterize the nature of the state vector are the time-dependent generalizations of the initial condition ratios $R_0$ and $C_0$ defined in (\ref{eq:rtdef}).

These quantities measure the relative importance of the turbulent viscosity perturbation
to the coherent velocity perturbation. 
A state vector having a high value of $C$ (or equivalently $R \simeq 1$) will be identified with  a ``turbulent'' state, while a 
state vector with a low value of $C$ (or equivalently $R \simeq 0$) will be associated with a ``laminar'' state. 

We plot $G_N(T)$ (as defined in (\ref{eq:total_gain})) against optimization  time interval $T$ 
 for the optimal and first sub-optimal state vectors  
(i.e. the two first singular values of the evolution operator, as discussed in appendix (\ref{app:svd})) in figure (\ref{fig:gain_svd}a). In this case, two modes are competing in order to define the overall optimal perturbation: a transient mode (plotted with a black line) which has strong transient growth of the value of the total norm at short times, and the least stable mode (plotted with a grey line) responsible for the weakest possible long time decay. For the sake of simplicity, we will denote these two modes by STO (short time optimal) perturbation  and LTO (long time optimal) perturbation. The STO perturbation reaches its maximum ($G_{N_{opt}}=3.46$) for $T_{N_{opt}}=0.38$ and the switching time for which the two modes have the same (total) gain is  $T_s\simeq1.25$.
The main result of this SVD analysis is that there is a competition between two modes for which a clear transient growth is observed, meaning that two perturbation growth mechanisms are relevant. Even if this dynamics cannot be associated exclusively with coherent velocity perturbation energy production (as in the LAM and FROZ cases discussed above), we can now say that the presence of the second perturbation evolution equation  (for $\tilde{\nu}$ in
(\ref{eq:pert_linearized}))
introduces new dynamics to the system's behaviour. Indeed, the term $P_{E_2}$ of equation (\ref{eq:pert_ener}) can now be a new source of energy. This term is directly proportional to both the magnitude of the slope of the coherent velocity perturbation and to the perturbation  turbulent viscosity, and  is thus responsible for  much richer dynamics of the system.

\begin{figure}
\begin{center}
%
%
%
\providelength{\AxesLineWidth}       \setlength{\AxesLineWidth}{0.5pt}%
\providelength{\plotwidth}           \setlength{\plotwidth}{5cm}
\providelength{\LineWidth}           \setlength{\LineWidth}{0.7pt}%
\providelength{\MarkerSize}          \setlength{\MarkerSize}{4pt}%
\newrgbcolor{GridColor}{0.8 0.8 0.8}%
%
\psset{xunit=0.500000\plotwidth,yunit=0.197177\plotwidth}%
\begin{pspicture}(-0.420000,-0.760736)(2.000000,4.000000)%


\psline[linewidth=\AxesLineWidth,linecolor=GridColor](0.000000,0.000000)(0.000000,0.060859)
\psline[linewidth=\AxesLineWidth,linecolor=GridColor](1.000000,0.000000)(1.000000,0.060859)
\psline[linewidth=\AxesLineWidth,linecolor=GridColor](2.000000,0.000000)(2.000000,0.060859)
\psline[linewidth=\AxesLineWidth,linecolor=GridColor](0.000000,0.000000)(0.024000,0.000000)
\psline[linewidth=\AxesLineWidth,linecolor=GridColor](0.000000,1.000000)(0.024000,1.000000)
\psline[linewidth=\AxesLineWidth,linecolor=GridColor](0.000000,2.000000)(0.024000,2.000000)
\psline[linewidth=\AxesLineWidth,linecolor=GridColor](0.000000,3.000000)(0.024000,3.000000)
\psline[linewidth=\AxesLineWidth,linecolor=GridColor](0.000000,4.000000)(0.024000,4.000000)

{ \footnotesize 
\rput[t](0.000000,-0.060859){$0$}
\rput[t](1.000000,-0.060859){$1$}
\rput[t](2.000000,-0.060859){$2$}
\rput[r](-0.024000,0.000000){$0$}
\rput[r](-0.024000,1.000000){$1$}
\rput[r](-0.024000,2.000000){$2$}
\rput[r](-0.024000,3.000000){$3$}
\rput[r](-0.024000,4.000000){$4$}
} 

\psframe[linewidth=\AxesLineWidth,dimen=middle](0.000000,0.000000)(2.000000,4.000000)

{ \small 
\rput[b](1.000000,-0.760736){
\begin{tabular}{c}
$T$\\
\end{tabular}
}

\rput[t]{90}(-0.420000,2.000000){
\begin{tabular}{c}
$G_N$\\
\end{tabular}
}
} 

\newrgbcolor{color206.0306}{0  0  0}
\psline[plotstyle=line,linejoin=1,linestyle=solid,linewidth=\LineWidth,linecolor=color206.0306]
(0.005000,1.044982)(0.015000,1.139611)(0.030000,1.289035)(0.070000,1.696572)(0.090000,1.894305)
(0.110000,2.085219)(0.125000,2.223547)(0.140000,2.357373)(0.155000,2.486231)(0.170000,2.609508)
(0.185000,2.726498)(0.195000,2.800634)(0.205000,2.871435)(0.215000,2.938702)(0.225000,3.002254)
(0.235000,3.061933)(0.245000,3.117606)(0.255000,3.169168)(0.260000,3.193382)(0.265000,3.216542)
(0.270000,3.238641)(0.275000,3.259676)(0.280000,3.279646)(0.285000,3.298550)(0.290000,3.316389)
(0.295000,3.333166)(0.300000,3.348885)(0.305000,3.363551)(0.310000,3.377172)(0.315000,3.389756)
(0.320000,3.401312)(0.325000,3.411850)(0.330000,3.421381)(0.335000,3.429919)(0.340000,3.437477)
(0.345000,3.444068)(0.350000,3.449708)(0.355000,3.454411)(0.360000,3.458195)(0.365000,3.461077)
(0.370000,3.463072)(0.375000,3.464200)(0.380000,3.464477)(0.385000,3.463924)(0.390000,3.462557)
(0.395000,3.460397)(0.400000,3.457463)(0.405000,3.453773)(0.410000,3.449347)(0.415000,3.444204)
(0.420000,3.438365)(0.425000,3.431848)(0.430000,3.424673)(0.435000,3.416860)(0.440000,3.408427)
(0.445000,3.399394)(0.450000,3.389779)(0.455000,3.379603)(0.460000,3.368882)(0.465000,3.357637)
(0.475000,3.333642)(0.485000,3.307763)(0.495000,3.280137)(0.505000,3.250900)(0.515000,3.220181)
(0.530000,3.171596)(0.545000,3.120358)(0.560000,3.066841)(0.580000,2.992539)(0.600000,2.915560)
(0.625000,2.816599)(0.665000,2.654602)(0.725000,2.410377)(0.760000,2.270508)(0.790000,2.153434)
(0.815000,2.058328)(0.840000,1.965744)(0.865000,1.875882)(0.890000,1.788893)(0.910000,1.721439)
(0.930000,1.655922)(0.950000,1.592364)(0.970000,1.530771)(0.990000,1.471141)(1.010000,1.413464)
(1.030000,1.357721)(1.050000,1.303886)(1.070000,1.251930)(1.090000,1.201818)(1.110000,1.153512)
(1.130000,1.106970)(1.150000,1.062150)(1.170000,1.019006)(1.190000,0.977493)(1.210000,0.937563)
(1.230000,0.899169)(1.250000,0.862262)(1.270000,0.826796)(1.290000,0.792723)(1.310000,0.759995)
(1.330000,0.728568)(1.350000,0.698394)(1.370000,0.669430)(1.390000,0.641631)(1.410000,0.614956)
(1.430000,0.589362)(1.450000,0.564808)(1.470000,0.541256)(1.490000,0.518666)(1.510000,0.497003)
(1.530000,0.476229)(1.550000,0.456311)(1.570000,0.437213)(1.590000,0.418905)(1.610000,0.401354)
(1.630000,0.384531)(1.655000,0.364480)(1.680000,0.345465)(1.705000,0.327434)(1.730000,0.310337)
(1.755000,0.294127)(1.780000,0.278759)(1.805000,0.264189)(1.830000,0.250377)(1.855000,0.237284)
(1.880000,0.224873)(1.905000,0.213109)(1.935000,0.199798)(1.965000,0.187317)(1.995000,0.175613)
(2.000000,0.173735)

\newrgbcolor{color207.0302}{0.6         0.6         0.6}
\psline[plotstyle=line,linejoin=1,linestyle=solid,linewidth=\LineWidth,linecolor=color207.0302]
(0.005000,1.034095)(0.020000,1.137992)(0.030000,1.204360)(0.035000,1.235846)(0.040000,1.266013)
(0.045000,1.294790)(0.050000,1.322144)(0.055000,1.348067)(0.060000,1.372566)(0.065000,1.395657)
(0.070000,1.417360)(0.075000,1.437694)(0.080000,1.456677)(0.085000,1.474325)(0.090000,1.490651)
(0.095000,1.505668)(0.100000,1.519387)(0.105000,1.531820)(0.110000,1.542982)(0.115000,1.552889)
(0.120000,1.561559)(0.125000,1.569015)(0.130000,1.575283)(0.135000,1.580394)(0.140000,1.584382)
(0.145000,1.587284)(0.150000,1.589143)(0.155000,1.590003)(0.160000,1.589913)(0.165000,1.588922)
(0.170000,1.587084)(0.175000,1.584451)(0.180000,1.581079)(0.185000,1.577024)(0.190000,1.572339)
(0.195000,1.567080)(0.200000,1.561301)(0.205000,1.555055)(0.210000,1.548392)(0.220000,1.534011)
(0.230000,1.518527)(0.245000,1.493936)(0.285000,1.426187)(0.300000,1.401573)(0.315000,1.377906)
(0.330000,1.355343)(0.345000,1.333960)(0.360000,1.313772)(0.375000,1.294753)(0.390000,1.276855)
(0.405000,1.260011)(0.420000,1.244150)(0.435000,1.229197)(0.450000,1.215079)(0.465000,1.201726)
(0.480000,1.189071)(0.495000,1.177053)(0.515000,1.161924)(0.535000,1.147707)(0.555000,1.134298)
(0.575000,1.121604)(0.600000,1.106621)(0.625000,1.092494)(0.650000,1.079110)(0.680000,1.063897)
(0.710000,1.049478)(0.745000,1.033509)(0.780000,1.018321)(0.820000,1.001769)(0.865000,0.984011)
(0.910000,0.967018)(0.960000,0.948889)(1.015000,0.929720)(1.075000,0.909587)(1.140000,0.888558)
(1.210000,0.866694)(1.285000,0.844057)(1.365000,0.820712)(1.450000,0.796726)(1.535000,0.773516)
(1.625000,0.749731)(1.715000,0.726712)(1.810000,0.703204)(1.905000,0.680472)(2.000000,0.658484)

\end{pspicture}%
  \vspace{0.5cm}
  \include{contribution}
	\caption{\label{fig:gain_svd}(a) The variation of $G_N(T)$ (defined in (\ref{eq:total_gain})) with optimization time $T$   of the short time
optimal (STO) perturbation (plotted with a black line) and long time optimal (LTO) perturbation (grey line) total norm  for the FULL  perturbation equations case, when the total norm $\|\cdot\|_N$ defined in (\ref{eq:wndef}) is optimized using an SVD analysis.
(b) The variation of  the ratio $C_0$ 
as defined in (\ref{eq:ratio_0}) (plotted with solid lines)
and $C(T)$ 
as defined in (\ref{eq:rtdef}) (plotted with dashed lines)
with optimization time $T$ of the STO perturbation (black lines)
and the LTO perturbation (grey lines).
and final in dashed lines) for both the optimal (black) and sub-optimal (grey) mode. We notice that there is a competition between two modes: the short time optimal (STO) perturbation and the long time optimal (LTO) perturbation. 
The STO perturbation is typically associated with  large values of $C_0$, meaning 
that the STO perturbation is initially ``turbulent'',
although it 
evolves to have $C(T) \ll 1$,
meaning that ultimately the perturbation is almost exclusively composed of coherent perturbation velocity.
On the other hand, the LTO perturbation is dominated at all times by the coherent perturbation velocity ($C(T)\ll C_0\ll1$) and thus we refer to it as a   ``laminar'' perturbation.}
\end{center}
\end{figure}
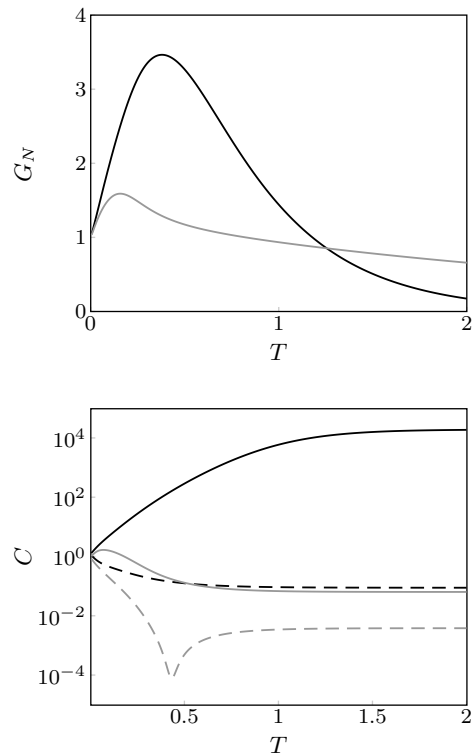

However, it is legitimate to question whether if this transient growth is associated with transient growth of the kinetic energy semi-norm of the perturbation or merely with  growth of the semi-norm of the turbulent viscosity perturbation. We can determine the value of $C_0$ 
(and hence the relative magnitudes of the perturbation velocity and the turbulent viscosity)  
for each optimization interval for both the initial state vector $C_0=C(0)$ and for the final state vector $C(T)$, as plotted in  figure (\ref{fig:gain_svd}b). We notice that the (short-time) STO perturbation is associated with ``turbulent'' ($C_0\gg1$) initial perturbations for sufficiently long optimization intervals, which evolves towards a more laminar ($C(T) \lesssim 1$) state, with a larger contribution from the coherent perturbation velocity $\tilde{u}$. This implies that this transient growth is due to perturbation kinetic energy production. The (late-time) LTO perturbation (which is actually optimal starting from $T=T_s \simeq 1.25$) is on the contrary initially mostly laminar (with low values of $C_0$) and evolves toward an almost completely laminar state, even more dominated by perturbation velocity (i.e. with $C(T)\ll1$).

\begin{figure}
	 \centering	
%
%
%
\providelength{\AxesLineWidth}       \setlength{\AxesLineWidth}{0.5pt}%
\providelength{\plotwidth}           \setlength{\plotwidth}{5cm}
\providelength{\LineWidth}           \setlength{\LineWidth}{0.7pt}%
\providelength{\MarkerSize}          \setlength{\MarkerSize}{4pt}%
\newrgbcolor{GridColor}{0.8 0.8 0.8}%
%
\psset{xunit=0.500000\plotwidth,yunit=0.112673\plotwidth}%
\begin{pspicture}(-0.420000,-3.331288)(2.000000,5.000000)%


\psline[linewidth=\AxesLineWidth,linecolor=GridColor](0.000000,-2.000000)(0.000000,-1.893497)
\psline[linewidth=\AxesLineWidth,linecolor=GridColor](1.000000,-2.000000)(1.000000,-1.893497)
\psline[linewidth=\AxesLineWidth,linecolor=GridColor](2.000000,-2.000000)(2.000000,-1.893497)
\psline[linewidth=\AxesLineWidth,linecolor=GridColor](0.000000,0.000000)(0.024000,0.000000)
\psline[linewidth=\AxesLineWidth,linecolor=GridColor](0.000000,5.000000)(0.024000,5.000000)


{ \footnotesize 
\rput[t](0.000000,-2.106503){$0$}
\rput[t](1.000000,-2.106503){$1$}
\rput[t](2.000000,-2.106503){$2$}
\rput[r](-0.024000,0.000000){$10^{0}$}
\rput[r](-0.024000,5.000000){$10^{5}$}
} 

\psframe[linewidth=\AxesLineWidth,dimen=middle](0.000000,-2.000000)(2.000000,5.000000)

{ \small 
\rput[b](1.000000,-3.331288){
\begin{tabular}{c}
$T$\\
\end{tabular}
}

\rput[t]{90}(-0.420000,1.500000){
\begin{tabular}{c}
$G_E$\\
\end{tabular}
}
} 

\newrgbcolor{color194.027}{0  0  0}
\psline[plotstyle=line,linejoin=1,linestyle=solid,linewidth=\LineWidth,linecolor=color194.027]
(0.005000,0.057989)(0.015000,0.171520)(0.025000,0.279156)(0.035000,0.379883)(0.045000,0.474089)
(0.055000,0.562645)(0.065000,0.646496)(0.075000,0.726520)(0.090000,0.841014)(0.105000,0.950572)
(0.125000,1.091148)(0.145000,1.226837)(0.165000,1.358182)(0.185000,1.485206)(0.205000,1.607761)
(0.225000,1.725696)(0.240000,1.811060)(0.255000,1.893771)(0.270000,1.973846)(0.285000,2.051325)
(0.300000,2.126266)(0.315000,2.198737)(0.330000,2.268820)(0.345000,2.336599)(0.360000,2.402164)
(0.375000,2.465605)(0.390000,2.527014)(0.405000,2.586477)(0.420000,2.644080)(0.435000,2.699904)
(0.455000,2.771700)(0.475000,2.840639)(0.495000,2.906873)(0.515000,2.970540)(0.535000,3.031760)
(0.555000,3.090641)(0.575000,3.147276)(0.595000,3.201745)(0.615000,3.254114)(0.635000,3.304440)
(0.655000,3.352771)(0.675000,3.399143)(0.695000,3.443587)(0.710000,3.475668)(0.725000,3.506686)
(0.740000,3.536646)(0.755000,3.565552)(0.770000,3.593410)(0.785000,3.620221)(0.800000,3.645988)
(0.815000,3.670713)(0.830000,3.694398)(0.845000,3.717045)(0.860000,3.738656)(0.875000,3.759234)
(0.890000,3.778782)(0.905000,3.797303)(0.920000,3.814804)(0.935000,3.831289)(0.950000,3.846767)
(0.965000,3.861245)(0.980000,3.874732)(0.995000,3.887240)(1.010000,3.898780)(1.025000,3.909367)
(1.040000,3.919014)(1.055000,3.927738)(1.070000,3.935555)(1.085000,3.942485)(1.100000,3.948546)
(1.115000,3.953758)(1.130000,3.958144)(1.145000,3.961723)(1.160000,3.964520)(1.175000,3.966557)
(1.190000,3.967857)(1.205000,3.968444)(1.220000,3.968343)(1.235000,3.967576)(1.250000,3.966169)
(1.265000,3.964145)(1.280000,3.961528)(1.295000,3.958341)(1.310000,3.954607)(1.325000,3.950350)
(1.340000,3.945592)(1.360000,3.938505)(1.380000,3.930616)(1.400000,3.921973)(1.420000,3.912623)
(1.440000,3.902611)(1.465000,3.889228)(1.490000,3.874954)(1.515000,3.859864)(1.540000,3.844024)
(1.570000,3.824117)(1.600000,3.803324)(1.630000,3.781738)(1.665000,3.755661)(1.700000,3.728734)
(1.740000,3.697056)(1.785000,3.660427)(1.835000,3.618692)(1.890000,3.571746)(1.950000,3.519532)
(2.000000,3.475378)

\newrgbcolor{color195.0265}{0.6         0.6         0.6}
\psline[plotstyle=line,linejoin=1,linestyle=solid,linewidth=\LineWidth,linecolor=color195.0265]
(0.005000,0.052512)(0.010000,0.103714)(0.015000,0.152522)(0.020000,0.198188)(0.025000,0.240285)
(0.030000,0.278639)(0.035000,0.313259)(0.040000,0.344267)(0.045000,0.371863)(0.050000,0.396284)
(0.055000,0.417784)(0.060000,0.436617)(0.065000,0.453026)(0.070000,0.467238)(0.075000,0.479458)
(0.080000,0.489866)(0.085000,0.498625)(0.090000,0.505875)(0.095000,0.511739)(0.100000,0.516325)
(0.105000,0.519727)(0.110000,0.522030)(0.115000,0.523310)(0.120000,0.523635)(0.125000,0.523070)
(0.130000,0.521674)(0.135000,0.519503)(0.140000,0.516611)(0.145000,0.513049)(0.150000,0.508868)
(0.155000,0.504115)(0.160000,0.498838)(0.165000,0.493083)(0.170000,0.486896)(0.175000,0.480319)
(0.185000,0.466166)(0.195000,0.450951)(0.210000,0.426787)(0.250000,0.360314)(0.265000,0.336383)
(0.280000,0.313614)(0.295000,0.292216)(0.305000,0.278765)(0.315000,0.265980)(0.325000,0.253857)
(0.335000,0.242385)(0.345000,0.231541)(0.355000,0.221302)(0.365000,0.211639)(0.380000,0.198154)
(0.395000,0.185787)(0.410000,0.174430)(0.425000,0.163983)(0.440000,0.154350)(0.455000,0.145445)
(0.470000,0.137191)(0.485000,0.129516)(0.505000,0.120079)(0.525000,0.111433)(0.545000,0.103468)
(0.570000,0.094332)(0.595000,0.085965)(0.620000,0.078244)(0.650000,0.069689)(0.685000,0.060514)
(0.725000,0.050877)(0.770000,0.040881)(0.820000,0.030579)(0.880000,0.019050)(0.950000,0.006427)
(1.035000,-0.008080)(1.145000,-0.026000)(1.300000,-0.050322)(1.550000,-0.088513)(2.000000,-0.156297)

\end{pspicture}%
		\caption{\label{fig:gain_eff_svd}Variation of energy gain $G_E(T)$ as defined in (\ref{eq:gain_u}) 
with the optimization time interval $T$ for the full linearized 
perturbation system of equations
(\ref{eq:pert_linearized}) for perturbations which  optimize  the total norm gain $G_N(T)$, as defined in (\ref{eq:total_gain}). The STO perturbations (plotted with a black line) achieves very large values of $G_E(T)$, while the LTO perturbation (plotted with a grey line) is associated with substantially smaller values of $G_E(T)$, comparable to those obtained in the FROZ case. The STO perturbation has a larger energy gain than the LTO perturbation in the plotted time window but eventually has a larger decay rate and is thus less efficient in preserving a large energy gain.}
\end{figure}
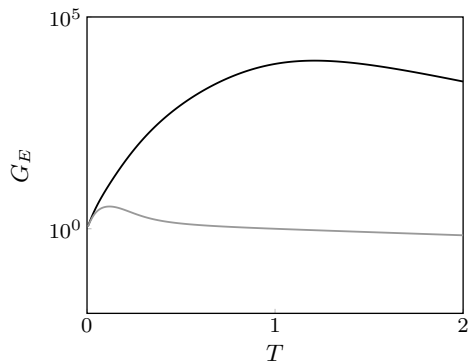

In order to identify the associated transient growth of the perturbation kinetic energy, we 
now consider the evolution of the perturbation velocity kinetic energy gain $G_E(T)$ as
defined in (\ref{eq:gain_u}). We plot this quantity (for both the STO perturbation and the LTO perturbation)  in figure (\ref{fig:gain_eff_svd}). We can see that this gain (which is not the optimal energy gain) can have very large values compared to the previous studies. Indeed, the perturbation kinetic energy is amplified by approximately five orders of magnitude and reaches its maximum for $T_{E_{opt}}=1.21$ with a gain of $G_{E_{opt}}=9.3\ 10^4$. We consider these two quantities as the ``optimal'' time and gain values extracted from a full-norm gain optimization. Since $G_E(T)$ is not optimized, but is just derived from the results of the full-norm optimal perturbation
for the entire perturbation velocity-viscosity state vector, it is not surprising to see that the STO perturbation remains the 
 mode with larger $G_E(T)$ after $T=T_s\simeq 1.25$. For sufficiently long times, the LTO perturbation will ultimately become dominant with respect to kinetic energy gain as well. Besides, this figure clearly underlines the difference in the order of magnitude between the ``laminar'' perturbations of the LTO perturbations, and the ``turbulent'' perturbations of the STO perturbations. Indeed, the turbulent STO perturbation is associated with very large values of energy gain while the laminar LTO perturbation shows an  energy gain which can be compared most closely to the LAM case  optimal gain shown in figure (\ref{fig:gain_lam_fr}).

\begin{figure}
\begin{center}
%
%
%
\providelength{\AxesLineWidth}       \setlength{\AxesLineWidth}{0.5pt}%
\providelength{\plotwidth}           \setlength{\plotwidth}{5cm}
\providelength{\LineWidth}           \setlength{\LineWidth}{0.7pt}%
\providelength{\MarkerSize}          \setlength{\MarkerSize}{4pt}%
\newrgbcolor{GridColor}{0.8 0.8 0.8}%
%
\psset{xunit=1.000000\plotwidth,yunit=0.098589\plotwidth}%
\begin{pspicture}(-0.210000,-5.521472)(1.000000,4.000000)%


\psline[linewidth=\AxesLineWidth,linecolor=GridColor](0.000000,-4.000000)(0.000000,-3.878282)
\psline[linewidth=\AxesLineWidth,linecolor=GridColor](0.500000,-4.000000)(0.500000,-3.878282)
\psline[linewidth=\AxesLineWidth,linecolor=GridColor](1.000000,-4.000000)(1.000000,-3.878282)
\psline[linewidth=\AxesLineWidth,linecolor=GridColor](0.000000,-4.000000)(0.012000,-4.000000)
\psline[linewidth=\AxesLineWidth,linecolor=GridColor](0.000000,-2.000000)(0.012000,-2.000000)
\psline[linewidth=\AxesLineWidth,linecolor=GridColor](0.000000,0.000000)(0.012000,0.000000)
\psline[linewidth=\AxesLineWidth,linecolor=GridColor](0.000000,2.000000)(0.012000,2.000000)
\psline[linewidth=\AxesLineWidth,linecolor=GridColor](0.000000,4.000000)(0.012000,4.000000)

{ \footnotesize 
\rput[t](0.000000,-4.121718){$0$}
\rput[t](0.500000,-4.121718){$0.5$}
\rput[t](1.000000,-4.121718){$1$}
\rput[r](-0.012000,-4.000000){$-4$}
\rput[r](-0.012000,-2.000000){$-2$}
\rput[r](-0.012000,0.000000){$0$}
\rput[r](-0.012000,2.000000){$2$}
\rput[r](-0.012000,4.000000){$4$}
} 

\psframe[linewidth=\AxesLineWidth,dimen=middle](0.000000,-4.000000)(1.000000,4.000000)

{ \small 
\rput[b](0.500000,-5.521472){
\begin{tabular}{c}
$x$\\
\end{tabular}
}

\rput[t]{90}(-0.210000,0.000000){
\begin{tabular}{c}
$\tilde{u}$\\
\end{tabular}
}
} 

\newrgbcolor{color181.0371}{0  0  0}
\psline[plotstyle=line,linejoin=1,linestyle=solid,linewidth=\LineWidth,linecolor=color181.0371]
(0.000000,-0.001675)(0.030303,-0.007639)(0.060606,-0.013572)(0.080808,-0.017006)(0.101010,-0.019872)
(0.121212,-0.022133)(0.141414,-0.023796)(0.161616,-0.024894)(0.181818,-0.025475)(0.202020,-0.025592)
(0.232323,-0.025013)(0.262626,-0.023675)(0.292929,-0.021726)(0.333333,-0.018392)(0.383838,-0.013374)
(0.454545,-0.005396)(0.585859,0.010052)(0.636364,0.015473)(0.676768,0.019293)(0.717172,0.022435)
(0.747475,0.024196)(0.777778,0.025298)(0.797980,0.025592)(0.818182,0.025475)(0.838384,0.024894)
(0.858586,0.023796)(0.878788,0.022133)(0.898990,0.019872)(0.919192,0.017006)(0.939394,0.013572)
(0.969697,0.007639)(1.000000,0.001675)

\newrgbcolor{color182.0366}{0  0  0}
\psline[plotstyle=line,linejoin=1,linestyle=dashed,linewidth=\LineWidth,linecolor=color182.0366]
(0.000000,-0.038181)(0.010101,-0.081249)(0.020202,-0.129604)(0.030303,-0.183634)(0.040404,-0.243695)
(0.050505,-0.310093)(0.060606,-0.383069)(0.070707,-0.462769)(0.080808,-0.549231)(0.090909,-0.642353)
(0.101010,-0.741880)(0.111111,-0.847379)(0.121212,-0.958237)(0.131313,-1.073645)(0.151515,-1.313960)
(0.171717,-1.558377)(0.181818,-1.678442)(0.191919,-1.794963)(0.202020,-1.906331)(0.212121,-2.010969)
(0.222222,-2.107370)(0.232323,-2.194136)(0.242424,-2.270007)(0.252525,-2.333877)(0.262626,-2.384818)
(0.272727,-2.422084)(0.282828,-2.445114)(0.292929,-2.453528)(0.303030,-2.447123)(0.313131,-2.425858)
(0.323232,-2.389846)(0.333333,-2.339334)(0.343434,-2.274695)(0.353535,-2.196409)(0.363636,-2.105050)
(0.373737,-2.001274)(0.383838,-1.885803)(0.393939,-1.759421)(0.404040,-1.622957)(0.414141,-1.477281)
(0.424242,-1.323297)(0.434343,-1.161932)(0.444444,-0.994136)(0.454545,-0.820875)(0.474747,-0.461881)
(0.525253,0.461881)(0.545455,0.820875)(0.555556,0.994136)(0.565657,1.161932)(0.575758,1.323297)
(0.585859,1.477281)(0.595960,1.622957)(0.606061,1.759421)(0.616162,1.885803)(0.626263,2.001274)
(0.636364,2.105050)(0.646465,2.196409)(0.656566,2.274695)(0.666667,2.339334)(0.676768,2.389846)
(0.686869,2.425858)(0.696970,2.447123)(0.707071,2.453528)(0.717172,2.445114)(0.727273,2.422084)
(0.737374,2.384818)(0.747475,2.333877)(0.757576,2.270007)(0.767677,2.194136)(0.777778,2.107370)
(0.787879,2.010969)(0.797980,1.906331)(0.808081,1.794963)(0.818182,1.678442)(0.838384,1.436367)
(0.858586,1.192610)(0.868687,1.073645)(0.878788,0.958237)(0.888889,0.847379)(0.898990,0.741880)
(0.909091,0.642353)(0.919192,0.549231)(0.929293,0.462769)(0.939394,0.383069)(0.949495,0.310093)
(0.959596,0.243695)(0.969697,0.183634)(0.979798,0.129604)(0.989899,0.081249)(1.000000,0.038181)

\end{pspicture}%
  \vspace{0.5cm}
%
%
%
\providelength{\AxesLineWidth}       \setlength{\AxesLineWidth}{0.5pt}%
\providelength{\plotwidth}           \setlength{\plotwidth}{5cm}
\providelength{\LineWidth}           \setlength{\LineWidth}{0.7pt}%
\providelength{\MarkerSize}          \setlength{\MarkerSize}{4pt}%
\newrgbcolor{GridColor}{0.8 0.8 0.8}%
%
\psset{xunit=1.000000\plotwidth,yunit=0.394355\plotwidth}%
\begin{pspicture}(-0.210000,-0.380368)(1.000000,2.000000)%


\psline[linewidth=\AxesLineWidth,linecolor=GridColor](0.000000,0.000000)(0.000000,0.030429)
\psline[linewidth=\AxesLineWidth,linecolor=GridColor](0.500000,0.000000)(0.500000,0.030429)
\psline[linewidth=\AxesLineWidth,linecolor=GridColor](1.000000,0.000000)(1.000000,0.030429)
\psline[linewidth=\AxesLineWidth,linecolor=GridColor](0.000000,0.000000)(0.012000,0.000000)
\psline[linewidth=\AxesLineWidth,linecolor=GridColor](0.000000,0.500000)(0.012000,0.500000)
\psline[linewidth=\AxesLineWidth,linecolor=GridColor](0.000000,1.000000)(0.012000,1.000000)
\psline[linewidth=\AxesLineWidth,linecolor=GridColor](0.000000,1.500000)(0.012000,1.500000)
\psline[linewidth=\AxesLineWidth,linecolor=GridColor](0.000000,2.000000)(0.012000,2.000000)

{ \footnotesize 
\rput[t](0.000000,-0.030429){$0$}
\rput[t](0.500000,-0.030429){$0.5$}
\rput[t](1.000000,-0.030429){$1$}
\rput[r](-0.012000,0.000000){$0$}
\rput[r](-0.012000,0.500000){$0.5$}
\rput[r](-0.012000,1.000000){$1$}
\rput[r](-0.012000,1.500000){$1.5$}
\rput[r](-0.012000,2.000000){$2$}
} 

\psframe[linewidth=\AxesLineWidth,dimen=middle](0.000000,0.000000)(1.000000,2.000000)

{ \small 
\rput[b](0.500000,-0.380368){
\begin{tabular}{c}
$x$\\
\end{tabular}
}

\rput[t]{90}(-0.210000,1.000000){
\begin{tabular}{c}
$\tilde{\nu}$\\
\end{tabular}
}
} 

\newrgbcolor{color194.04}{0  0  0}
\psline[plotstyle=line,linejoin=1,linestyle=solid,linewidth=\LineWidth,linecolor=color194.04]
(0.000000,0.201702)(0.010101,0.377957)(0.020202,0.531946)(0.030303,0.666457)(0.040404,0.783930)
(0.050505,0.886503)(0.060606,0.976046)(0.070707,1.054195)(0.080808,1.122384)(0.090909,1.181866)
(0.101010,1.233740)(0.111111,1.278963)(0.121212,1.318377)(0.131313,1.352716)(0.141414,1.382622)
(0.151515,1.408656)(0.161616,1.431311)(0.171717,1.451015)(0.181818,1.468144)(0.191919,1.483026)
(0.202020,1.495949)(0.212121,1.507162)(0.222222,1.516885)(0.232323,1.525309)(0.242424,1.532602)
(0.252525,1.538908)(0.262626,1.544356)(0.272727,1.549057)(0.282828,1.553107)(0.292929,1.556592)
(0.303030,1.559585)(0.313131,1.562151)(0.323232,1.564346)(0.333333,1.566220)(0.343434,1.567815)
(0.353535,1.569170)(0.373737,1.571283)(0.393939,1.572775)(0.424242,1.574191)(0.454545,1.574948)
(0.494949,1.575307)(0.535354,1.575096)(0.565657,1.574503)(0.595960,1.573340)(0.616162,1.572095)
(0.636364,1.570316)(0.646465,1.569170)(0.656566,1.567815)(0.666667,1.566220)(0.676768,1.564346)
(0.686869,1.562151)(0.696970,1.559585)(0.707071,1.556592)(0.717172,1.553107)(0.727273,1.549057)
(0.737374,1.544356)(0.747475,1.538908)(0.757576,1.532602)(0.767677,1.525309)(0.777778,1.516885)
(0.787879,1.507162)(0.797980,1.495949)(0.808081,1.483026)(0.818182,1.468144)(0.828283,1.451015)
(0.838384,1.431311)(0.848485,1.408656)(0.858586,1.382622)(0.868687,1.352716)(0.878788,1.318377)
(0.888889,1.278963)(0.898990,1.233740)(0.909091,1.181866)(0.919192,1.122384)(0.929293,1.054195)
(0.939394,0.976046)(0.949495,0.886503)(0.959596,0.783930)(0.969697,0.666457)(0.979798,0.531946)
(0.989899,0.377957)(1.000000,0.201702)

\newrgbcolor{color195.0396}{0  0  0}
\psline[plotstyle=line,linejoin=1,linestyle=dashed,linewidth=\LineWidth,linecolor=color195.0396]
(0.000000,0.009048)(0.010101,0.019161)(0.020202,0.030417)(0.030303,0.042884)(0.040404,0.056627)
(0.050505,0.071696)(0.060606,0.088129)(0.070707,0.105943)(0.080808,0.125133)(0.090909,0.145669)
(0.101010,0.167491)(0.111111,0.190509)(0.121212,0.214601)(0.131313,0.239613)(0.151515,0.291648)
(0.181818,0.371350)(0.191919,0.397384)(0.202020,0.422748)(0.212121,0.447220)(0.222222,0.470601)
(0.232323,0.492716)(0.242424,0.513424)(0.252525,0.532615)(0.262626,0.550211)(0.272727,0.566168)
(0.282828,0.580474)(0.292929,0.593146)(0.303030,0.604224)(0.313131,0.613774)(0.323232,0.621878)
(0.333333,0.628635)(0.343434,0.634153)(0.353535,0.638550)(0.363636,0.641950)(0.373737,0.644475)
(0.383838,0.646250)(0.393939,0.647397)(0.404040,0.648031)(0.414141,0.648264)(0.424242,0.648197)
(0.454545,0.647089)(0.484848,0.646028)(0.505051,0.645889)(0.525253,0.646293)(0.565657,0.647925)
(0.575758,0.648197)(0.585859,0.648264)(0.595960,0.648031)(0.606061,0.647397)(0.616162,0.646250)
(0.626263,0.644475)(0.636364,0.641950)(0.646465,0.638550)(0.656566,0.634153)(0.666667,0.628635)
(0.676768,0.621878)(0.686869,0.613774)(0.696970,0.604224)(0.707071,0.593146)(0.717172,0.580474)
(0.727273,0.566168)(0.737374,0.550211)(0.747475,0.532615)(0.757576,0.513424)(0.767677,0.492716)
(0.777778,0.470601)(0.787879,0.447220)(0.797980,0.422748)(0.808081,0.397384)(0.828283,0.344885)
(0.858586,0.265365)(0.868687,0.239613)(0.878788,0.214601)(0.888889,0.190509)(0.898990,0.167491)
(0.909091,0.145669)(0.919192,0.125133)(0.929293,0.105943)(0.939394,0.088129)(0.949495,0.071696)
(0.959596,0.056627)(0.969697,0.042884)(0.979798,0.030417)(0.989899,0.019161)(1.000000,0.009048)

\end{pspicture}%
	\caption{\label{fig:initial_final_full_sys1}Variation with $x$ of 
the 
initial perturbation (plotted with solid lines) and final perturbation (at $T=T_{E_{opt}}$, plotted
with a dashed line) of: 
(a) coherent perturbation velocity $\tilde{u}$ and (b) perturbation
turbulent viscosity $\tilde{\nu}$ for the 
STO perturbation. 
We notice a decay of the turbulent viscosity perturbation (through diffusion, destruction and other mechanisms described by equation (\ref{eq:Keq})) giving rise to large perturbation velocities in the final perturbation state vector. This perturbation  is not present in either the LAM case or the FROZ case, and arises from the richer dynamics in the FULL case
where the turbulent viscosity evolves in space and time.}
		\end{center}
\end{figure}
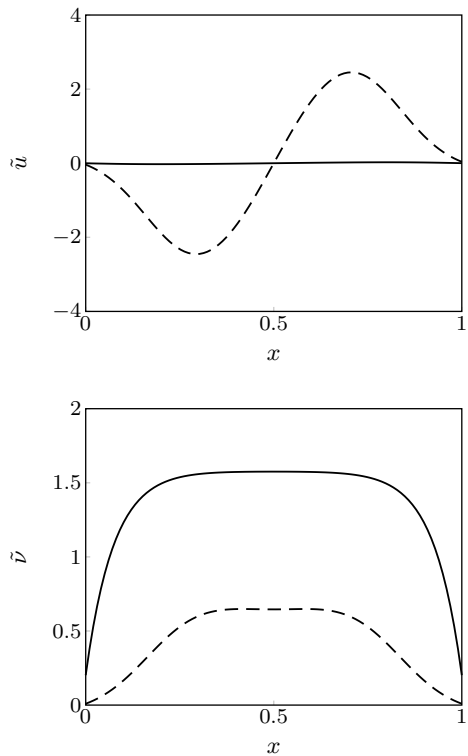

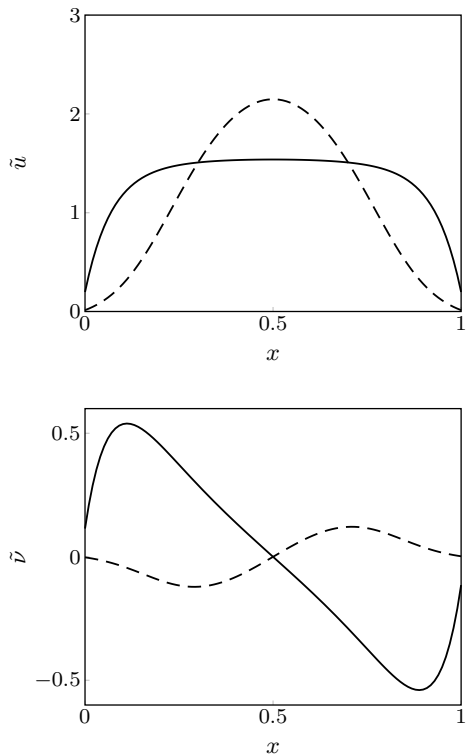
\begin{figure}
\begin{center}
%
%
%
\providelength{\AxesLineWidth}       \setlength{\AxesLineWidth}{0.5pt}%
\providelength{\plotwidth}           \setlength{\plotwidth}{5cm}
\providelength{\LineWidth}           \setlength{\LineWidth}{0.7pt}%
\providelength{\MarkerSize}          \setlength{\MarkerSize}{4pt}%
\newrgbcolor{GridColor}{0.8 0.8 0.8}%
%
\psset{xunit=1.000000\plotwidth,yunit=0.262903\plotwidth}%
\begin{pspicture}(-0.210000,-0.570552)(1.000000,3.000000)%


\psline[linewidth=\AxesLineWidth,linecolor=GridColor](0.000000,0.000000)(0.000000,0.045644)
\psline[linewidth=\AxesLineWidth,linecolor=GridColor](0.500000,0.000000)(0.500000,0.045644)
\psline[linewidth=\AxesLineWidth,linecolor=GridColor](1.000000,0.000000)(1.000000,0.045644)
\psline[linewidth=\AxesLineWidth,linecolor=GridColor](0.000000,0.000000)(0.012000,0.000000)
\psline[linewidth=\AxesLineWidth,linecolor=GridColor](0.000000,1.000000)(0.012000,1.000000)
\psline[linewidth=\AxesLineWidth,linecolor=GridColor](0.000000,2.000000)(0.012000,2.000000)
\psline[linewidth=\AxesLineWidth,linecolor=GridColor](0.000000,3.000000)(0.012000,3.000000)

{ \footnotesize 
\rput[t](0.000000,-0.045644){$0$}
\rput[t](0.500000,-0.045644){$0.5$}
\rput[t](1.000000,-0.045644){$1$}
\rput[r](-0.012000,0.000000){$0$}
\rput[r](-0.012000,1.000000){$1$}
\rput[r](-0.012000,2.000000){$2$}
\rput[r](-0.012000,3.000000){$3$}
} 

\psframe[linewidth=\AxesLineWidth,dimen=middle](0.000000,0.000000)(1.000000,3.000000)

{ \small 
\rput[b](0.500000,-0.570552){
\begin{tabular}{c}
$x$\\
\end{tabular}
}

\rput[t]{90}(-0.210000,1.500000){
\begin{tabular}{c}
$\tilde{u}$\\
\end{tabular}
}
} 

\newrgbcolor{color181.0378}{0  0  0}
\psline[plotstyle=line,linejoin=1,linestyle=solid,linewidth=\LineWidth,linecolor=color181.0378]
(0.000000,0.197046)(0.010101,0.368259)(0.020202,0.517050)(0.030303,0.646382)(0.040404,0.758826)
(0.050505,0.856616)(0.060606,0.941689)(0.070707,1.015728)(0.080808,1.080193)(0.090909,1.136353)
(0.101010,1.185306)(0.111111,1.228008)(0.121212,1.265286)(0.131313,1.297858)(0.141414,1.326347)
(0.151515,1.351291)(0.161616,1.373159)(0.171717,1.392355)(0.181818,1.409229)(0.191919,1.424085)
(0.202020,1.437186)(0.212121,1.448757)(0.222222,1.458997)(0.232323,1.468073)(0.242424,1.476132)
(0.252525,1.483301)(0.262626,1.489689)(0.272727,1.495391)(0.282828,1.500487)(0.292929,1.505049)
(0.303030,1.509136)(0.313131,1.512802)(0.323232,1.516093)(0.333333,1.519046)(0.343434,1.521697)
(0.353535,1.524075)(0.363636,1.526206)(0.373737,1.528110)(0.393939,1.531317)(0.414141,1.533817)
(0.434343,1.535703)(0.454545,1.537038)(0.474747,1.537868)(0.494949,1.538219)(0.515152,1.538103)
(0.535354,1.537515)(0.555556,1.536436)(0.575758,1.534832)(0.595960,1.532649)(0.616162,1.529809)
(0.626263,1.528110)(0.636364,1.526206)(0.646465,1.524075)(0.656566,1.521697)(0.666667,1.519046)
(0.676768,1.516093)(0.686869,1.512802)(0.696970,1.509136)(0.707071,1.505049)(0.717172,1.500487)
(0.727273,1.495391)(0.737374,1.489689)(0.747475,1.483301)(0.757576,1.476132)(0.767677,1.468073)
(0.777778,1.458997)(0.787879,1.448757)(0.797980,1.437186)(0.808081,1.424085)(0.818182,1.409229)
(0.828283,1.392355)(0.838384,1.373159)(0.848485,1.351291)(0.858586,1.326347)(0.868687,1.297858)
(0.878788,1.265286)(0.888889,1.228008)(0.898990,1.185306)(0.909091,1.136353)(0.919192,1.080193)
(0.929293,1.015728)(0.939394,0.941689)(0.949495,0.856616)(0.959596,0.758826)(0.969697,0.646382)
(0.979798,0.517050)(0.989899,0.368259)(1.000000,0.197046)

\newrgbcolor{color182.0374}{0  0  0}
\psline[plotstyle=line,linejoin=1,linestyle=dashed,linewidth=\LineWidth,linecolor=color182.0374]
(0.000000,0.013642)(0.010101,0.029086)(0.020202,0.046522)(0.030303,0.066142)(0.040404,0.088144)
(0.050505,0.112723)(0.060606,0.140067)(0.070707,0.170352)(0.080808,0.203733)(0.090909,0.240339)
(0.101010,0.280263)(0.111111,0.323560)(0.121212,0.370234)(0.131313,0.420239)(0.141414,0.473473)
(0.151515,0.529778)(0.161616,0.588941)(0.171717,0.650697)(0.181818,0.714737)(0.191919,0.780712)
(0.212121,0.916938)(0.252525,1.195183)(0.272727,1.330970)(0.282828,1.396816)(0.292929,1.460908)
(0.303030,1.522989)(0.313131,1.582833)(0.323232,1.640244)(0.333333,1.695049)(0.343434,1.747102)
(0.353535,1.796278)(0.363636,1.842475)(0.373737,1.885603)(0.383838,1.925594)(0.393939,1.962387)
(0.404040,1.995937)(0.414141,2.026207)(0.424242,2.053166)(0.434343,2.076793)(0.444444,2.097070)
(0.454545,2.113983)(0.464646,2.127524)(0.474747,2.137685)(0.484848,2.144461)(0.494949,2.147850)
(0.505051,2.147850)(0.515152,2.144461)(0.525253,2.137685)(0.535354,2.127524)(0.545455,2.113983)
(0.555556,2.097070)(0.565657,2.076793)(0.575758,2.053166)(0.585859,2.026207)(0.595960,1.995937)
(0.606061,1.962387)(0.616162,1.925594)(0.626263,1.885603)(0.636364,1.842475)(0.646465,1.796278)
(0.656566,1.747102)(0.666667,1.695049)(0.676768,1.640244)(0.686869,1.582833)(0.696970,1.522989)
(0.707071,1.460908)(0.717172,1.396816)(0.727273,1.330970)(0.747475,1.195183)(0.787879,0.916938)
(0.808081,0.780712)(0.818182,0.714737)(0.828283,0.650697)(0.838384,0.588941)(0.848485,0.529778)
(0.858586,0.473473)(0.868687,0.420239)(0.878788,0.370234)(0.888889,0.323560)(0.898990,0.280263)
(0.909091,0.240339)(0.919192,0.203733)(0.929293,0.170352)(0.939394,0.140067)(0.949495,0.112723)
(0.959596,0.088144)(0.969697,0.066142)(0.979798,0.046522)(0.989899,0.029086)(1.000000,0.013642)

\end{pspicture}%
  \vspace{0.5cm}
%
%
%
\providelength{\AxesLineWidth}       \setlength{\AxesLineWidth}{0.5pt}%
\providelength{\plotwidth}           \setlength{\plotwidth}{5cm}
\providelength{\LineWidth}           \setlength{\LineWidth}{0.7pt}%
\providelength{\MarkerSize}          \setlength{\MarkerSize}{4pt}%
\newrgbcolor{GridColor}{0.8 0.8 0.8}%
%
\psset{xunit=1.000000\plotwidth,yunit=0.657258\plotwidth}%
\begin{pspicture}(-0.210000,-0.828221)(1.000000,0.600000)%


\psline[linewidth=\AxesLineWidth,linecolor=GridColor](0.000000,-0.600000)(0.000000,-0.581742)
\psline[linewidth=\AxesLineWidth,linecolor=GridColor](0.500000,-0.600000)(0.500000,-0.581742)
\psline[linewidth=\AxesLineWidth,linecolor=GridColor](1.000000,-0.600000)(1.000000,-0.581742)
\psline[linewidth=\AxesLineWidth,linecolor=GridColor](0.000000,-0.500000)(0.012000,-0.500000)
\psline[linewidth=\AxesLineWidth,linecolor=GridColor](0.000000,0.000000)(0.012000,0.000000)
\psline[linewidth=\AxesLineWidth,linecolor=GridColor](0.000000,0.500000)(0.012000,0.500000)

{ \footnotesize 
\rput[t](0.000000,-0.618258){$0$}
\rput[t](0.500000,-0.618258){$0.5$}
\rput[t](1.000000,-0.618258){$1$}
\rput[r](-0.012000,-0.500000){$-0.5$}
\rput[r](-0.012000,0.000000){$0$}
\rput[r](-0.012000,0.500000){$0.5$}
} 

\psframe[linewidth=\AxesLineWidth,dimen=middle](0.000000,-0.600000)(1.000000,0.600000)

{ \small 
\rput[b](0.500000,-0.828221){
\begin{tabular}{c}
$x$\\
\end{tabular}
}

\rput[t]{90}(-0.210000,0.000000){
\begin{tabular}{c}
$\tilde{\nu}$\\
\end{tabular}
}
} 

\newrgbcolor{color194.0408}{0  0  0}
\psline[plotstyle=line,linejoin=1,linestyle=solid,linewidth=\LineWidth,linecolor=color194.0408]
(0.000000,0.114433)(0.010101,0.209780)(0.020202,0.288580)(0.030303,0.353042)(0.040404,0.405097)
(0.050505,0.446425)(0.060606,0.478494)(0.070707,0.502586)(0.080808,0.519822)(0.090909,0.531179)
(0.101010,0.537512)(0.111111,0.539570)(0.121212,0.538004)(0.131313,0.533386)(0.141414,0.526210)
(0.151515,0.516910)(0.161616,0.505862)(0.171717,0.493388)(0.181818,0.479770)(0.191919,0.465247)
(0.202020,0.450024)(0.212121,0.434274)(0.232323,0.401755)(0.282828,0.318893)(0.313131,0.270237)
(0.343434,0.223108)(0.373737,0.177584)(0.404040,0.133513)(0.434343,0.090612)(0.474747,0.034635)
(0.545455,-0.062495)(0.575758,-0.104803)(0.606061,-0.148057)(0.636364,-0.192587)(0.666667,-0.238637)
(0.696970,-0.286299)(0.727273,-0.335384)(0.777778,-0.418144)(0.787879,-0.434274)(0.797980,-0.450024)
(0.808081,-0.465247)(0.818182,-0.479770)(0.828283,-0.493388)(0.838384,-0.505862)(0.848485,-0.516910)
(0.858586,-0.526210)(0.868687,-0.533386)(0.878788,-0.538004)(0.888889,-0.539570)(0.898990,-0.537512)
(0.909091,-0.531179)(0.919192,-0.519822)(0.929293,-0.502586)(0.939394,-0.478494)(0.949495,-0.446425)
(0.959596,-0.405097)(0.969697,-0.353042)(0.979798,-0.288580)(0.989899,-0.209780)(1.000000,-0.114433)

\newrgbcolor{color195.0403}{0  0  0}
\psline[plotstyle=line,linejoin=1,linestyle=dashed,linewidth=\LineWidth,linecolor=color195.0403]
(0.000000,-0.002142)(0.010101,-0.004526)(0.020202,-0.007171)(0.030303,-0.010094)(0.040404,-0.013310)
(0.050505,-0.016833)(0.060606,-0.020671)(0.070707,-0.024829)(0.080808,-0.029305)(0.090909,-0.034091)
(0.101010,-0.039171)(0.111111,-0.044519)(0.131313,-0.055881)(0.181818,-0.085621)(0.191919,-0.091248)
(0.202020,-0.096591)(0.212121,-0.101576)(0.222222,-0.106133)(0.232323,-0.110197)(0.242424,-0.113711)
(0.252525,-0.116624)(0.262626,-0.118897)(0.272727,-0.120496)(0.282828,-0.121399)(0.292929,-0.121590)
(0.303030,-0.121063)(0.313131,-0.119819)(0.323232,-0.117865)(0.333333,-0.115216)(0.343434,-0.111891)
(0.353535,-0.107916)(0.363636,-0.103318)(0.373737,-0.098130)(0.383838,-0.092388)(0.393939,-0.086129)
(0.404040,-0.079393)(0.414141,-0.072222)(0.424242,-0.064658)(0.434343,-0.056747)(0.454545,-0.040061)
(0.474747,-0.022531)(0.525253,0.022531)(0.545455,0.040061)(0.565657,0.056747)(0.575758,0.064658)
(0.585859,0.072222)(0.595960,0.079393)(0.606061,0.086129)(0.616162,0.092388)(0.626263,0.098130)
(0.636364,0.103318)(0.646465,0.107916)(0.656566,0.111891)(0.666667,0.115216)(0.676768,0.117865)
(0.686869,0.119819)(0.696970,0.121063)(0.707071,0.121590)(0.717172,0.121399)(0.727273,0.120496)
(0.737374,0.118897)(0.747475,0.116624)(0.757576,0.113711)(0.767677,0.110197)(0.777778,0.106133)
(0.787879,0.101576)(0.797980,0.096591)(0.808081,0.091248)(0.828283,0.079787)(0.868687,0.055881)
(0.888889,0.044519)(0.898990,0.039171)(0.909091,0.034091)(0.919192,0.029305)(0.929293,0.024829)
(0.939394,0.020671)(0.949495,0.016833)(0.959596,0.013310)(0.969697,0.010094)(0.979798,0.007171)
(0.989899,0.004526)(1.000000,0.002142)

\end{pspicture}%
	\caption{\label{fig:initial_final_full_sys2}
Variation with $x$ of 
the 
initial perturbation (plotted with solid lines) and final perturbation (at $T=T_{E_{opt}}$, plotted
with a dashed line) of: 
(a) coherent perturbation velocity $\tilde{u}$ and (b) perturbation
turbulent viscosity $\tilde{\nu}$ for the 
LTO perturbation. 
The turbulent viscosity perturbation has a minor role in the dynamics.
Indeed, the behaviour of the coherent perturbation velocity $\tilde{u}$ is very similar to the behaviour observed in the FROZ case (see figure (\ref{fig:pert_lam_fr}b)).}
\end{center}
\end{figure}

The linear optimal gain analysis of this small problem brings to light the fact that there is a competition between essentially laminar and turbulent  perturbations. The turbulent perturbation is strikingly associated with a very large transient growth of the perturbation kinetic energy (due to the term $P_{E_2}$ of equation (\ref{eq:pert_ener})), and is optimal for short times. On the other hand, the laminar perturbation is only optimal for longer times, and is in fact the least stable perturbation, responsible for the weakest possible decay.
We show the initial and final time (at $T=T_{E_{opt}}$) structure of the STO perturbations and LTO perturbations 
in figures (\ref{fig:initial_final_full_sys1}) and (\ref{fig:initial_final_full_sys2}) respectively. The  STO (turbulent) perturbations are antisymmetric in $\tilde{u}$ and symmetric in $\tilde{\nu}$ whereas the LTO  (laminar) perturbations have the opposite symmetries.

This SVD analysis used to find the optimal perturbation associated with the largest achievable total norm gain given by  (\ref{eq:total_gain}) shows the major difference  between the FROZ case and the FULL case (as defined in table (\ref{tab:cases})). Indeed, we see that a new type
of perturbation appears, for which the turbulent viscosity component $\tilde{\nu}$ of the state vector is much larger than the coherent perturbation velocity $\tilde{u}$, which leads to a substantially larger kinetic energy gain.
The main drawback of this method is that we only access the energy gain information through the optimization of a non-physical norm (the total norm). Therefore, the calculated energy gain is not optimal in any sense. Indeed, the optimized total gain defined in (\ref{eq:total_gain}) is a product between the energy gain and a term which is a nontrivial function of the parameter $C_0$. As a consequence, in this analysis this contribution parameter $C_0$ is an output of the optimization (see figure (\ref{fig:gain_svd}b)) instead of being an input parameter.

 The semi-norm framework introduced in section (\ref{sec:VAR})  allows us to consider the optimization of the energy gain directly,
and it is interesting to investigate what, if any, are the points
of connection between the results of calculations
based on the semi-norm framework, and the 
SVD analysis based around optimization of the gain expressed
in terms of the total norm $\| \cdot \|_N$.

\ \\
\paragraph{Energy optimization: semi-norm constraints}\ \\

\begin{figure}
		\centering
%
%
%
\providelength{\AxesLineWidth}       \setlength{\AxesLineWidth}{0.5pt}%
\providelength{\plotwidth}           \setlength{\plotwidth}{5cm}
\providelength{\LineWidth}           \setlength{\LineWidth}{0.7pt}%
\providelength{\MarkerSize}          \setlength{\MarkerSize}{4pt}%
\newrgbcolor{GridColor}{0.8 0.8 0.8}%
%
\psset{xunit=0.500000\plotwidth,yunit=0.262903\plotwidth}%
\begin{pspicture}(-0.420000,-0.871582)(2.000000,2.698970)%


\psline[linewidth=\AxesLineWidth,linecolor=GridColor](0.000000,-0.301030)(0.000000,-0.255386)
\psline[linewidth=\AxesLineWidth,linecolor=GridColor](0.500000,-0.301030)(0.500000,-0.255386)
\psline[linewidth=\AxesLineWidth,linecolor=GridColor](1.000000,-0.301030)(1.000000,-0.255386)
\psline[linewidth=\AxesLineWidth,linecolor=GridColor](1.500000,-0.301030)(1.500000,-0.255386)
\psline[linewidth=\AxesLineWidth,linecolor=GridColor](2.000000,-0.301030)(2.000000,-0.255386)
\psline[linewidth=\AxesLineWidth,linecolor=GridColor](0.000000,0.000000)(0.024000,0.000000)
\psline[linewidth=\AxesLineWidth,linecolor=GridColor](0.000000,1.000000)(0.024000,1.000000)
\psline[linewidth=\AxesLineWidth,linecolor=GridColor](0.000000,2.000000)(0.024000,2.000000)

\psline[linewidth=\AxesLineWidth,linecolor=GridColor](0.000000,0.301030)(0.016000,0.301030)
\psline[linewidth=\AxesLineWidth,linecolor=GridColor](0.000000,0.477121)(0.016000,0.477121)
\psline[linewidth=\AxesLineWidth,linecolor=GridColor](0.000000,0.602060)(0.016000,0.602060)
\psline[linewidth=\AxesLineWidth,linecolor=GridColor](0.000000,0.698970)(0.016000,0.698970)
\psline[linewidth=\AxesLineWidth,linecolor=GridColor](0.000000,0.778151)(0.016000,0.778151)
\psline[linewidth=\AxesLineWidth,linecolor=GridColor](0.000000,0.845098)(0.016000,0.845098)
\psline[linewidth=\AxesLineWidth,linecolor=GridColor](0.000000,0.903090)(0.016000,0.903090)
\psline[linewidth=\AxesLineWidth,linecolor=GridColor](0.000000,0.954243)(0.016000,0.954243)
\psline[linewidth=\AxesLineWidth,linecolor=GridColor](0.000000,1.301030)(0.016000,1.301030)
\psline[linewidth=\AxesLineWidth,linecolor=GridColor](0.000000,1.477121)(0.016000,1.477121)
\psline[linewidth=\AxesLineWidth,linecolor=GridColor](0.000000,1.602060)(0.016000,1.602060)
\psline[linewidth=\AxesLineWidth,linecolor=GridColor](0.000000,1.698970)(0.016000,1.698970)
\psline[linewidth=\AxesLineWidth,linecolor=GridColor](0.000000,1.778151)(0.016000,1.778151)
\psline[linewidth=\AxesLineWidth,linecolor=GridColor](0.000000,1.845098)(0.016000,1.845098)
\psline[linewidth=\AxesLineWidth,linecolor=GridColor](0.000000,1.903090)(0.016000,1.903090)
\psline[linewidth=\AxesLineWidth,linecolor=GridColor](0.000000,1.954243)(0.016000,1.954243)

{ \footnotesize 
\rput[t](0.000000,-0.346674){$0$}
\rput[t](0.500000,-0.346674){$0.5$}
\rput[t](1.000000,-0.346674){$1$}
\rput[t](1.500000,-0.346674){$1.5$}
\rput[t](2.000000,-0.346674){$2$}
\rput[r](-0.024000,0.000000){$10^{0}$}
\rput[r](-0.024000,1.000000){$10^{1}$}
\rput[r](-0.024000,2.000000){$10^{2}$}
} 

\psframe[linewidth=\AxesLineWidth,dimen=middle](0.000000,-0.301030)(2.000000,2.698970)

{ \small 
\rput[b](1.000000,-0.871582){
\begin{tabular}{c}
$T$\\
\end{tabular}
}

\rput[t]{90}(-0.420000,1.198970){
\begin{tabular}{c}
$G_E$\\
\end{tabular}
}
} 

\newrgbcolor{color232.0045}{0.50196     0.50196     0.50196}
\psline[plotstyle=line,linejoin=1,linestyle=solid,linewidth=\LineWidth,linecolor=color232.0045]
(1.999663,-0.208190)(2.000000,-0.208241)
\psline[plotstyle=line,linejoin=1,linestyle=solid,linewidth=\LineWidth,linecolor=color232.0045]
(0.010000,0.002505)(0.030991,0.012589)(0.050993,0.021352)(0.069987,0.028837)(0.087976,0.035115)
(0.105492,0.040416)(0.122474,0.044748)(0.139463,0.048290)(0.156992,0.051146)(0.174959,0.053263)
(0.193454,0.054652)(0.212988,0.055330)(0.233934,0.055260)(0.256986,0.054369)(0.282920,0.052537)
(0.312485,0.049602)(0.348435,0.045157)(0.394408,0.038596)(0.470416,0.026688)(0.518919,0.019480)
(0.653426,-0.002668)(0.817391,-0.028913)(0.959491,-0.051118)(1.022785,-0.060049)(1.193714,-0.086367)
(1.461459,-0.127167)(1.521700,-0.135511)(1.999663,-0.208190)

\newrgbcolor{color233.004}{0.50196     0.50196     0.50196}
\psline[plotstyle=line,linejoin=1,linestyle=solid,linewidth=\LineWidth,linecolor=color233.004]
(1.999663,-0.188693)(2.000000,-0.188744)
\psline[plotstyle=line,linejoin=1,linestyle=solid,linewidth=\LineWidth,linecolor=color233.004]
(0.010000,0.004539)(0.024493,0.016833)(0.038990,0.028142)(0.053495,0.038482)(0.067490,0.047538)
(0.081477,0.055690)(0.094979,0.062705)(0.108497,0.068883)(0.121975,0.074201)(0.135462,0.078707)
(0.149475,0.082562)(0.163489,0.085607)(0.177953,0.087945)(0.192953,0.089582)(0.208996,0.090534)
(0.225945,0.090742)(0.244443,0.090162)(0.264975,0.088697)(0.288421,0.086182)(0.315965,0.082349)
(0.350947,0.076560)(0.398928,0.067700)(0.461487,0.055495)(0.521400,0.044876)(0.592351,0.031344)
(0.686807,0.014348)(0.786779,-0.002745)(0.899343,-0.021178)(0.959491,-0.030819)(1.023279,-0.040020)
(1.137684,-0.057937)(1.460000,-0.107392)(1.522191,-0.116034)(1.960482,-0.182742)(1.999663,-0.188693)

\newrgbcolor{color234.004}{0  0  0}
\psline[plotstyle=line,linejoin=1,linestyle=dashed,linewidth=\LineWidth,linecolor=color234.004]
(1.999663,-0.015483)(2.000000,-0.015549)
\psline[plotstyle=line,linejoin=1,linestyle=dashed,linewidth=\LineWidth,linecolor=color234.004]
(0.010000,0.073395)(0.014998,0.120417)(0.019995,0.160808)(0.025493,0.199231)(0.030991,0.232627)
(0.036490,0.261950)(0.042490,0.290125)(0.048492,0.315000)(0.054996,0.338852)(0.061498,0.359999)
(0.068489,0.380218)(0.075481,0.398254)(0.082976,0.415564)(0.090977,0.432127)(0.099484,0.447977)
(0.108497,0.463199)(0.118481,0.478595)(0.128965,0.493358)(0.139463,0.506868)(0.150477,0.519835)
(0.161994,0.532225)(0.173961,0.543975)(0.185948,0.554696)(0.198464,0.564860)(0.210996,0.574065)
(0.223950,0.582630)(0.236934,0.590299)(0.249959,0.597118)(0.262985,0.603097)(0.276431,0.608427)
(0.289922,0.612955)(0.303466,0.616713)(0.317457,0.619799)(0.331907,0.622182)(0.346929,0.623848)
(0.361986,0.624741)(0.377404,0.624891)(0.392903,0.624285)(0.408993,0.622867)(0.425897,0.620554)
(0.446405,0.616864)(0.474392,0.610892)(0.495389,0.605724)(0.512971,0.600660)(0.531353,0.594504)
(0.553437,0.586192)(0.581838,0.574575)(0.610000,0.562211)(0.639327,0.548445)(0.673354,0.531532)
(0.711973,0.511391)(0.756959,0.486952)(0.813447,0.455257)(0.908983,0.400482)(0.948321,0.377621)
(0.967864,0.367026)(1.011472,0.344795)(1.035708,0.331400)(1.076235,0.308458)(1.136681,0.275739)
(1.186166,0.249656)(1.236152,0.224213)(1.283138,0.201158)(1.329641,0.179216)(1.370748,0.160639)
(1.431601,0.134322)(1.455895,0.124461)(1.478114,0.116338)(1.519253,0.102019)(1.568756,0.083913)
(1.602803,0.072391)(1.645130,0.058942)(1.682029,0.048002)(1.719220,0.037797)(1.755873,0.028548)
(1.791540,0.020351)(1.826544,0.013106)(1.860000,0.006966)(1.900193,0.000545)(1.963856,-0.009163)
(1.995051,-0.014602)(1.999663,-0.015483)

\newrgbcolor{color235.004}{0  0  0}
\psline[plotstyle=line,linejoin=1,linestyle=solid,linewidth=\LineWidth,linecolor=color235.004]
(1.999663,1.223864)(2.000000,1.223562)
\psline[plotstyle=line,linejoin=1,linestyle=solid,linewidth=\LineWidth,linecolor=color235.004]
(0.010000,1.053295)(0.012499,1.153054)(0.015498,1.248171)(0.018996,1.337352)(0.022994,1.420353)
(0.026992,1.489529)(0.031491,1.555639)(0.036490,1.618474)(0.041990,1.678039)(0.047492,1.730040)
(0.053495,1.780125)(0.060000,1.828259)(0.066991,1.874374)(0.074482,1.918666)(0.082477,1.961254)
(0.090977,2.002272)(0.099984,2.041859)(0.109499,2.080150)(0.119479,2.117099)(0.129465,2.151150)
(0.139463,2.182592)(0.149475,2.211642)(0.159499,2.238471)(0.169471,2.263084)(0.179451,2.285806)
(0.189449,2.306799)(0.199466,2.326177)(0.209498,2.344027)(0.219463,2.360305)(0.229937,2.375957)
(0.239936,2.389597)(0.249959,2.402069)(0.260000,2.413418)(0.270451,2.424075)(0.280922,2.433646)
(0.291425,2.442204)(0.301959,2.449800)(0.312485,2.456455)(0.322931,2.462186)(0.333906,2.467332)
(0.344921,2.471654)(0.355975,2.475196)(0.367451,2.478084)(0.379398,2.480280)(0.391399,2.481692)
(0.403454,2.482345)(0.415955,2.482246)(0.429385,2.481341)(0.444396,2.479518)(0.462478,2.476450)
(0.480369,2.472665)(0.496395,2.468534)(0.511485,2.463871)(0.526868,2.458270)(0.543870,2.451210)
(0.563958,2.441949)(0.585835,2.430987)(0.607978,2.419013)(0.631325,2.405462)(0.656459,2.389919)
(0.683807,2.372042)(0.712960,2.352007)(0.744326,2.329459)(0.778794,2.303659)(0.815419,2.275220)
(0.860000,2.239433)(0.898331,2.207880)(0.935736,2.175839)(0.957965,2.157138)(0.978244,2.141216)
(1.004910,2.120587)(1.020319,2.107737)(1.049316,2.081876)(1.069326,2.064360)(1.112938,2.027905)
(1.136681,2.006899)(1.200815,1.948838)(1.286138,1.871348)(1.407948,1.759799)(1.447208,1.723279)
(1.464380,1.708162)(1.490103,1.687000)(1.510486,1.670093)(1.528596,1.653950)(1.581062,1.604939)
(1.793062,1.410643)(1.909482,1.304896)(1.999663,1.223864)

\end{pspicture}%
		\caption{\label{fig:gain_semi_norm}Variation of the energy semi-norm gain $G_E(T)$ as defined
in (\ref{eq:gain_u})  with optimization time interval $T$ for values of $C_0=K_0/E_0$ ranging from $10^{-2}$ to $10^{1}$. Gray curves corresponds to $C_0=10^{-2}$ and $C_0=10^{-1}$, the black dashed curve corresponds to the balanced case $C_0=1$ and the black curve corresponds to $C_0=10$.
We notice that when the contribution of the turbulent viscosity perturbation increases, higher gains are achieved. A whole family of black curves exists with gains evolving linearly with $C_0$ for larger values of $C_0$.}
\end{figure}
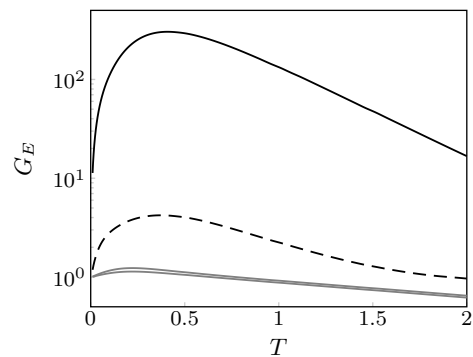

The perturbation kinetic energy gain $G_E$ defined in (\ref{eq:gain_u}) is a highly relevant quantity in the dynamics of the system. As a consequence we will now use the semi-norm framework developed in section (\ref{sec:VAR}) to optimize this quantity over finite time
intervals.
We present in figure (\ref{fig:gain_semi_norm}) the variation of  $G_E(T) $ with optimization time interval $T$ for different values of $C_0$ (as defined in (\ref{eq:ratio_0}). Gray curves represent energy gains for $C_0<1$ (i.e. for ``laminar'' perturbations) whereas the black curves represent energy gains for $C_0>1$, (i.e. for ``turbulent'' perturbations)  while the black curve corresponds to the balanced case $C_0=1$. We can clearly see that laminar perturbations corresponding to low values of $C_0$ are associated with very low gain values, while the turbulent perturbations (with $C_0 > 1$) can lead to much higher gains. We do not  plot on the figure the optimal curves for large values of $C_0$ because we find that the gain increases linearly with $C_0$ for large $C_0$.


\begin{figure}
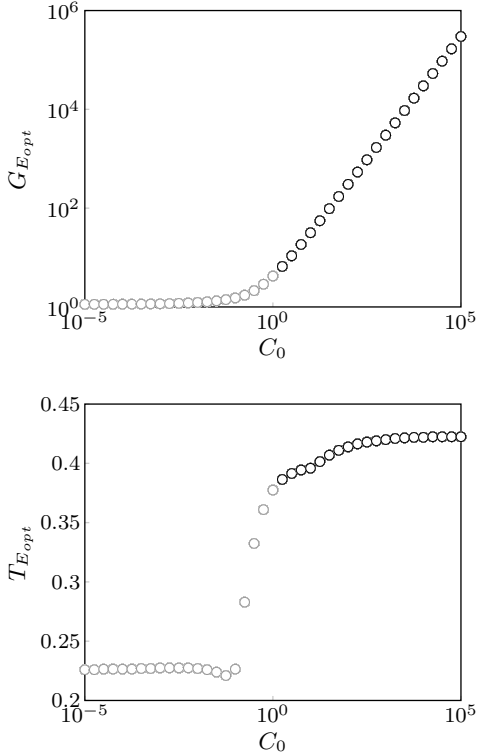

  \include{max_gain}
  \vspace{0.5cm}
	\include{time_opt}		
	\caption{\label{fig:max_gain_time_opt}Variation with $C_0=K_0/E_0$ as defined
in (\ref{eq:ratio_0}) of (a) optimal gain $G_{E_{opt}}$; (b) optimal times $T_{E_{opt}}$ for the FULL case. We see that the optimal gain increases linearly with $C_0$, meaning that we can linearly extract as much energy as we want from the base flow and turbulent viscosity perturbation. The optimal time is also increasing with $C_0$, but not in such a neat fashion. The line types are the same as in figure (\ref{fig:gain_semi_norm}).}
\end{figure}

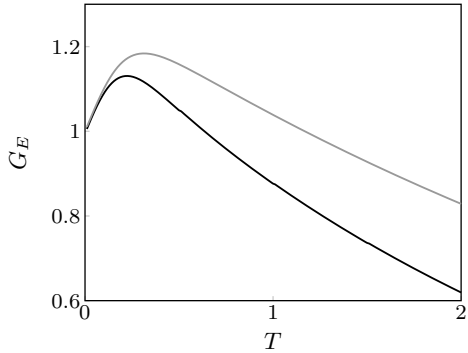
\begin{figure}
		\centering 
%
%
%
\providelength{\AxesLineWidth}       \setlength{\AxesLineWidth}{0.5pt}%
\providelength{\plotwidth}           \setlength{\plotwidth}{5cm}
\providelength{\LineWidth}           \setlength{\LineWidth}{0.7pt}%
\providelength{\MarkerSize}          \setlength{\MarkerSize}{4pt}%
\newrgbcolor{GridColor}{0.8 0.8 0.8}%
%
\psset{xunit=0.500000\plotwidth,yunit=1.126728\plotwidth}%
\begin{pspicture}(-0.420000,0.466871)(2.000000,1.300000)%


\psline[linewidth=\AxesLineWidth,linecolor=GridColor](0.000000,0.600000)(0.000000,0.610650)
\psline[linewidth=\AxesLineWidth,linecolor=GridColor](1.000000,0.600000)(1.000000,0.610650)
\psline[linewidth=\AxesLineWidth,linecolor=GridColor](2.000000,0.600000)(2.000000,0.610650)
\psline[linewidth=\AxesLineWidth,linecolor=GridColor](0.000000,0.600000)(0.024000,0.600000)
\psline[linewidth=\AxesLineWidth,linecolor=GridColor](0.000000,0.800000)(0.024000,0.800000)
\psline[linewidth=\AxesLineWidth,linecolor=GridColor](0.000000,1.000000)(0.024000,1.000000)
\psline[linewidth=\AxesLineWidth,linecolor=GridColor](0.000000,1.200000)(0.024000,1.200000)

{ \footnotesize 
\rput[t](0.000000,0.589350){$0$}
\rput[t](1.000000,0.589350){$1$}
\rput[t](2.000000,0.589350){$2$}
\rput[r](-0.024000,0.600000){$0.6$}
\rput[r](-0.024000,0.800000){$0.8$}
\rput[r](-0.024000,1.000000){$1$}
\rput[r](-0.024000,1.200000){$1.2$}
} 

\psframe[linewidth=\AxesLineWidth,dimen=middle](0.000000,0.600000)(2.000000,1.300000)

{ \small 
\rput[b](1.000000,0.466871){
\begin{tabular}{c}
$T$\\
\end{tabular}
}

\rput[t]{90}(-0.420000,0.950000){
\begin{tabular}{c}
$G_E$\\
\end{tabular}
}
} 

\newrgbcolor{color218.0262}{0  0  0}
\psline[plotstyle=line,linejoin=1,linestyle=solid,linewidth=\LineWidth,linecolor=color218.0262]
(1.999881,0.619338)(2.000000,0.619312)
\psline[plotstyle=line,linejoin=1,linestyle=solid,linewidth=\LineWidth,linecolor=color218.0262]
(0.010000,1.005483)(0.031695,1.028580)(0.049297,1.046266)(0.064882,1.060877)(0.079194,1.073267)
(0.092580,1.083858)(0.105370,1.093008)(0.117678,1.100873)(0.129696,1.107641)(0.141481,1.113392)
(0.153163,1.118231)(0.164853,1.122230)(0.176657,1.125437)(0.188678,1.127881)(0.200982,1.129568)
(0.213743,1.130503)(0.227049,1.130664)(0.241076,1.130015)(0.256032,1.128495)(0.272052,1.126034)
(0.289589,1.122495)(0.309075,1.117698)(0.331164,1.111378)(0.357020,1.103073)(0.388965,1.091869)
(0.433488,1.075224)(0.491147,1.052895)(0.498539,1.049932)(0.503470,1.048779)(0.510286,1.047484)
(0.515962,1.045416)(0.527187,1.040871)(0.609156,1.009623)(0.676948,0.984608)(0.744590,0.960501)
(0.814470,0.936458)(0.887532,0.912183)(0.963740,0.887719)(0.992581,0.878581)(0.999454,0.876478)
(1.011000,0.874577)(1.019017,0.872043)(1.095194,0.848835)(1.180446,0.823714)(1.268444,0.798629)
(1.358878,0.773690)(1.451575,0.748966)(1.493216,0.738036)(1.500172,0.736344)(1.511125,0.734872)
(1.550086,0.724845)(1.646620,0.700771)(1.745012,0.677065)(1.845190,0.653757)(1.947218,0.630847)
(1.999881,0.619338)

\newrgbcolor{color219.0258}{0.6         0.6         0.6}
\psline[plotstyle=line,linejoin=1,linestyle=solid,linewidth=\LineWidth,linecolor=color219.0258]
(1.999881,0.828672)(2.000000,0.828649)
\psline[plotstyle=line,linejoin=1,linestyle=solid,linewidth=\LineWidth,linecolor=color219.0258]
(0.010000,1.010622)(0.044887,1.047367)(0.066884,1.069544)(0.085376,1.087155)(0.102082,1.102032)
(0.117577,1.114817)(0.132275,1.125955)(0.146463,1.135741)(0.160295,1.144339)(0.173953,1.151909)
(0.187462,1.158504)(0.201080,1.164275)(0.214839,1.169244)(0.228878,1.173466)(0.243240,1.176952)
(0.258160,1.179746)(0.273728,1.181836)(0.290097,1.183212)(0.307438,1.183851)(0.326114,1.183712)
(0.346311,1.182729)(0.368452,1.180812)(0.393007,1.177840)(0.420866,1.173611)(0.453189,1.167828)
(0.491917,1.160003)(0.541046,1.149148)(0.611134,1.132656)(0.876798,1.068057)(1.008260,1.036866)
(1.135694,1.007445)(1.263371,0.978790)(1.392401,0.950654)(1.523114,0.922975)(1.655658,0.895730)
(1.790167,0.868903)(1.926666,0.842501)(1.999881,0.828672)

\end{pspicture}%
		\caption{\label{fig:feva_vs_fla}Variation of coherent perturbation kinetic energy gain $G_E(T)$, as defined
in (\ref{eq:gain_u}) for the FROZ case (plotted with a grey line) and the FULL  case in the limit $C_0 \rightarrow 0$ ( plotted with a black line).
This results has been obtained with the same set of parameters values, and especially with $r=0.5$. Even in this limit, the results for the two cases  differ quantitatively, both in optimal time and optimal gain predictions.}
\end{figure}

We plot in figure (\ref{fig:max_gain_time_opt}) the maximum gain $G_{E_{opt}}$ as a function of the ratio
$C_0$ as defined in (\ref{eq:ratio_0}). The perturbation velocity energy gain in the limit of low $C_0$ is constant ($G_{E_{opt}}\simeq 1.13$). In order to compare the limit of the FULL case and the frozen turbulent viscosity FROZ case, (using the same
values for $\nu=0.05$ and $r=0.5$, where $r$ is defined in (\ref{eq:ratio}))
as we plot the two corresponding optimal gain curves in figure (\ref{fig:feva_vs_fla}). The first observation is that the FULL case does not converge toward the FROZ case when the contribution from the turbulent viscosity perturbation is very small compared to the coherent velocity
perturbation (i.e. in the limit $C_0 \rightarrow 0$). This means that no matter how small $C_0$ is, the contribution of the viscosity perturbation to the dynamics is never negligible. Indeed, the maximum gain we obtained for FROZ case was  approximately $G_{E_{opt}} = G_{N_{opt}} \simeq 1.18$. 
This means that the stability analysis of mean flows is a singular problem, since in the limit of low values of the turbulent viscosity perturbation, we do not recover the results of the FROZ case.

Moreover, in the limit of large $C_0$ (turbulent perturbations), $G_{E_{opt}}$ evolves linearly with $C_0$:
\begin{equation}
G_{E_{opt}} \propto C_0,
\end{equation}
with a proportionality coefficient of about $3$ in our case. 
This is the signature of the linear relation involved in the energy production mechanism (term $P_{E_2}$ of equation (\ref{eq:pert_ener})). As a consequence, we can linearly extract energy from the base flow via the  perturbation turbulent viscosity perturbation as the ratio $C_0$ increases. 
Indeed, the growth of the turbulent perturbations is larger for initially large values of $K$ because for large values of $\tilde{\nu}$, we consequently have a large production term $P_{E_2}$ in (\ref{eq:energy}). In fact, $\tilde{\nu}$ catalyses very strong growth of $E$ for the turbulent modes. For the laminar modes however, this mechanism exists but is not proportionally significant. (It is totally absent for the FROZ case).

Turning our attention to time dependence, figure (\ref{fig:max_gain_time_opt}b) shows that turbulent perturbations are associated with larger optimal times than laminar perturbations,  though once again, the optimal time as $C_0\rightarrow 0$ for the FULL case is
still different from the optimal time for the FROZ case.
In summary, 
we present the results 
of the three different analyses in  table (\ref{tab:sumup}). We distinguish
between laminar perturbations ($C_0 \ll 1$) and  turbulent perturbations ($C_0 \gg 1$) 
for the FULL case, i.e. solutions of the full linearized equation
system defined by 
(\ref{eq:pert_linearized}).

\begin{table}
\begin{center}
\begin{tabular}{l c c c c c c c c}
 	&	\phantom{1} &   \phantom{1}{LAM}\phantom{1} 	  & \phantom{1}{FROZ}\phantom{12}				& \multicolumn{5}{c}{{FULL}}\\
 	     				  \cline{5-9}
 	&	\phantom{1} &	   			  	&	    				& \multicolumn{2}{c}{\footnotesize{SVD}} & \phantom{1}	& \multicolumn{2}{c}{\footnotesize{SN}}  \\
 	     					\cline{5-6}\cline{8-9}
 	&	\phantom{1} &							&		& \footnotesize{LTO}  & \footnotesize{STO}	& & \footnotesize{LAM}	& \footnotesize{TURB}	\\
  \\
  \hline  
  \hline
  \\
  $r$ 				  & \phantom{1} & $0$				   & $0.5$				 & $0.5$								& $0.5$						&	   	 & $0.5$			&	$0.5$		 	\\
  $C_0$					& \phantom{1} & $0$				   & $0$				 	 & $\mathbf{\lesssim1}$			& $\mathbf{\gg1}$						&	   	 & $\ll1$			&	$\gg1$		 	\\
  $T_{E_{opt}}$	& \phantom{1} & $\mathbf{1.15}$  & $\mathbf{0.31}$ 	   & $0.12$							& $1.21$					&			& $\mathbf{0.22}$ &	$\mathbf{0.42}$	\\
  $T_{N_{opt}}$ & \phantom{1} & $\mathbf{1.15}$   & $\mathbf{0.31}$ 		 & $\mathbf{0.16}$					& $\mathbf{0.38}$	&			& $0.22$					& $0.39$	\\
  $G_{E_{opt}}$ &	\phantom{1} & $\mathbf{2.91}$ 	& $\mathbf{1.18}$ 		 & $3.34$			& $9.3\ 10^3$			&			& $\mathbf{1.13}$		& $\mathbf{\propto C_0}$\\
  $G_{N_{opt}}$ &	\phantom{1} & $\mathbf{2.91}$    & $\mathbf{1.18}$ 		 & $\mathbf{1.59}$			& $\mathbf{3.46}$ &		 & $1.13$			& $3.43$ \\
  \\
  \hline  
  \hline
  \\
\end{tabular}
\caption{\label{tab:sumup}Summary of the results section. LAM: Laminar case using 
the frozen turbulent viscosity assumption $\tilde{\nu}=0$
 with $r=0=\overline{\nu_t}$; FROZ: Frozen turbulent viscosity case with $\tilde{\nu}=0$, FULL: Full linearized case solving (\ref{eq:pert_linearized}), SVD: Singular Value Decomposition, SN: Semi-Norm, STO: Short Time Optimal, LTO: Long Time Optimal, LAM: Laminar ($C_0\ll 1)$, TURB: Turbulent ($C_0\gg 1$). The bold font means that this quantity is optimal.}
\end{center}
\end{table}

\begin{figure}
	 \centering	\input{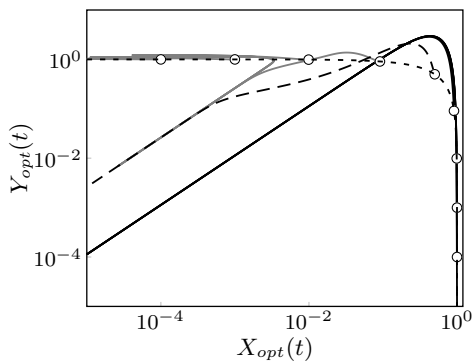}
		\caption{\label{fig:parametrized} Evolution diagram with time of scaled perturbation turbulent viscosity  $X_{opt}(t)$ and perturbation velocity $Y_{opt}(t)$, defined by (\ref{eq:rkt}) and (\ref{eq:ret}) respectively, parameterized by  time $t$ for an optimal perturbation with optimization interval $T=0.3$, 
for different values of the initial ratio $C_0$. We can clearly see the two different types of perturbation: laminar perturbations ($C_0\ll1$) are plotted with gray lines; and turbulent perturbations ($C_0\gg1$) are plotted with black lines which are the two limit cases. Initial conditions are represented with black circles, and lie on the curve $x+y=1$ (plotted with a short dashed line). The laminar perturbations do not have significant  transient growth, while the turbulent perturbations in black have very large transient growth of the energy. An intermediate state (plotted with a dashed line) exists for $C_0=1$, corresponding to the point $(0.5,0.5)$ on this figure.}
\end{figure}

Another way to understand the mechanism of energy production is to consider the plot of the kinetic energy of the coherent perturbation velocity as a function of the semi-norm of the  perturbation turbulent viscosity, parameterized by time $t$. We consider all the optimal perturbations for a given optimization time interval $T=0.3$ for different values of the ratio $C_0=K_0/E_0$ (defined in (\ref{eq:ratio_0})) and evolve the corresponding optimal perturbations in time. In figure (\ref{fig:parametrized}) we plot parametric curves
defined by  $X_{opt}(t),Y_{opt}(t)$ where $X_{opt}(t)$ and
$Y_{opt}(t)$ are defined as
\begin{eqnarray}
X_{opt}(t)= \frac{K(t)}{E_0+K_0} &=&G_E(t) C(t) (1-R_0),\label{eq:rkt} \\  
Y_{opt}(t)=\frac{E(t)}{E_0+K_0} &=&G_E(t) (1-R_0)\label{eq:ret}.
\end{eqnarray}

On the figure initial conditions 
are shown as black dots, which from the definitions
(\ref{eq:ret})-(\ref{eq:rkt}) clearly must lie on the curve $x+y=1$, which is plotted with a dashed line.
Using the same line types as figure (\ref{fig:gain_semi_norm}),
gray curves denote ``laminar'' perturbations ($C_0 \ll 1$)
and
black curves denote ``turbulent'' perturbations ($C_0 \gg 1$)
while a black dashed curve denotes the marginal perturbation with $C_0=1$.

The gain $G_E$ can be retrieved from this figure by evaluating the difference in $y-$coordinate from an initial condition to the maximum of the corresponding curve, due to the logarithmic scalings.  The gray curves (for the laminar perturbations) are associated with very low gains since there is little vertical variation while the black curves (turbulent perturbations) show a very large energy gain since the distance from an initial condition to the maximum of the curves is getting larger and larger as the amount of perturbation 
turbulent viscosity in the initial perturbation increases.
Moreover, we can see that the energy production for the turbulent perturbations does not immediately lead to a decrease of the turbulent viscosity perturbation norm, implying (of course) that the energy has been taken from the mean flow. Indeed, the optimal turbulent perturbation has the best symmetry and shape for both $\tilde{u}$ and $\tilde{\nu}$ in order to maximize this production, through the large size of the catalytic term $P_{E_2}$ in (\ref{eq:pert_ener}).

\begin{figure}
	 \centering	
%
%
%
\providelength{\AxesLineWidth}       \setlength{\AxesLineWidth}{0.5pt}%
\providelength{\plotwidth}           \setlength{\plotwidth}{5cm}
\providelength{\LineWidth}           \setlength{\LineWidth}{0.7pt}%
\providelength{\MarkerSize}          \setlength{\MarkerSize}{4pt}%
\newrgbcolor{GridColor}{0.8 0.8 0.8}%
%
\psset{xunit=0.196882\plotwidth,yunit=0.131452\plotwidth}%
\begin{pspicture}(-6.066628,-6.141104)(0.079181,1.000000)%


\psline[linewidth=\AxesLineWidth,linecolor=GridColor](-4.000000,-5.000000)(-4.000000,-4.908712)
\psline[linewidth=\AxesLineWidth,linecolor=GridColor](-2.000000,-5.000000)(-2.000000,-4.908712)
\psline[linewidth=\AxesLineWidth,linecolor=GridColor](0.000000,-5.000000)(0.000000,-4.908712)
\psline[linewidth=\AxesLineWidth,linecolor=GridColor](-5.000000,-4.000000)(-4.939050,-4.000000)
\psline[linewidth=\AxesLineWidth,linecolor=GridColor](-5.000000,-2.000000)(-4.939050,-2.000000)
\psline[linewidth=\AxesLineWidth,linecolor=GridColor](-5.000000,0.000000)(-4.939050,0.000000)


{ \footnotesize 
\rput[t](-4.000000,-5.091288){$10^{-4}$}
\rput[t](-2.000000,-5.091288){$10^{-2}$}
\rput[t](0.000000,-5.091288){$10^{0}$}
\rput[r](-5.060950,-4.000000){$10^{-4}$}
\rput[r](-5.060950,-2.000000){$10^{-2}$}
\rput[r](-5.060950,0.000000){$10^{0}$}
} 

\psframe[linewidth=\AxesLineWidth,dimen=middle](-5.000000,-5.000000)(0.079181,1.000000)

{ \small 
\rput[b](-2.460409,-6.141104){
\begin{tabular}{c}
$X_{opt}(t)$\\
\end{tabular}
}

\rput[t]{90}(-6.066628,-2.000000){
\begin{tabular}{c}
$Y_{opt}(t)$\\
\end{tabular}
}
} 

\newrgbcolor{color305.0062}{0.50196     0.50196     0.50196}
\psline[plotstyle=line,linejoin=1,linestyle=solid,linewidth=\LineWidth,linecolor=color305.0062]
(-4.998960,-2.580640)(-5.000000,-2.581681)
\psline[plotstyle=line,linejoin=1,linestyle=solid,linewidth=\LineWidth,linecolor=color305.0062]
(-1.212247,-0.027492)(-1.261852,-0.022876)(-1.307465,-0.015192)(-1.351463,-0.005469)(-1.395368,0.005498)
(-1.485042,0.029022)(-1.531548,0.040764)(-1.579497,0.052095)(-1.629039,0.062808)(-1.680334,0.072751)
(-1.733493,0.081828)(-1.788641,0.089980)(-1.845881,0.097186)(-1.905336,0.103449)(-1.967128,0.108794)
(-2.031405,0.113264)(-2.098334,0.116907)(-2.168118,0.119781)(-2.240995,0.121949)(-2.317260,0.123470)
(-2.397261,0.124408)(-2.481429,0.124820)(-2.570282,0.124765)(-2.664458,0.124295)(-2.764735,0.123460)
(-2.987599,0.120873)(-3.901551,0.108338)(-4.038452,0.105839)(-4.095011,0.103276)(-4.051572,0.100664)
(-3.942730,0.098017)(-3.687184,0.092661)(-3.572813,0.089969)(-3.471440,0.087279)(-3.382089,0.084596)
(-3.303222,0.081926)(-3.233323,0.079272)(-3.171067,0.076639)(-3.115340,0.074029)(-3.065221,0.071444)
(-3.019950,0.068887)(-2.978898,0.066359)(-2.907434,0.061392)(-2.847555,0.056549)(-2.796898,0.051831)
(-2.753715,0.047234)(-2.716677,0.042755)(-2.684759,0.038390)(-2.657149,0.034131)(-2.633201,0.029973)
(-2.612387,0.025908)(-2.594276,0.021931)(-2.578509,0.018035)(-2.564785,0.014213)(-2.552850,0.010460)
(-2.542487,0.006770)(-2.533513,0.003137)(-2.525767,-0.000442)(-2.519110,-0.003973)(-2.513423,-0.007459)
(-2.508600,-0.010906)(-2.504550,-0.014315)(-2.501190,-0.017691)(-2.498450,-0.021036)(-2.496266,-0.024353)
(-2.494581,-0.027645)(-2.492883,-0.032540)(-2.492050,-0.037390)(-2.491959,-0.042201)(-2.492504,-0.046978)
(-2.493597,-0.051725)(-2.495773,-0.058015)(-2.498636,-0.064268)(-2.502996,-0.072040)(-2.509126,-0.081317)
(-2.517205,-0.092088)(-2.528627,-0.105878)(-2.543619,-0.122675)(-2.566587,-0.147040)(-2.606737,-0.188077)
(-2.735460,-0.317134)(-4.998960,-2.580640)

\newrgbcolor{color306.0057}{0  0  0}
\psline[plotstyle=line,linejoin=1,linestyle=solid,linewidth=\LineWidth,linecolor=color306.0057]
(-4.998604,-3.945010)(-5.000000,-3.946406)
\psline[plotstyle=line,linejoin=1,linestyle=solid,linewidth=\LineWidth,linecolor=color306.0057]
(-0.000040,-4.034040)(-0.008660,-2.042887)(-0.016333,-1.490718)(-0.023524,-1.159723)(-0.030532,-0.924038)
(-0.037471,-0.742005)(-0.044432,-0.594582)(-0.051459,-0.471459)(-0.058588,-0.366409)(-0.065838,-0.275389)
(-0.073227,-0.195613)(-0.080761,-0.125084)(-0.088448,-0.062317)(-0.096290,-0.006170)(-0.104285,0.044249)
(-0.112433,0.089660)(-0.120728,0.130654)(-0.129166,0.167719)(-0.137739,0.201270)(-0.146440,0.231662)
(-0.155262,0.259201)(-0.164195,0.284153)(-0.173231,0.306752)(-0.182362,0.327204)(-0.191579,0.345694)
(-0.200875,0.362384)(-0.210242,0.377421)(-0.219672,0.390936)(-0.229159,0.403047)(-0.238695,0.413862)
(-0.248276,0.423477)(-0.257896,0.431981)(-0.267549,0.439452)(-0.277231,0.445965)(-0.286937,0.451585)
(-0.296664,0.456374)(-0.306408,0.460387)(-0.316167,0.463675)(-0.325936,0.466286)(-0.335714,0.468263)
(-0.345499,0.469645)(-0.355287,0.470469)(-0.365079,0.470770)(-0.374871,0.470579)(-0.384663,0.469925)
(-0.394453,0.468835)(-0.404240,0.467335)(-0.414024,0.465448)(-0.423804,0.463195)(-0.433578,0.460598)
(-0.443347,0.457674)(-0.453110,0.454442)(-0.462866,0.450919)(-0.472615,0.447120)(-0.482358,0.443058)
(-0.492093,0.438749)(-0.501821,0.434205)(-0.521255,0.424458)(-0.540658,0.413905)(-0.560032,0.402625)
(-0.579376,0.390686)(-0.598693,0.378150)(-0.617982,0.365073)(-0.646867,0.344548)(-0.675698,0.323060)
(-0.704479,0.300735)(-0.733215,0.277680)(-0.761909,0.253987)(-0.800111,0.221541)(-0.838256,0.188262)
(-0.885867,0.145692)(-0.933415,0.102240)(-0.990402,0.049168)(-1.056813,-0.013746)(-1.132634,-0.086630)
(-1.227330,-0.178780)(-1.350344,-0.299718)(-1.520574,-0.468405)(-1.785271,-0.732132)(-2.456315,-1.402747)
(-4.998604,-3.945010)

\newrgbcolor{color307.0057}{0  0  0}
\psline[plotstyle=line,linejoin=1,showpoints=true,dotstyle=Bo,dotsize=\MarkerSize,linestyle=none,linewidth=\LineWidth,linecolor=color307.0057]
(-1.212256,-0.027473)(-1.212256,-0.027473)

\newrgbcolor{color308.0057}{0  0  0}
\psline[plotstyle=line,linejoin=1,showpoints=true,dotstyle=Bo,dotsize=\MarkerSize,linestyle=none,linewidth=\LineWidth,linecolor=color308.0057]
(-0.000043,-4.034046)(-0.000043,-4.034046)

\newrgbcolor{color309.0057}{0  0  0}
\psline[plotstyle=line,linejoin=1,linestyle=dashed,dash=2pt 3pt,linewidth=\LineWidth,linecolor=color309.0057]
(-0.000009,-4.698970)(-0.000004,-5.000000)
\psline[plotstyle=line,linejoin=1,linestyle=dashed,dash=2pt 3pt,linewidth=\LineWidth,linecolor=color309.0057]
(-5.000000,-0.000004)(-3.259637,-0.000239)(-2.698970,-0.000869)(-2.360514,-0.001898)(-2.118045,-0.003322)
(-1.929224,-0.005142)(-1.774691,-0.007358)(-1.644166,-0.009968)(-1.531357,-0.012969)(-1.432268,-0.016356)
(-1.343998,-0.020129)(-1.264481,-0.024288)(-1.192330,-0.028826)(-1.126389,-0.033741)(-1.065754,-0.039030)
(-1.009750,-0.044688)(-0.957779,-0.050712)(-0.909354,-0.057104)(-0.864104,-0.063858)(-0.821714,-0.070970)
(-0.781885,-0.078438)(-0.744366,-0.086266)(-0.708964,-0.094447)(-0.675512,-0.102978)(-0.643821,-0.111865)
(-0.613751,-0.121111)(-0.585194,-0.130710)(-0.558022,-0.140670)(-0.532140,-0.150992)(-0.507449,-0.161686)
(-0.483861,-0.172760)(-0.461313,-0.184216)(-0.439735,-0.196065)(-0.419052,-0.208323)(-0.399212,-0.221003)
(-0.380166,-0.234115)(-0.361860,-0.247683)(-0.344257,-0.261719)(-0.327311,-0.276249)(-0.310984,-0.291299)
(-0.295241,-0.306898)(-0.280056,-0.323069)(-0.265392,-0.339856)(-0.251231,-0.357288)(-0.237547,-0.375409)
(-0.224324,-0.394253)(-0.211535,-0.413886)(-0.199173,-0.434341)(-0.187220,-0.455684)(-0.175666,-0.477973)
(-0.164500,-0.501276)(-0.153706,-0.525682)(-0.143289,-0.551247)(-0.133234,-0.578084)(-0.123540,-0.606284)
(-0.114204,-0.635956)(-0.105224,-0.667218)(-0.096601,-0.700210)(-0.088331,-0.735088)(-0.080415,-0.772036)
(-0.072856,-0.811240)(-0.065653,-0.852942)(-0.058807,-0.897429)(-0.052321,-0.945004)(-0.046202,-0.996023)
(-0.040448,-1.050952)(-0.035067,-1.110306)(-0.030058,-1.174769)(-0.025424,-1.245193)(-0.021167,-1.322667)
(-0.017295,-1.408490)(-0.013806,-1.504594)(-0.010706,-1.613501)(-0.007995,-1.738975)(-0.005678,-1.886391)
(-0.003755,-2.064997)(-0.002229,-2.290730)(-0.001100,-2.596879)(-0.000365,-3.075721)(-0.000030,-4.154902)
(-0.000009,-4.698970)

\end{pspicture}%
		\caption{\label{fig:parametrized_svd} Evolution diagram 
with time of  scaled perturbation turbulent viscosity  $X_{opt}(t)$ and perturbation velocity $Y_{opt}(t)$, defined
by (\ref{eq:rkt}) and (\ref{eq:ret}) respectively, parameterized by time $t$ for an full-norm optimal perturbation 
(for $T=T_{E_{opt}}$, $T=1.21$ for the STO perturbation and $T=0.12$ for the LTO perturbation) for the optimization
of the total norm $\| \cdot \|_N$ defined
in (\ref{eq:wndef})  using the SVD analysis as described in section (\ref{sec:fullsvd}). The gray curve
plots  the time evolution of the LTO perturbation and the black curve plots the time evolution of the STO perturbation. 
By comparison with figure (\ref{fig:parametrized}), there is apparently a close
relationship between the STO perturbation and the turbulent perturbation from the FULL case, and
the LTO perturbation and laminar perturbation from the FULL case.
}
\end{figure}
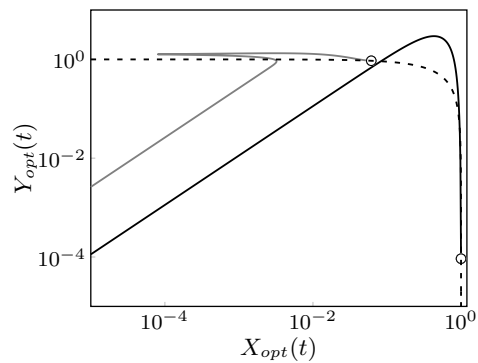

In order to compare these results with the SVD analysis presented 
in section (\ref{sec:fullsvd}), we plot the equivalent parametric curves for an STO perturbation and an LTO perturbation (for the same values of the parameters as before) in figure (\ref{fig:parametrized_svd}). We can see that these two perturbations correspond 
closely to what we have identified using  the optimization of the FULL case using semi-norm constraints.
The STO perturbations appear to  be appropriately classified as ``turbulent'' perturbations, while the LTO
perturbations appear
to be appropriately classified as ``laminar'' perturbations. 

However, even if the SVD analysis appears to give some relevant information, it is completely unable to predict the optimal gain and time. Indeed, the SVD also finds an (unphysical) optimal value for the ratio $C_0$ for all optimization time intervals (e.g. the initial condition used to produce figure (\ref{fig:parametrized_svd})), which prevents the calculated $G_E$ gain from being ``optimal'' in any meaningful sense.

The addition of the dynamic constraint on the turbulent viscosity perturbation brings new dynamical properties to the system of equations, in terms of both optimal time and gain value, compared to the  simpler FROZ case
with a  frozen turbulent viscosity.
Such a method misses a fundamental physical 
process, the possibility of extracting energy from the mean flow due to spatial and temporal variations
in the turbulent viscosity. 
Indeed,  the frozen turbulent viscosity model yields a different result from the full linearized analysis even with an initial ratio $C_0=K_0/E_0 \rightarrow 0$. In other words, in order to capture the true dynamics of the system, analysis of the full linearized system
of equations seems to be indispensable, because of the new source
of perturbation kinetic energy associated with fluctuating turbulent viscosity perturbation $\tilde{\nu}$.

Applying our  semi-norm framework allows us to identify two qualitatively
different perturbations associated with two different dynamics. The first perturbation is driven principally by coherent perturbation velocity 
and hence energy extraction from the mean flow via the term $P_{E_1}$ in (\ref{eq:pert_ener}). 
Such perturbations
do not typically have a large perturbation kinetic energy gain, and we refer to them
as 
``laminar'' perturbations since they typically have small turbulent viscosity perturbation. The other perturbation  is on the contrary essentially driven by a dominant turbulent viscosity perturbation, and is associated with very large energy production. Indeed, the energy gain
 increases linearly with the ratio $C_0=K_0/E_0$ describing the relative contribution of the initial turbulent viscosity to
the initial kinetic energy in the perturbation vector. In essence,  we have performed  a multiscale stability analysis where we are able to control the type of perturbation we imposed to the system, and we find that
small-scale perturbations (parameterized by spatially varying turbulent viscosity) 
are much more efficient
at driving the growth of velocity perturbations (through the term $P_{E_2}$). We can in fact extract as much energy as we want from the base mean flow, given that the magnitude  of the term $\tilde{\nu}\partial_x\tilde{u}$ is large and has the appropriate symmetry.

However, nonlinear effects as well as feedback on the mean flow will certainly 
lead to saturation of this energy production and then lead to an identification of an optimal $C_0$ defining an energy gain ($G_E$) curve which will be different from that which was obtained using the singular value decomposition. We indeed think nonlinear saturation is needed in this particular case to find the optimal initial ratio $C_0(T)$, but another set of linear equations describing a whole different problem might well be suitable for the identification of an optimal initial ratio $C_0(T)$ associated with an optimal energy gain $G_E(T)$ which will be different from the couple ($G_E(T),C_0(T)$) obtained when optimizing $G_N(T)$ through a SVD analysis. Moreover, 
it is reasonable to suppose that, in some particular problems, physical arguments 
may lead to estimates of the appropriate size of the initial turbulent viscosity (or more generally the respective size of the different components of the state vector), and hence
give us a physically acceptable range for the parameter $C_0$.

 The underlying physics described is of course consistent with  SVD analysis optimizing the (non-physical) gain
of the total 2-norm of the perturbation state vector, since we showed that the dynamics of the calculated short time optimal (STO)
perturbation and long time optimal (LTO) perturbation corresponds respectively to the turbulent perturbations
and laminar perturbations dynamics of the full linearized system of equations. 
The new framework yields
detailed
information on the problem and, in fact allowed us to have a full understanding thanks to the possibility of separation of scales in the perturbation vector. Furthermore, 
the framework allows us to approach the  problem from a physical point of view by choosing the type of perturbation 
to impose and then identifying 
the associated dynamics.

\subsection{Sensitivity analysis}

Naturally, we are also able to conduct a 
sensitivity analysis,
 which  will give us some information about how a change in the mean flow $\overline{\mathbf{q}}$ or any of the three parameters $\nu$, $c_1$ or $c_2$ influences the optimal value of the objective functional $\mathcal{J}$.
We focus on the most interesting case 
where $C_0\gg 1$, meaning that we have a large transient growth due to a draining of energy from the velocity mean flow to the velocity perturbation through the catalysis allowed by the optimal symmetry chosen by both $\tilde{u}$ and $\tilde{\nu}$.

\begin{figure}
\includegraphics[width=8cm]{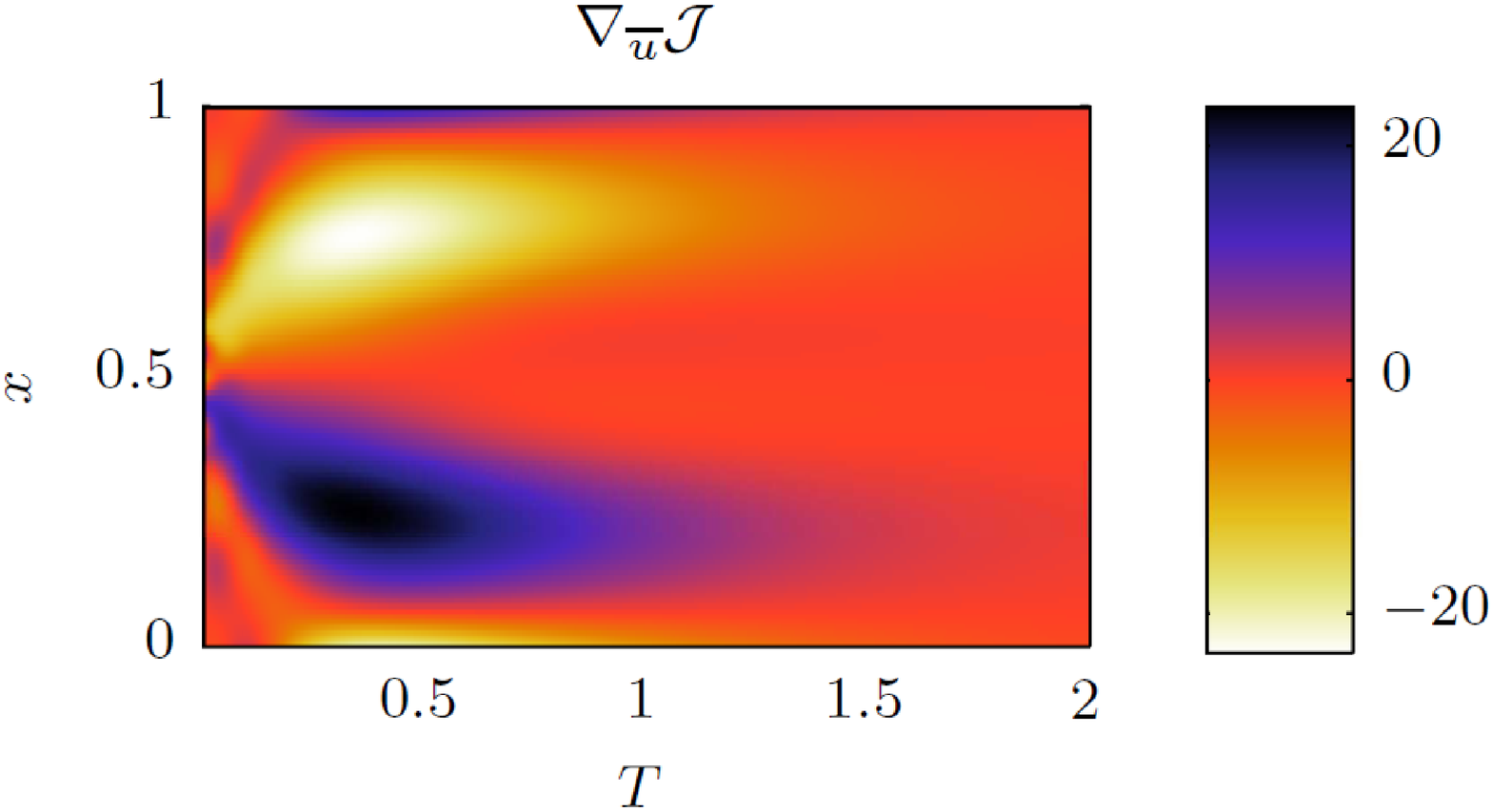}
\includegraphics[width=8cm]{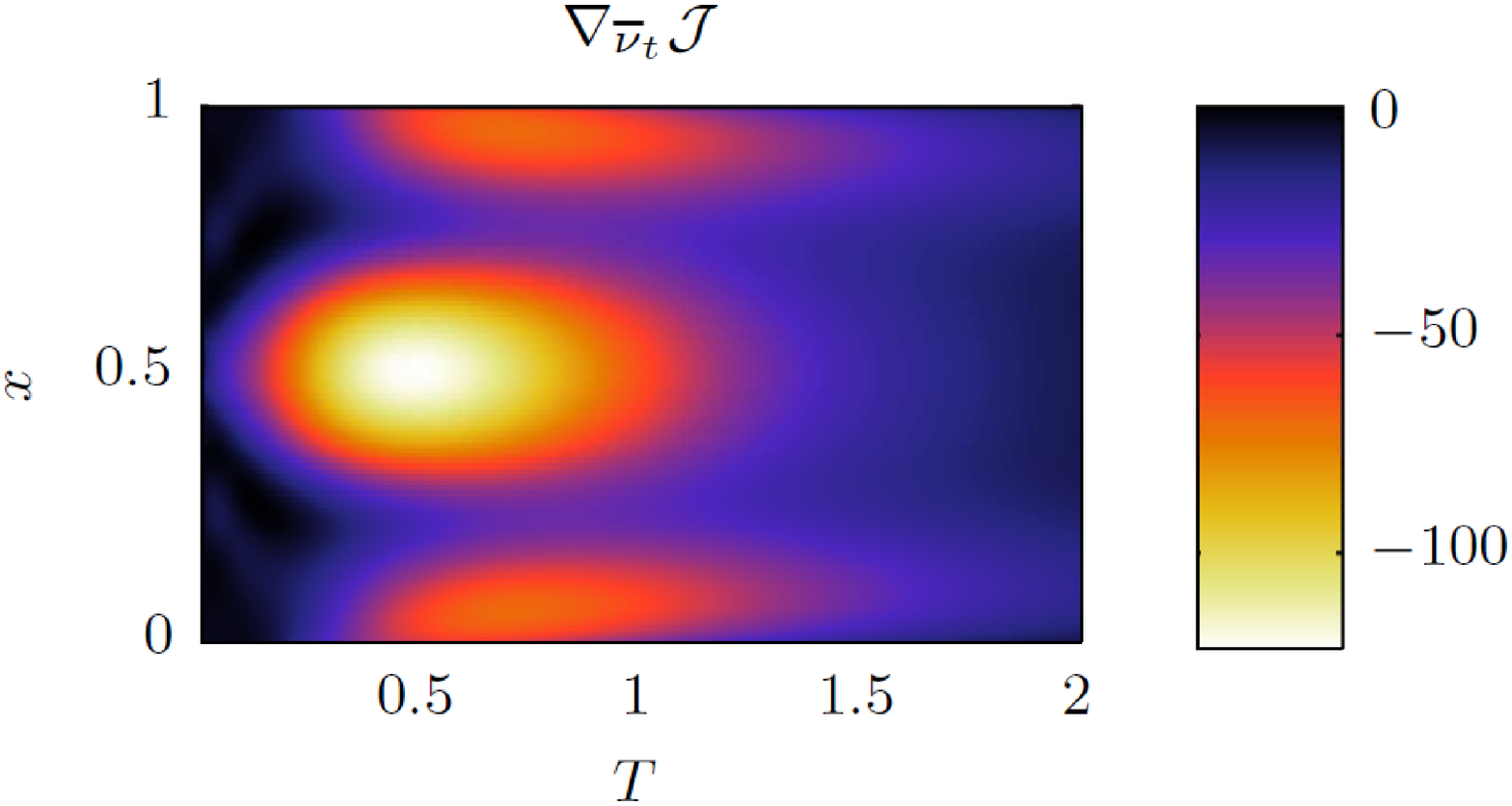}
\caption{\label{fig:sens_base}Space-time plots of the sensitivity of the final optimized energy $E_{opt}(T)$ for varying 
optimization time intervals $T$ with respect to:
(a) the mean flow velocity $\nabla_{\overline{u}}\mathcal{J}$; (b) the mean flow turbulent viscosity $\nabla_{\overline{\nu}_t}\mathcal{J}$ for $C_0 \gg 1$.}
\end{figure}

We plot on figure (\ref{fig:sens_base}a) the sensitivity of the final optimal energy $E_{opt}(T)$
to the mean flow velocity $\nabla_{\overline{u}}\mathcal{J}$ 
and on figure (\ref{fig:sens_base}b) the sensitivity of the final optimal energy
 to the mean flow turbulent viscosity $\nabla_{\overline{\nu}_t}\mathcal{J}$.

The general structure of these space-time diagrams (where time is the optimizing time interval $T$) is useful to understand the role of the base mean flow structure in the production of energy. Indeed, we can see by looking at the general trend of these two figures that the zones of the domain which give the largest sensitivity magnitude are situated on both sides of the middle of the domain for the base mean flow velocity and centred in the middle of the domain for the mean  turbulent viscosity. This can be simply explained by the fact that an increase of the slope of the mean flow velocity (increase of the mean velocity for $0 < x < 1/2$, or decrease for $1/2 < x <1$) will give rise to a larger energy production  (see for instance figure (\ref{fig:gain_semi_norm})). Moreover, since the perturbation is mostly localized in the middle of the domain, an increase in the mean turbulent viscosity will be more dramatic in this zone compared to the edges of the domain.
However, when looking more carefully at the sensitivity functions, we notice that a zone of high sensitivity appears on each side of the domain for $T\simeq0.2$ for the mean flow velocity. These layers have an opposite sign to what we would simply expect to increase the negative slope amplitude $\partial_x \overline{u}$.
In the same way, two other zones of sensitivity appear for times $T\simeq0.4$ in the $\nabla_{\nu_t}\mathcal{J}$ plot and are situated close to the boundaries of the domain. These zones are also associated with negative sensitivity but a smaller absolute value compared to what is observed in the middle of the domain. This is consistent with the zones of high sensitivities at the boundaries in $\nabla_{\overline{u}}\mathcal{J}$ and suggests that the steeper the positive slope at the boundary, the larger the effect on the energy production.

\begin{figure}
\input{grad_nu}
\vspace{0.5cm}
\input{grad_c1}
\vspace{0.5cm}
%
%
%
\providelength{\AxesLineWidth}       \setlength{\AxesLineWidth}{0.5pt}%
\providelength{\plotwidth}           \setlength{\plotwidth}{5cm}
\providelength{\LineWidth}           \setlength{\LineWidth}{0.7pt}%
\providelength{\MarkerSize}          \setlength{\MarkerSize}{4pt}%
\newrgbcolor{GridColor}{0.8 0.8 0.8}%
%
\psset{xunit=0.500000\plotwidth,yunit=0.927894\plotwidth}%
\begin{pspicture}(-0.420000,-0.561656)(2.000000,0.450000)%


\psline[linewidth=\AxesLineWidth,linecolor=GridColor](0.000000,-0.400000)(0.000000,-0.387067)
\psline[linewidth=\AxesLineWidth,linecolor=GridColor](1.000000,-0.400000)(1.000000,-0.387067)
\psline[linewidth=\AxesLineWidth,linecolor=GridColor](2.000000,-0.400000)(2.000000,-0.387067)
\psline[linewidth=\AxesLineWidth,linecolor=GridColor](0.000000,-0.400000)(0.024000,-0.400000)
\psline[linewidth=\AxesLineWidth,linecolor=GridColor](0.000000,-0.200000)(0.024000,-0.200000)
\psline[linewidth=\AxesLineWidth,linecolor=GridColor](0.000000,0.000000)(0.024000,0.000000)
\psline[linewidth=\AxesLineWidth,linecolor=GridColor](0.000000,0.200000)(0.024000,0.200000)
\psline[linewidth=\AxesLineWidth,linecolor=GridColor](0.000000,0.400000)(0.024000,0.400000)

{ \footnotesize 
\rput[t](0.000000,-0.412933){$0$}
\rput[t](1.000000,-0.412933){$1$}
\rput[t](2.000000,-0.412933){$2$}
\rput[r](-0.024000,-0.400000){$-0.4$}
\rput[r](-0.024000,-0.200000){$-0.2$}
\rput[r](-0.024000,0.000000){$0$}
\rput[r](-0.024000,0.200000){$0.2$}
\rput[r](-0.024000,0.400000){$0.4$}
} 

\psframe[linewidth=\AxesLineWidth,dimen=middle](0.000000,-0.400000)(2.000000,0.450000)

{ \small 
\rput[b](1.000000,-0.561656){
\begin{tabular}{c}
$T$\\
\end{tabular}
}

\rput[t]{90}(-0.420000,0.025000){
\begin{tabular}{c}
$\nabla_{c_2}\mathcal{J}$\\
\end{tabular}
}
} 

\newrgbcolor{color219.0328}{0  0  0}
\psline[plotstyle=line,linejoin=1,linestyle=dashed,dash=3pt 2pt 1pt 2pt,linewidth=\LineWidth,linecolor=color219.0328]
(1.999663,-0.079743)(2.000000,-0.079700)
\psline[plotstyle=line,linejoin=1,linestyle=dashed,dash=3pt 2pt 1pt 2pt,linewidth=\LineWidth,linecolor=color219.0328]
(0.010000,-0.000073)(0.024493,-0.000735)(0.038490,-0.002137)(0.052494,-0.004300)(0.066491,-0.007231)
(0.080478,-0.010937)(0.094479,-0.015434)(0.108497,-0.020736)(0.122973,-0.027053)(0.137962,-0.034467)
(0.153483,-0.043034)(0.169970,-0.053072)(0.187448,-0.064663)(0.206989,-0.078608)(0.229937,-0.096045)
(0.260995,-0.120851)(0.323430,-0.171682)(0.351449,-0.193641)(0.375410,-0.211482)(0.399431,-0.228387)
(0.422413,-0.243662)(0.440383,-0.254786)(0.455968,-0.263585)(0.470416,-0.270871)(0.488365,-0.278940)
(0.524878,-0.293874)(0.547894,-0.302683)(0.565443,-0.308648)(0.582836,-0.313734)(0.601917,-0.318465)
(0.622373,-0.322696)(0.642841,-0.326130)(0.663457,-0.328790)(0.684306,-0.330682)(0.705949,-0.331854)
(0.728807,-0.332284)(0.752402,-0.331923)(0.776805,-0.330758)(0.802396,-0.328744)(0.829776,-0.325773)
(0.860000,-0.321651)(0.890766,-0.316674)(0.920830,-0.311011)(1.006945,-0.293385)(1.034707,-0.286860)
(1.120794,-0.264744)(1.305394,-0.215623)(1.375175,-0.197689)(1.447208,-0.179813)(1.475151,-0.173610)
(1.521700,-0.163835)(1.580566,-0.150737)(1.647168,-0.137014)(1.706907,-0.125483)(1.768725,-0.114366)
(1.830993,-0.103984)(1.894564,-0.094202)(1.959482,-0.085027)(1.999663,-0.079743)

\newrgbcolor{color220.0323}{0  0  0}
\psline[plotstyle=line,linejoin=1,linestyle=dashed,linewidth=\LineWidth,linecolor=color220.0323]
(1.999663,0.037249)(2.000000,0.037225)
\psline[plotstyle=line,linejoin=1,linestyle=dashed,linewidth=\LineWidth,linecolor=color220.0323]
(0.010000,0.009167)(0.021495,0.019935)(0.034490,0.033269)(0.048992,0.049352)(0.066491,0.070046)
(0.093478,0.103586)(0.132962,0.152940)(0.197962,0.232948)(0.220958,0.260211)(0.239936,0.281604)
(0.256986,0.299722)(0.272444,0.315076)(0.287420,0.328886)(0.301457,0.340818)(0.314971,0.351331)
(0.328414,0.360814)(0.341412,0.369062)(0.354466,0.376440)(0.367451,0.382885)(0.380396,0.388440)
(0.393405,0.393170)(0.406475,0.397086)(0.419431,0.400168)(0.432878,0.402577)(0.446908,0.404302)
(0.461983,0.405345)(0.477378,0.405623)(0.493380,0.405116)(0.509495,0.403825)(0.526370,0.401684)
(0.544373,0.398585)(0.563463,0.394475)(0.583835,0.389268)(0.606462,0.382629)(0.631824,0.374305)
(0.660494,0.363994)(0.694337,0.350907)(0.738798,0.332721)(0.865906,0.279044)(0.940250,0.248634)
(0.962455,0.240235)(0.993258,0.229757)(1.022785,0.219630)(1.070803,0.202460)(1.109490,0.189604)
(1.155407,0.175162)(1.197766,0.162660)(1.242186,0.150388)(1.286639,0.138932)(1.332625,0.127903)
(1.377150,0.118021)(1.435094,0.106087)(1.464867,0.100599)(1.541605,0.088161)(1.585552,0.081176)
(1.645639,0.072616)(1.703304,0.065193)(1.765325,0.058032)(1.830497,0.051328)(1.899680,0.045036)
(1.973076,0.039182)(1.999663,0.037249)

\newrgbcolor{color221.0323}{0  0  0}
\psline[plotstyle=line,linejoin=1,linestyle=solid,linewidth=\LineWidth,linecolor=color221.0323]
(1.999663,-0.042494)(2.000000,-0.042475)
\psline[plotstyle=line,linejoin=1,linestyle=solid,linewidth=\LineWidth,linecolor=color221.0323]
(0.010000,0.009094)(0.026493,0.024039)(0.050993,0.047624)(0.082976,0.078712)(0.100485,0.094813)
(0.114490,0.106762)(0.128965,0.118085)(0.143966,0.128839)(0.160498,0.139709)(0.176955,0.149641)
(0.192452,0.158114)(0.206487,0.164957)(0.219962,0.170712)(0.233434,0.175625)(0.246448,0.179551)
(0.258995,0.182551)(0.271447,0.184749)(0.283920,0.186178)(0.296438,0.186845)(0.308994,0.186754)
(0.321936,0.185880)(0.335405,0.184176)(0.349440,0.181605)(0.364470,0.178026)(0.379897,0.173529)
(0.395412,0.168200)(0.411488,0.161856)(0.431380,0.153021)(0.482365,0.129214)(0.505964,0.117933)
(0.525375,0.107760)(0.581838,0.076350)(0.621381,0.055324)(0.654942,0.038266)(0.684306,0.024189)
(0.712467,0.011537)(0.740303,-0.000114)(0.767398,-0.010610)(0.794314,-0.020198)(0.821343,-0.028997)
(0.848340,-0.036965)(0.875788,-0.044234)(0.904919,-0.051107)(0.933736,-0.057149)(0.958474,-0.061592)
(0.984721,-0.065424)(1.038715,-0.071978)(1.070803,-0.075224)(1.104901,-0.077810)(1.143726,-0.079919)
(1.183665,-0.081275)(1.227188,-0.081913)(1.274705,-0.081774)(1.326668,-0.080789)(1.387614,-0.078768)
(1.488089,-0.074104)(1.544137,-0.071042)(1.695628,-0.061463)(1.896095,-0.048633)(1.999663,-0.042494)

\end{pspicture}%
\center
\includegraphics[scale=0.5]{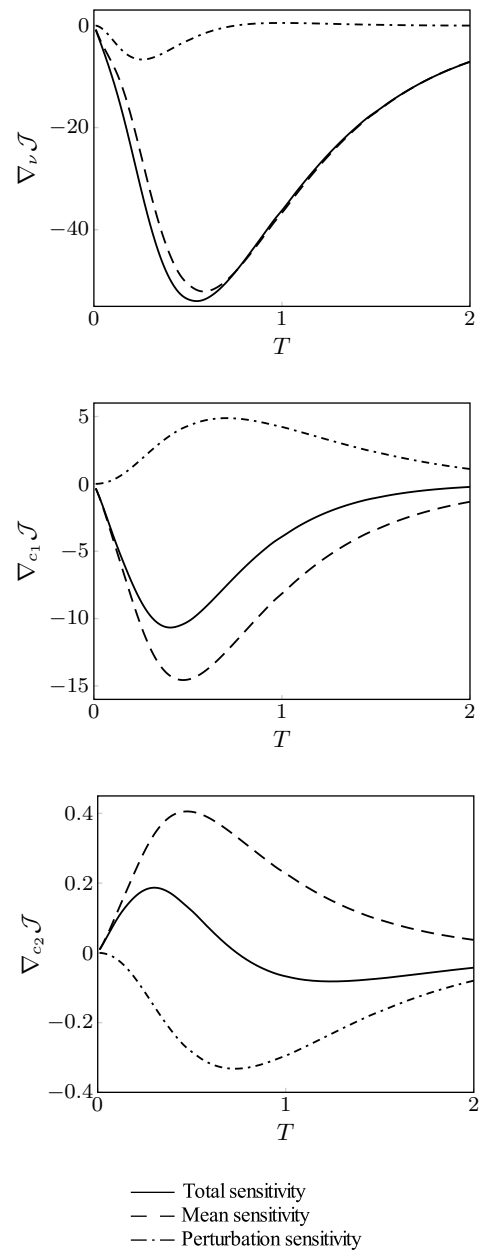}
\caption{\label{fig:sens_param}Variation of the sensitivity of the final optimal energy $E_{opt}(T)$ with optimization time interval $T$ to:
(a)  the viscosity $\nabla_{\nu}\mathcal{J}$;
(b)  the turbulent viscosity production coefficient $\nabla_{c_1}\mathcal{J}$;
(c) the turbulent viscosity destruction coefficient $\nabla_{c_2}\mathcal{J}$. 
Solid lines show the variation
with optimization time interval of the total sensitivities, dashed lines 
show the variation of the mean flow 
sensitivities, and 
dashed-dotted lines show the variation
of the perturbation
sensitivities.
Increasing $\nu$ decreases the value of the objective functional $\mathcal{J}$, and mean and perturbation contributions act in the same way, with a larger sensitivity due to the perturbation. A positive variation of $c_1$ also leads to a decrease of the objective functional,
 although the perturbation and mean flow sensitivity act  oppositely,
with the perturbation sensitivity being
positively correlated with the total sensitivity. The change in $\mathcal{J}$ due to a variation in $c_2$ is opposite (in both mean and perturbation sensitivity) to the one observed with $c_1$, but is an order of magnitude smaller.}
\end{figure}

In figure (\ref{fig:sens_param}), we plot  the sensitivities to the constraint parameters of the problem. For the mean flow, both  $\nabla_{\nu}\mathcal{J}$ and $\nabla_{c_1}\mathcal{J}$ are negative since they are respectively associated with dissipation and production of viscosity (indirect dissipation), while for $\nabla_{c_2}\mathcal{J}$, this term is positive since the destruction of viscosity decreases the energy dissipation of the mean flow.
Interestingly, there is qualitatively 
different behaviour when we look at the perturbation contribution. Indeed, except for $\nu$ for which the sign of the sensitivity is the same for mean and perturbation sensitivities, we observe an opposite trend between these two contributions for $c_1$ and $c_2$ sensitivities. As we have seen, the base flow total viscosity (composed of both laminar and turbulent viscosity) is only responsible for decay while the perturbation turbulent viscosity
actually  triggers large energy growth. As a consequence, increasing base flow turbulent viscosity will lead to a decrease of $\mathcal{J}$ whereas adding some perturbation turbulent viscosity will boost the energy growth. The trends can now be explained by noticing that this process of production and destruction of either base or perturbation turbulent viscosity is driven by the parameters $c_1$ and $c_2$.
However, it is important to notice that the total sensitivities are separated by an order of magnitude, with:
\begin{equation}
\nabla_{\nu}\mathcal{J} \gg \nabla_{c_1}\mathcal{J} \gg \nabla_{c_2}\mathcal{J},
\end{equation}
and so 
the turbulence modelling parameters actually have a relatively minor impact on the dynamics of the perturbation
compared to modifications in the viscosity.

\subsection{Discussion}

Interestingly, although the conventional SVD analysis is  not 
a formally  correct method to study optimal energy growth in this system where the energy is a semi-norm of the state vector, it still yields valuable insight, in particular  that the frozen turbulent viscosity assumption is not appropriate. Indeed, we find a mode associated with a large turbulent viscosity perturbation amplitude which triggers, as we expected from the energy analysis carried out in section (\ref{sec:RAB}), a new energy production mechanism and gives rise to a very large transient growth of the energy.
The interpretation and validation of the results of the SVD analysis however relies on the use of the semi-norm 
framework presented in this paper. 
The semi-norm framework allows us to conclude that we can linearly extract energy from the turbulent viscosity component, and the larger its amplitude, the larger the optimal energy at the end of the optimization interval. We can identify two qualitatively
different types of behaviour depending on the initial value $C_0$ of the ratio of the turbulent viscosity perturbation amplitude to the velocity perturbation amplitude. When this ratio is small, the energy production is small and is only due to the interaction between the mean flow velocity and the perturbation velocity, and the perturbation is essentially ``laminar''. However, for large values of this parameter, a new behaviour  appears, and such ``turbulent'' perturbations exhibit very large
perturbation kinetic energy gain for which the gain evolves linearly with $C_0$ suggesting some universal behaviour.
Finally, from  a sensitivity analysis,  the influence of the modelling parameters can be shown to be very small compared to the influence of the laminar viscosity parameter on the flow evolution.	
	
\section{Extensions}\label{sec:EXT}

This paper is, to the best of our knowledge, the first attempt to treat formally the semi-norm gain optimization problem. 
We address this problem by using a Lagrangian variational framework, which instead of having a single initial normalization constraint, has complementary  semi-norm  constraints allowing us to control the relative contributions of the different components in the perturbation vector. The gain is then trivially computed by forming the ratio of the semi-norm at initial time (chosen through normalization), to its value at the final time (resulting from the optimization). The optimization procedure can then be repeated for different values of the parameter 
quantifying the relative size of the initial amplitudes of the different components of the state vector.


This kind of multiscale nonmodal stability analysis is of a particular interest for systems where the definition of the energy can come from different physical contributions. Our framework then provides a systematic procedure to separate the different energy contributions and optimize any desired gain, even defined with semi-norms. 
 A few of the problems where nonmodal stability analysis is being applied, and where the multiscale stability analysis presented here could be appropriate include turbulent mean flows (\cite{crouch}, \cite{cossu}), compressible flows and thermoacoustics (\cite{hanifi}, \cite{juniper}, \cite{sujith}, \cite{Selimefendigil}), Rayleigh-Bénard type flows with density gradient due to temperature effects (\cite{soundar}), coupled fluid and electric field systems (\cite{Castellanos}, \cite{fulvio}), magnetohydrodynamics (\cite{chen}), and irreversible mixing in density stratified flows (\cite{ivey}, \cite{caulfield}).

Moreover, more than just being a way to calculate semi-norms gain, the method can be used in other kinds of problems. Indeed, in flow control, the optimal placing and type of action of actuators could be derived from the optimization of conventional objective functionals (energy, drag, etc) with constraints on semi-norms defined on a compact support of the domain.  Constraining the other part of the domain, choosing a large value for the semi-norm defined on the area of interest and a small value for the complementary semi-norm, we would be able to find the optimal localized forcing.
Finally, 
although in this paper we focussed on semi-norms
being used to define the objective functional, 
 properties (\ref{eq:norm1}) and (\ref{eq:norm2}) of semi-norms were actually not used in the development of our framework. We developed our  framework in terms  of semi-norms because of their relevance
to fluid dynamics problems. Indeed, these two properties
are not essential to the framework, and it is straightforward
to generalize the framework for arbitrary functionals $f$,
developing appropriate semi-norms to constrain the magnitude of elements of the entire state
vector space (in particular the kernel of $f$).

\section{Conclusion}\label{sec:CON}
	
	In this paper, we  develop a  general Lagrangian variational framework 
for optimization problems using semi-norm constraints.
We present 
a systematic way to study optimal gains defined in terms of the state vector, with the introduction of new parameters setting the ratio between the different components of the perturbation state vector. This framework is a way to perform multi-scale stability analysis, where the different components of the perturbation state vector do not have (necessarily) the same amplitude (which is relevant for multiphysics problems). 
To demonstrate the utility of this framework,  we consider a simple idealized problem with 
a coupled set of two Burgers equations describing the evolution of flow velocity and a transport equation for a turbulent viscosity for which the production and destruction are controlled through two modelling parameters. This  constitutes a minimal set to describe much of the key physics underlying the Reynolds-Averaged Navier Stokes equations and also some properties of mean flows in closed nonlinear dissipative systems. The Reynolds-Averaged Burgers (RAB) equations are a simple one-dimensional approximation of a RANS equation for which a Boussinesq hypothesis of turbulence is considered, and a new turbulent viscosity is introduced, governed by a transport equation, reminiscent of the Spalart-Allmaras turbulence model \cite{spalart}. The nonmodal stability analysis of this system 
also allows us to investigate
the usefulness of the 
assumption that the turbulent viscosity is ``frozen'' at a constant value.

After deriving the perturbation equations for this system, we  perform the stability analysis in three different cases: the laminar case (``LAM'') where the turbulent viscosity was zero, the frozen turbulent viscosity case (``FROZ'')
where the turbulent viscosity was set at a constant value, and the fully linearized analysis case (``FULL'') where a turbulent viscosity perturbation was considered. The results obtained in the first two cases show that the only effect of the frozen turbulent viscosity is to add spatially varying damping in the system and thus a straightforward decrease of the energy gain.
For the full linearized problem,  the analysis of the total gain optimal (i.e. optimizing the 2-norm of the state vector) using SVD analysis establishes the presence of a new type of perturbation driven largely by turbulent viscosity perturbation effects, and associated with substantially larger gains than  was found in the two simpler cases. The sub-optimal perturbation
identified by the SVD analysis is mainly driven by  perturbation velocity and is very close in terms of perturbation structure and gain to the optimal perturbation identified for the frozen turbulent viscosity case. 

The use of the semi-norm gain optimization framework developed in this paper allows us to investigate the dependence of the different types of perturbations possible on the optimization method. Indeed, instead of obtaining the relative contribution of turbulent viscosity and mean flow in the perturbation vector as a result of the optimization of the most obvious 2-norm of the initial perturbations, we can, thanks to the new framework, consider this as an input of the optimization problem and then investigate it in a much deeper way. 

In the limit of very low turbulent viscosity perturbation to mean flow perturbation ratio,
(i.e. $C_0=K_0/E_0 \ll 1$) the results are (in order of magnitude) similar to the frozen turbulent viscosity case, though the fully linearized model never converges toward the frozen turbulent viscosity case's behaviour, 
even as this ratio tends to zero,  clearly illustrating the singularity of the problem. On the other hand, in the limit of very ``turbulent'' perturbations, (i.e. when $C_0 \gg 1$)   we show that a second type of behaviour arises, leading to a large transfer of energy from the base mean flow to the  perturbation velocity, catalysed by substantial variation in the turbulent viscosity. This transfer of energy is possible because the  turbulent viscosity perturbation adopts an optimal shape in order to extract energy from the mean flow. More precisely, we show that the perturbation energy gain evolves linearly with the ratio between the two components of the state vector which seems to be an universal behaviour. 

These results show that the semi-norm framework is an interesting way to retrieve the physics given by the modal decomposition of a SVD analysis, through physics considerations, rather than mathematical arrangements. Besides, a sensitivity analysis of the system shows that the influence of modelling parameters (in particular the production and destruction of turbulent viscosity) is smaller than the influence of physical parameters (the viscosity). Moreover, for turbulent perturbations,
we show that the mean and perturbation sensitivities with respect to modeling parameters have a different sign, meaning they are competing.

The main conclusion of our
investigation of this model problem is that the frozen turbulent viscosity assumption might be  relevant if the perturbation in turbulent viscosity is very low compared to the magnitude of the mean flow perturbation,
although there still
appears to be nontrivial quantitative
differences. In the other limit,
when the perturbation
turbulent viscosity has significant initial 
magnitude,  
we clearly conclude that  frozen turbulent viscosity  is unable to describe the real dynamics of a perturbation governed by the full perturbation RAB equations (\ref{eq:pert_linearized}). Therefore, we believe frozen turbulent
viscosity is highly unlikely to describe correctly more complicated systems such as the Reynolds-Averaged Navier-Stokes (RANS) equations since our one-dimensional model (\ref{eq:RAB_sys}) possesses many of the key features of the RANS equations: i.e. time dependence, advection effects, a dissipative nature and a closed nonlinearity. It is important to stress that the variational framework employed here also gives as an output the sensitivity with respect to the constraints of the problem, and thus offers a powerful analysis tool.
Finally, we wish to reiterate  that the problem chosen here (optimal perturbation gain defined in terms of a semi-norm of the state vector) is one of the simplest we could have imagined and was chosen in order to present this method in a (hopefully) pedagogical way. However, non-linearity, time averaging norms, non-autonomous operators and more can be added to the framework with minor impact on the algorithmic approach.


	\appendix

\section{Linearized RAB operators}\label{app:op}

\subsection{Direct operator}

The direct linearized Reynolds Averaged Burgers operator defined in section (\ref{sec:RAB}) in equation (\ref{eq:RAB_sys}) is defined as follows:
\begin{equation}
\tilde{\mathbf{L}}=\left(
\begin{array}{cc}
\tilde{L}_{11} & \tilde{L}_{12} \\
\tilde{L}_{21} & \tilde{L}_{22} 
\end{array}
\right),
\end{equation}
with the following corresponding block matrices:
\begin{equation}
\begin{array}{lcl}
\tilde{L}_{11} & = &    -  \overline{u}\partial_x  - \partial_x\overline{u}   + (\nu+\nu_t)\partial_{xx} + \partial_x\nu_t\partial_x,\\
\tilde{L}_{12} & = & \partial_x\overline{u}\partial_x + \partial_{xx}\overline{u},\\
\tilde{L}_{21} & = &  c_1 \sgn(\partial_x\overline{u})\nu_t\partial_x - \partial_x\nu_t,\\
\tilde{L}_{22} & = & - \overline{u}\partial_x + \partial_{xx} \nu_t + 2\partial_x\nu_t\partial_x +(\nu+\nu_t)\partial_{xx}\\
& ~ & +c_1\left|\partial_x\overline{u}\right| - 2c_2\nu_t.
\end{array}
\end{equation}
We notice that the matrix $\tilde{L}_{11}$ corresponds to the frozen turbulent viscosity equation.

\subsection{Adjoint operator}

The linear adjoint Reynolds Averaged Burgers operator defined in section (\ref{sec:RAB}) in equation (\ref{eq:adjoint}) is defined as follows:
\begin{equation}
\tilde{\mathbf{L}}^\dagger = \left(
\begin{array}{cc}
\tilde{L}^\dagger_{11} & \tilde{L}^\dagger_{12} \\
\tilde{L}^\dagger_{21} & \tilde{L}^\dagger_{22} 
\end{array}
\right),
\end{equation}
with the following corresponding block matrices:
\begin{equation}
\begin{array}{lcl}
\tilde{L}^\dagger_{11}			& = &      \overline{u}\partial_x  + (\nu+\nu_t)\partial_{xx} + \partial_x\nu_t\partial_x,\\
\tilde{L}^\dagger_{12} 	  & = & 		 - \partial_x\nu_t - c_1\sgn(\partial_x\overline{u})(\nu_t\partial_x+\partial_x\nu_t),\\
\tilde{L}^\dagger_{21}     & = & 		 - \partial_x\overline{u}\partial_x,\\
\tilde{L}^\dagger_{22}			& = & 		 \overline{u}\partial_x + \partial_x\overline{u}  + (\nu+\nu_t)\partial_{xx} + c_1\left|\partial_x\overline{u}\right| \\ 
& ~ & - 2c_2\nu_t.
\end{array}
\end{equation}

\section{Expression of sensitivity functions and matrices}\label{app:sens}

\subsection{Base flow sensitivity functions}

The sensitivity functions defined in (\ref{eq:sens_bf}), describing
the sensitivity of the objective functional 
${\mathcal{J}}$ defined in equation (\ref{eq:objfdef}) to a change in
the base mean flow
$\overline{\mathbf{q}}=(\overline{u},\overline{\nu}_t)$ are defined as:
\begin{equation}
\begin{array}{ll}
\mathbf{S}_{\overline{u}}(\tilde{\mathbf{q}}) = & \tilde{u}\partial_x \tilde{u}^\dagger + \tilde{\nu}\partial_{xx}\tilde{u}^\dagger + \partial_x\tilde{\nu}\partial_x\tilde{u}^\dagger - \tilde{\nu}^\dagger \partial_x \tilde{\nu}\\
 & + c_1\sgn\left( \partial_x \overline{u}\right) \left(\tilde{\nu}^\dagger\partial_x \tilde{\nu} + \tilde{\nu}\partial_x \tilde{\nu}^\dagger\right).
\end{array}
\end{equation}

\begin{equation}	
\begin{array}{ll}
\mathbf{S}_{\nu_t}(\tilde{\mathbf{q}}) =  & -\partial_x \tilde{u} \partial_x\tilde{u}^\dagger + \tilde{\nu}\partial_{xx}\tilde{\nu}^\dagger  - 2c_2 \tilde{\nu}^\dagger \tilde{\nu} + \tilde{u}\partial_x\tilde{\nu}^\dagger \\
& + \tilde{\nu}^\dagger\partial_x\tilde{u} - c_1\sgn\left(\partial_x\overline{u}\right)\tilde{\nu}^\dagger\partial_x\tilde{u}.
\end{array}
\end{equation}

\subsection{Parameter sensitivity vectors and matrices}

\paragraph{Base mean flow contribution.}

The sensitivity vectors defined in (\ref{eq:sens_param}), describing
the change of the objective functional 
${\mathcal{J}}$ defined in equation (\ref{eq:objfdef}) 
under a change
of the parameters in the base mean flow equations (\ref{eq:RAB_sys}) are defined as:
\begin{equation}
\overline{\mathbf{S}}_{\nu}(\overline{\mathbf{q}}) = \left(
\begin{array}{c}
\partial_{xx} \overline{u} \\
\partial_{xx} \nu_t
\end{array}
\right).
\end{equation}

\begin{equation}
\overline{\mathbf{S}}_{c_1}(\overline{\mathbf{q}}) = \left(
\begin{array}{c}
0\\
-\left|\partial_x \overline{u} \right| \nu_t
\end{array}
\right),
\end{equation}

\begin{equation}
\overline{\mathbf{S}}_{c_2}(\overline{\mathbf{q}}) = \left(
\begin{array}{c}
0\\
\nu_t^2
\end{array}
\right).
\end{equation}

\paragraph{Perturbation contribution.}

The sensitivity vectors defined in (\ref{eq:sens_param}), describing
the change of the objective functional
${\mathcal{J}}$ defined in equation (\ref{eq:objfdef}) 
  under a change of the parameters in the perturbation equations
  (\ref{eq:pert_linearized}) are defined as:
\begin{equation}
\tilde{\mathbf{S}}_{\nu} = \left(
\begin{array}{cc}
\partial_{xx} & 0 \\
0 					 & \partial_{xx}
\end{array}
\right),
\end{equation}

\begin{equation}
\tilde{\mathbf{S}}_{c_1} = \left(
\begin{array}{cc}
0 & 0\\
- \sgn\left(\partial_x \overline{u}\right)\nu_t \partial_x 					 &  - \left|\partial_x\overline{u}\right|
\end{array}
\right),
\end{equation}

\begin{equation}
\tilde{\mathbf{S}}_{c_2} =  \left(
\begin{array}{cc}
0 & 0\\
0					 &  2\nu_t
\end{array}
\right).
\end{equation}

\section{Optimal perturbation with Singular Value Decomposition (SVD)}\label{app:svd}

Given that the flow is stable for any value of the viscosity (the flow
tends toward neutral stability when $\nu$ becomes small), we expect to
be able to capture the properties of the dynamics  by focussing on the transient growth mechanisms involved.
For a general linear equation (discretized) of the type $d_t
\mathbf{q} = \mathbf{L} q$, we can derive the exact solution at time
$T$, which will from now on be called the horizon time. This solution
can simply be expressed in term of the evolution operator $\mathbf{M}$
(which is the matrix exponential of $\mathbf{L}T$) and the initial
condition $\mathbf{q}(0)=\mathbf{q}_0$, i.e.
\begin{equation}
\mathbf{q}(T)=\mathbf{M}\mathbf{q}_0=e^{T\mathbf{L}}\mathbf{q}_0.
\end{equation}
Let us now define the energy of a state vector as the following weighted scalar product:
\begin{equation}
\left\| \mathbf{q} \right\|_E^2 = \mathbf{q}^H\mathbf{W}\mathbf{q}.
\end{equation}
The maximum gain we can achieve for a time $T$ is simply expressed as an optimization problem:
\begin{equation}
G(T) = \max_{\mathbf{q}_0\neq0} \dfrac{\|\mathbf{q}(T)\|_E^2}{\|\mathbf{q}_0\|_E^2}= \max_{\mathbf{q}_0\neq0} \dfrac{\|e^{T\mathbf{L}}\mathbf{q}_0\|_E^2}{\|\mathbf{q}_0\|_E^2}.
\end{equation}
Now, assuming that $\mathbf{W}$ is a symmetric positive definite matrix, we use a Cholesky decomposition to write:
\begin{equation}
\mathbf{W} = \mathbf{F}^H\mathbf{F}.
\end{equation}
We then change the variable from $\mathbf{q}$ to $\mathbf{q}'$ defined as:
\begin{equation}
\mathbf{q}' = \mathbf{F} \mathbf{q}.
\end{equation}
With this small transformation, we are able to write the gain in the following way:
\begin{equation}
G(T) = \max_{\mathbf{q}_0'\neq0} \dfrac{\left\|\mathbf{F}e^{T\mathbf{L}}\mathbf{F}^{-1}\mathbf{q}_0'\right\|_2}{\left\|\mathbf{q}_0'\right\|_2},
\end{equation}
which is simply the 2-norm of the matrix $\mathbf{M}' =
\mathbf{F}e^{T\mathbf{L}}\mathbf{F}^{-1}$, which  can be computed through a singular value decomposition.
The output of the SVD is a diagonal matrix $\mathbf{\Sigma}$, consisting of all the singular values which are real positive numbers, and two orthogonal matrices $\mathbf{U}'$ and $\mathbf{V}'$:
\begin{equation}
\mathbf{M}'=\mathbf{\mathbf{U}' \Sigma \mathbf{V}'^H},
\end{equation}
The norm of the matrix (and as a consequence the gain) is given by the
largest singular value (i.e. the first coefficient in $\mathbf{\Sigma}$), while the first column of $\mathbf{V}=\mathbf{F}^{-1}\mathbf{V}'$ gives the optimal perturbation $\mathbf{q}_0$ and the first column of $\mathbf{U}=\mathbf{F}^{-1}\mathbf{U}'$ give the optimal state at the horizon time $T$.
With this method, we are able to find the optimal gain for different
horizon times, as well as the associated  optimal perturbation.


\bibliographystyle{ieeetr} 
\bibliography{biblio} 

\end{document}